\documentclass[twocolumn, twocolappendix]{aastex631} 

\usepackage{mathtools}
\usepackage{float}
\graphicspath{{./}{figures/}}

\begin{document}

\title{Evolution, structure and topology of self-generated turbulent reconnection layers}

\author{Raheem Beg}
\affiliation{Division of Mathematics, University of Dundee, Dundee DD1 4HN, United Kingdom}

\author{Alexander J. B. Russell}
\affiliation{Division of Mathematics, University of Dundee, Dundee DD1 4HN, United Kingdom}

\author{Gunnar Hornig}
\affiliation{Division of Mathematics, University of Dundee, Dundee DD1 4HN, United Kingdom}




\begin{abstract}
We present a 3D MHD simulation of two merging flux ropes exhibiting self-generated and self-sustaining turbulent reconnection (SGTR) that is fully 3D and fast. The exploration of SGTR is crucial for understanding the relationship between MHD turbulence and magnetic reconnection in astrophysical contexts including the solar corona. We investigate the pathway towards SGTR and apply novel tools to analyse the structure and topology of the reconnection layer. The simulation proceeds from 2.5D Sweet-Parker reconnection to 2.5D nonlinear tearing, followed by a dynamic transition to a final SGTR phase that is globally quasi-stationary. The transition phase is dominated by a kink instability of a large ``cat-eye'' flux rope and the proliferation of a broad stochastic layer. The reconnection layer has two general characteristic thickness scales which correlate with the reconnection rate and differ by a factor of approximately six: an inner scale corresponding with current and vorticity densities, turbulent fluctuations, and outflow jets, and an outer scale associated with field line stochasticity. The effective thickness of the reconnection layer is the inner scale of the effective reconnection electric field produced by turbulent fluctuations, not the stochastic thickness. The dynamics within the reconnection layer are closely linked with flux rope structures that are highly topologically complicated. Explorations of the flux rope structures and distinctive intermediate regions between the inner core and stochastic separatrices (``SGTR wings'') are potentially key to understanding SGTR. The study concludes with a discussion on the apparent dualism between plasmoid-mediated and stochastic perspectives on SGTR.
\end{abstract}



\section{Introduction} \label{sec:intro}

Magnetic reconnection is the fundamental process where field lines in a magnetised plasma change their topology \citep{PriestForbes2000}, and is widely agreed to be a crucial phenomenon behind explosive dynamic events in the solar corona, such as solar flares and coronal heating \citep[e.g.,][]{Sturrock1966, KarpenEA2012, Klimchuk2015, WyperEA2017}. This has been the topic of intense research in recent years, as state-of-the-art numerical simulations have provided mounting evidence that reconnection in high Lundquist number plasma is an intrinsically three-dimensional (3D) process where magnetohydrodynamic (MHD) turbulence plays a crucial role \citep[e.g.,][]{MatthaeusLamkin1985, MatthaeusLamkin1986, LazarianVishniac1999, LoureiroEA2009, ServidioEA2009, ServidioEA2010, ServidioEA2011a, EyinkEA2011, EyinkEA2013, KowalEA2009, KowalEA2012}. The connections between turbulence and magnetic reconnection are currently a highly active field, in which the inherent coupling between two highly nonlinear, dynamic and multiscale 3D processes make investigations challenging, both analytically and numerically \citep[see reviews by][and references therein]{LazarianEA2015, ZweibelYamada2016, LazarianEA2020, JiEA2022}.

Reconnection research was initially centred around the Sweet-Parker model \citep{Parker1957, Sweet1958}, which considered a simple laminar two-dimensional (2D) configuration involving a thin current sheet of length $L$ and uniform thickness $\delta$. The reconnection process is assumed to achieve a laminar steady state, with clear inflow and outflow regions, and the reconnection rate and aspect ratio are found to scale as $V_{\mathrm{rec}} \approx v_A S^{-1/2}$ and $\delta/L \approx S^{-1/2}$, respectively, where $v_A$ is the Alfv\'en speed and $S = Lv_A/\eta$ is the local Lundquist number \citep{PriestForbes2000}. While the Sweet-Parker model offers a useful analytic framework for a basic laminar reconnection process, it has many well known limitations. Lundquist numbers found in astrophysical environments are huge, with typical values of order $S \sim 10^{12}\mbox{--}10^{14}$ in the solar corona \citep{HuangEA2017} and $S \sim 10^{16}$ in the interstellar medium (ISM) \citep{KowalEA2009}. Observations of the solar corona reveal much faster reconnection rates estimated up to at least $V_{\mathrm{rec}} \sim 10^{-3}v_A$  \citep{PriestForbes2000}, so the Sweet-Parker prediction has been argued to be unrealistic in most astrophysical contexts \citep[e.g.,][]{JafariEA2018}. At the same time, many solar flare observations are consistent with the global magnetic topology of Sweet-Parker reconnection, which leads to the goal of retaining this magnetic topology at the largest scales, while exploring solutions to the rate problem.

Research therefore shifted towards the development of a ``fast'' reconnection model which could yield a reconnection rate that is a significant enhancement from the ``slow'' reconnection rate predicted by Sweet-Parker. Here, ``fast'' means that the reconnection rate is independent of, or weakly dependent on, the Lundquist number \citep{PriestForbes2000}. Turbulence has long been viewed as a promising explanation for how reconnection becomes fast, and this idea was formalised by \citet{LazarianVishniac1999} who generalised the Sweet-Parker model by combining its large-scale topology with injected 3D MHD turbulence consistent with \citet{GoldreichSridhar1995} theory. Here, reconnection was conjectured to be significantly enhanced via field line wandering or stochasticity, which allows multiple small scale reconnection events to occur simultaneously, and which broadens of the effective thickness of the reconnection layer enabling the efficient ejection of reconnected magnetic flux. The Lazarian-Vishniac model was subsequently tested with the use of  numerical simulations with driven weak turbulence in \citet{KowalEA2009, KowalEA2012}, finding strong agreement with theoretical predictions. Important refinements to the theory of turbulent reconnection have further been proposed to account for the break down of the \citet{Alfven1942} magnetic flux freezing theorem under the influence of turbulence. In particular, the concept of ``spontaneous stochasticity'' of Lagrangian particle trajectories was investigated mathematically by \citet{EyinkEA2011} and later confirmed numerically by \citet{EyinkEA2013}, whereas the original Lazarian-Vishniac model only theorised the spontaneous stochasticity of magnetic field lines.

An important topic that has been attracting considerable attention recently is whether turbulent reconnection can be \textit{self-generated}, i.e.,\ initiated by instabilities in the absence of imposed turbulent driving. This has also been accompanied by an associated question of whether turbulent reconnection can be \textit{self-sustaining}, i.e.,\ induces a steady state through renewal of MHD turbulence generated as a by-product of the reconnection process \citep{Strauss1988, KowalEA2009, LazarianVishniac2009}. Early turbulent reconnection simulations by \citet{KowalEA2009, KowalEA2012} and \citet{LoureiroEA2009} made use of imposed turbulent driving, meaning that the dynamics were dependent on the input injection power. Hence, effectively simulating and studying self-generated (and self-sustaining) turbulent reconnection (SGTR) is crucial for effective comparisons with astrophysical observations. SGTR poses several additional barriers to testing, but was successfully demonstrated in a kinetic simulation by \citet{DaughtonEA2011} \citep[also see][]{BowersLi2007} and an incompressible MHD simulation by Beresnyak \citep[2013, republished as][]{Beresnyak2017}. Since then, SGTR has been reported in numerous MHD simulations \citep{OishiEA2015, HuangBhattacharjee2016, StrianiEA2016, KowalEA2017, KowalEA2020, YangEA2020} and kinetic simulations for nonrelativistic \citep{LiuEA2013, Pritchett2013, NakamuraEA2013, DaughtonEA2014, DahlinEA2015, DahlinEA2017, NakamuraEA2017, LeEA2018, StanierEA2019, LiEA2019, AgudeloRuedaEA2021, Zhang2EA2021} and relativistic \citep{LiuEA2011, GuoEA2015, GuoEA2021, ZhangEA2021} plasmas.

The dominant theme in the majority of the previous MHD studies was the investigation of how SGTR properties, such as the reconnection rate, scale with the Lundquist number and other parameters, and comparing the dynamics with Lazarian-Vishniac, particularly the turbulent statistics. It is also important to note that most of these studies have been fast turbulent analogues of 1D magnetic annihilation. To the best of our knowledge, only one study has previously been reported that explicitly modelled the fast turbulent analogue of 2D Sweet-Parker reconnection including outflow jets \citep{HuangBhattacharjee2016}, even though this scenario is arguably most relevant to many applications including solar flares.

Simulations of SGTR reveal a highly energetic complex process with significant temporal variation. Initial current sheets are observed to develop 2D and 3D instabilities that generate random perturbations throughout the reconnection layer. Simultaneously, the thickness of the reconnection layer rapidly expands, resulting in a broad region of MHD turbulence. These turbulent regions contain numerous coherent structures over a broad range of length scales, fragmented current and vorticity layers threaded by stochastic field lines \citep{DaughtonEA2011, DaughtonEA2014} and anisotropic turbulent eddies \citep{HuangBhattacharjee2016}. Once the process saturates, the evolution continues to be dynamic with coherent structures being subject to various instabilities and coalescing in a chaotic manner. This leads to a number outstanding problems, including:  how do instabilities seed stochasticity in the first place, what are the dominant onset and driving mechanisms behind SGTR, and what is the internal structure of turbulent reconnection layers? This paper aims to advance knowledge of these issues, for 3D SGTR in a Sweet-Parker-type global magnetic topology. 

Efforts to understand SGTR build on extensive previous work on MHD instabilities. The growth and nonlinear interaction of tearing modes have been frequently identified as a crucial component of the turbulent reconnection onset and continual generation of turbulence and stochasticity. In 2D systems, the secondary tearing or plasmoid instability has been extensively studied \citep{HuangBhattacharjee2010, HuangEA2011, KarlickyEA2012, HuangBhattacharjee2012, WanEA2013, HuangEA2017, DongEA2018, HuangEA2019,  PotterEA2019}, leading to the popular ``plasmoid-mediated'' perspective. The plasmoid instability has also been investigated in detail in 3D systems that permit oblique modes, which form on resonance surfaces where $\boldsymbol{k \cdot B} = 0$ \citep{EdmondsonEA2010, BaalrudEA2012, EdmondsonLynch2017, ComissoEA2017, LingamComisso2018, ComissoEA2018, LeakeEA2020}. Unlike in 2D systems, where neat chains of magnetic islands or plasmoids are formed with nested flux surfaces, nonlinear  tearing in 3D appears to completely disrupt the initial laminar current sheet, producing highly filamentary flux rope structures with turbulent interiors, where stochasticity dismantles any internal flux surfaces \citep{BowersLi2007, Beresnyak2017, DaughtonEA2014, HuangBhattacharjee2016, OishiEA2015}. 

Kink instabilities have also been found to be a significant mechanism in SGTR, particularly during the explosive onset of turbulence and fast reconnection \citep{LiuEA2011, OishiEA2015, DahlinEA2015, GuoEA2015, StrianiEA2016, StanierEA2019, LiEA2019, GuoEA2021, Zhang2EA2021}. Flux ropes generated by tearing have a tendency to kink \citep{DahlburgEA1992, DahlburgEA2003, DahlburgEA2005, LapentaBettarini2011, LeakeEA2020}, and the interchange between kinking and tearing modes has been recognised to contribute towards the generation of turbulence \citep{GuoEA2015} and chaotic field lines \citep{GuoEA2021, Zhang2EA2021}. \citet{OishiEA2015} noted that a 3D slab-type kink instability along the initial current layers can occur in the absence of tearing, yet still lead to fast reconnection. Similar phenomena such as the drift-kink instability \citep[e.g.,][]{ZenitaniHoshino2005, ZenitaniHoshino2007,ZenitaniHoshino2008,ZhangEA2021} and lower-hybrid drift instability (LHDI) \citep[e.g.,][]{Daughton2003, LeEA2018} have also been observed in SGTR simulations.

Finally, due to the substantial velocity shear within the reconnection layer and its boundary, Kelvin-Helmholtz instabilities have also been frequently proposed as another major driving process in SGTR \citep{LazarianEA2015}, especially for explaining MHD turbulence production \citep{Beresnyak2017, OishiEA2015, StrianiEA2016, KowalEA2017, KowalEA2020}. \citet{KowalEA2020} investigated the statistical influence of the tearing and Kelvin-Helmholtz instabilities separately, by detecting and analysing regions with intense magnetic or velocity shear. The authors concluded that while tearing instabilities made the major contribution to initiating turbulent reconnection in their simulation, the Kelvin-Helmholtz instability became the dominant driving component sustaining the turbulence once the turbulent layer was sufficiently mature. There has also been extensive research into ``vortex-induced'' SGTR for kinetic simulations of the magnetopause \citep[e.g.,][]{NakamuraEA2013, DaughtonEA2014, NakamuraEA2017}, where reconnection, turbulent mixing, and secondary tearing modes are coupled and driven by the compression of the current layer by Kelvin-Helmholtz instabilities.

To understand turbulent reconnection, an advanced understanding of all of these 3D mechanisms and the structure of the reconnection layer appears to be crucial. Several papers employing kinetic simulations have applied various tools from magnetic topology for further insight \citep{DaughtonEA2014, DahlinEA2017, StanierEA2019}, but these have only recently been applied to MHD simulations to a lesser extent \citep{YangEA2020}. The filling factor and multiscale nature of the thickening current and vorticity layers have been briefly investigated for MHD simulations, with the rate of change of the characteristic current thickness $d\delta/dt$ even being used as a proxy for the reconnection rate, for turbulent reconnection in global topologies analogous to 1D magnetic annihilation \citep{Beresnyak2017, OishiEA2015, KowalEA2017, YangEA2020}.

The main aim of this paper is to explore the production, evolution, structure and topology of the self-generated and self-sustaining turbulent reconnection layer. Within this, we want to investigate the pathway towards fast reconnection, including the instabilities responsible for its onset and generation, identify the characteristic thickness scales of the reconnection layer, and determine any relationships to the global reconnection rate. 

The approach of this paper is a numerical experiment in the style of \citet{HuangBhattacharjee2016}, from which we borrow their distinctive initial setup. One of the main challenges in simulating SGTR in a Sweet-Parker-like configuration is choosing a suitable initial condition and set of boundary conditions, to allow the formation of large-scale outflow and inflow regions within the model, and facilitate the development of an adequately stable reconnection layer. Many previous studies of SGTR have in fact started from initial conditions that, like a standard Harris sheet, are spatially invariant apart from in the one dimension across the current sheet. When this slab-type initial condition is combined with periodic boundary conditions along the reconnection layer parallel to the reconnecting field, large scale outflows are prevented and reconnected flux accumulates in the reconnection layer \citep{Beresnyak2017, OishiEA2015, KowalEA2017, KowalEA2020, YangEA2020}. Some slab-type studies for kinetic simulations have aimed to circumvent that problem by using outgoing boundary conditions, e.g., \citet{DaughtonEA2011} [borrowing a technique from \citet{DaughtonEA2006}] and \citet{ZhangEA2021}  [applying a method by \citet{SironiEA2016}]; both of these papers used absorbing boundary conditions that mimic open boundaries, where particles and magnetic flux are allowed to permanently escape and the electromagnetic fields are prescribed to minimise the reflection of waves leaving the system, to imitate larger systems.  A similar approach was used in closely related MHD simulations with driven turbulence by \citet{KowalEA2009} and \citet{KowalEA2012}, where the normal derivatives of the density and momentum were fixed at zero. However, these outgoing conditions have their own limitations and may influence how the reconnection layer and the outflows evolve. More generally, while slab models may in the best case approximate SGTR in the central portion of the reconnection layer, models that treat the outflow jets clearly provide a fuller picture, including the dynamics of the outflowing parts of the reconnection layer. We also comment that \citet{Beresnyak2017}, \citet{OishiEA2015} and \citet{YangEA2020} employed additional periodic boundaries across the current sheet, which causes the inflow regions to become disrupted and enables strong perpendicular fluctuations to influence the dynamics. While previous MHD studies have made significant contributions to improving understanding of SGTR, experience from laminar models suggests that global properties are likely to be sensitive to the global magnetic topology and to the absence or presence of reconnection outflow jets. As far as we are aware, \citet{HuangBhattacharjee2016} is the only MHD paper published so far that has investigated SGTR in a Sweet-Parker-type configuration including explicitly modelled reconnection outflows.

This paper is structured as follows. The numerical simulation setup is described in Section~\ref{sec:model}, followed by detailed results in Section~\ref{sec:results}. Section~\ref{sec:pathway} narrates the dynamic evolution towards fast reconnection, including the turbulent reconnection onset and various observed instabilities. In Section~\ref{sec:thickness}, we provide a detailed analysis on the mean thickness scales associated with the reconnection layer and their temporal evolution, and inspect the magnetic topology inside the SGTR layer. In Section~\ref{sec:meanprofiles}, we obtain mean-field properties of SGTR, which confirms the existence of ``inner'' and ``outer'' characteristic scales and existence of distinct regions that we refer to as the SGTR ``core'' and SGTR ``wings''. Important properties of the turbulent reconnection process are explored and discussed in Section~\ref{sec:discussion}, such as Sweet-Parker scalings (Section~\ref{sec:SPscalings}), the role and properties of flux rope structures (Section~\ref{sec:FRS}), plasmoid-mediated and stochastic perspectives (Section~\ref{sec:dichotomy}), and the pathway dependence towards SGTR (Section~\ref{sec:pathdependence}). The paper finishes with conclusions in Section~\ref{sec:conclusion}. 

\section{Simulation model}\label{sec:model}

To carry out our 3D compressive visco-resistive MHD numerical simulation, we used \textit{Lare3d} \citep{ArberEA2001}, which is a Lagrangian remap code employing a staggered spatial grid with shock viscosity and a numerical scheme accurate up to second order derivatives. The governing MHD equations, in nondimensionalised Lagrangian form, are 
\begin{eqnarray*}
\frac{D \rho}{D t} & = & - \boldsymbol{\nabla \cdot} \left( \rho \boldsymbol{v} \right),  \\
\frac{D\boldsymbol{v}}{Dt} & = & \frac{1}{\rho}\left( \boldsymbol{j \times B} - \boldsymbol{\nabla} p \right),  \\
\frac{D \boldsymbol{B}}{Dt} & = & \left( \boldsymbol{B \cdot \nabla} \right) \boldsymbol{v} - \left( \boldsymbol{\nabla \cdot v} \right) \boldsymbol{B} - \boldsymbol{\nabla \times} \left(\eta \boldsymbol{j} \right),  \\
\frac{D \epsilon}{Dt} & = & \frac{1}{\rho} \left( - p \boldsymbol{\nabla \cdot v} + \eta j^2 \right), 
\end{eqnarray*}
where
\begin{eqnarray*}
\epsilon & = & \frac{p}{\rho (\gamma - 1)} =  \frac{2T}{\gamma - 1}, \\
\boldsymbol{j} & = & \boldsymbol{\nabla \times B}, 
\end{eqnarray*}
using standard notations:  mass density $\rho$, plasma velocity $\boldsymbol{v}$, current density $\boldsymbol{j}$, magnetic field $\boldsymbol{B}$, pressure $p$, resistivity $\eta$, temperature $T$, specific internal energy density $\epsilon$, ratio of specific heats $\gamma=5/3$, and material derivative operator $D/Dt$. 

\begin{figure} 
\centering
\hspace*{-1.4cm}\includegraphics[width=1.24\columnwidth]{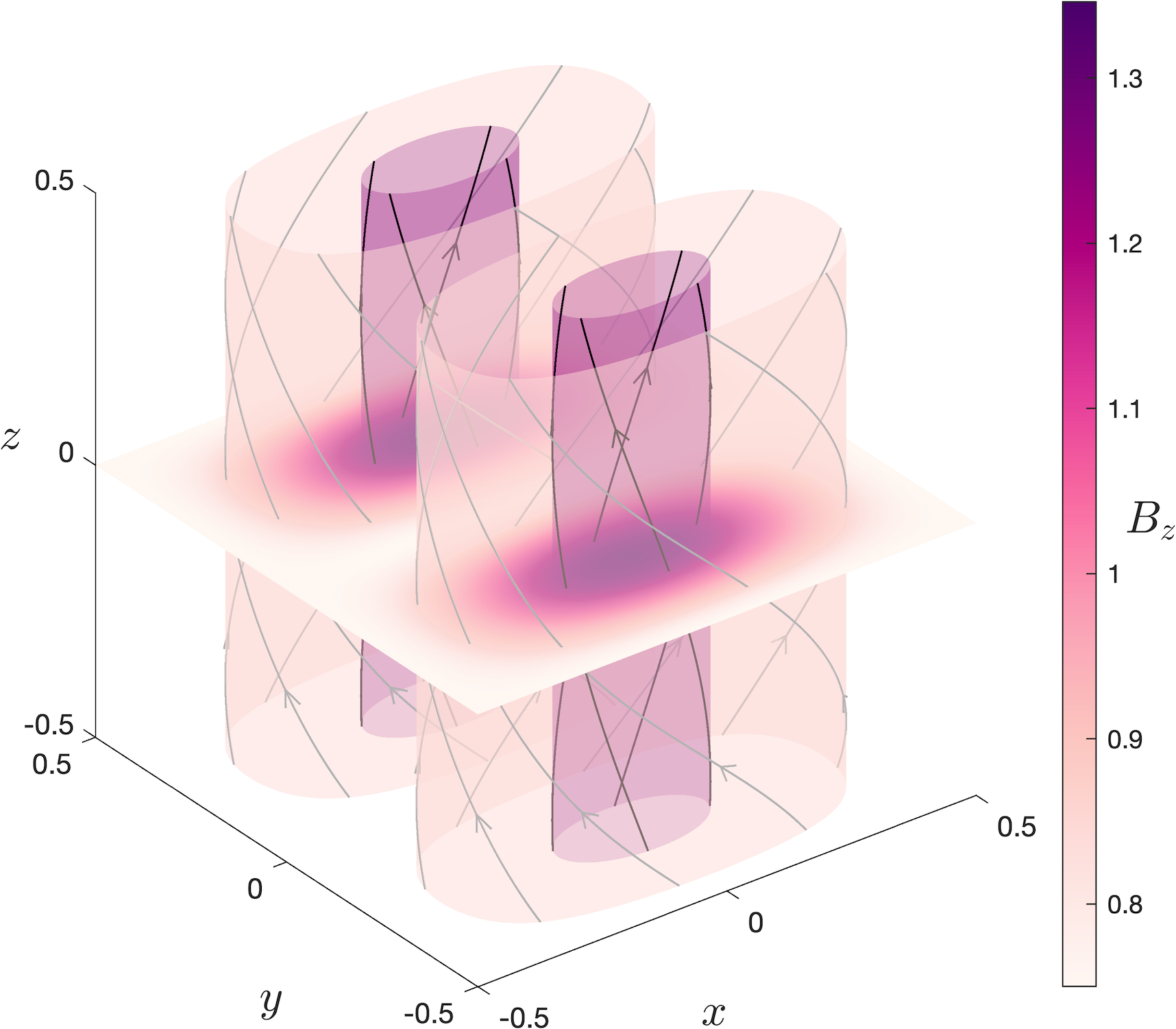}
\caption{3D diagram of the magnetic field at $t=0.0$, illustrating the position and left-handed orientation of the two flux ropes. The pseudocolour plot on the midplane $z=0.0$ indicates the strength of the guide field $B_z>0$, which is initially invariant in the $z$ direction. Isosurfaces of $B_z$ at two different values are compared along with some sample magnetic field lines that lie on the corresponding flux surfaces.  \label{fig:IC}}
\end{figure}

We replicated the initial setup from \citet{HuangBhattacharjee2016}, which considered a thin current sheet between two twisted flux ropes within a unit cube $(x,y,z) \in [-0.5,0.5]^3$. The flux rope merging setup used here is similar to the Gold-Hoyle solar flare model \citep{GoldHoyle1960}; it also has laboratory applications including merging compression startup in spherical tokamaks \citep[e.g.,][]{BrowningEA2014}. An illustration of this configuration is shown in Figure~\ref{fig:IC}. The current sheet is initially located on the midplane $y=0.0$, with finite length in the $x$ direction. The flux ropes are threaded by a guide-field in the periodic $z$ direction and have the same left-handed orientation. The two-and-a-half dimensional (2.5D) magnetic field at $t=0.0$ is set as
\begin{linenomath*}\begin{equation*}
\boldsymbol{B}(x,y) = \boldsymbol{e}_z \boldsymbol{\times \nabla} \psi + B_z \boldsymbol{e}_z, 
\end{equation*}\end{linenomath*}
where the flux function is
\begin{linenomath*}\begin{equation*}
\psi(x,y) = \frac{1}{2 \pi} \cos(\pi x) \sin\left(2 \pi y\right) \tanh\left(\frac{y}{h}\right),\label{eqn:psi}
\end{equation*}\end{linenomath*}
with current sheet thickness $h=1/300$. \citet{HuangBhattacharjee2016} set $B_z$ so that the setup was approximately forced-balanced. To achieve the same in our case, we set the guide-field as
\begin{linenomath*}\begin{equation*}
B_z(x,y)  = \sqrt{5 \pi^2 \psi^2 + B_{z,\text{min}}^2}, \label{eqn:Bz}
\end{equation*}\end{linenomath*}
which is an approximate solution derived using the Grad-Shafranov equation \citep{GradRubin1958, Shafranov1966} and asymptotic matching. The force-balance improves for regions further away from the current sheet around $y=0.0$. The value $B_{z,\text{min}} = 0.75$ was chosen to ensure that the guide-field strength within the reconnection layer, where it is at its minimum, was consistent with \citet{HuangBhattacharjee2016}. 

We initialised the plasma to have a uniform temperature $T = 1$ and density $\rho = 1$, which together give constant pressure $p$ at $t=0.0$. To aid the formation of current sheet instabilities and initiate the 3D turbulent reconnection process, random velocity noise of magnitude $v \sim 10^{-3}$ was applied as part of the initial condition. The background resistivity was set as $\eta = 5 \times 10^{-6}$, giving Lundquist number $S=\eta^{-1}= 2\times 10^5$, based on the unit box size and initial current sheet length $L\approx 1$. Some reconnection studies instead report $S$ values based on the current sheet half-length \citep[e.g.,][]{HuangEA2019, SinghEA2019}; under that convention, for comparison with such works, the Lundquist number rescales to $S=10^5$. In practice, this value is an upper bound on the effective Lundquist number, which may be reduced by numerical resistivity. We can be confident that the effective Lundquist number comfortably exceeds a lower bound of $S = 10^4$ \citep{LapentaLazarian2012}, since the current sheet was unstable to the tearing instability in production and lower resolution runs.

The simulation was performed on a uniform grid of resolution $1064^3$. Preliminary work was carried out at a lower resolution of $500^3$ and this displayed the same phenomena. The boundary conditions were set as periodic in the $z$ direction, and perfectly conducting and free slipping in the $x$ and $y$ directions, i.e.,\ $\boldsymbol{\hat{n} \cdot B} = 0$ and $\boldsymbol{\hat{n} \cdot v} = 0$ on $x = \pm 0.5$ and $y=\pm 0.5$. In \textit{Lare3d}, these side boundary conditions were imposed with appropriate symmetry arguments by pairing ghost cells with domain cells near the boundaries. The simulation was run up to $t=5.0$, with frames recorded every $\Delta t = 0.1$.

\section{Results}\label{sec:results}
In agreement with \citet{HuangBhattacharjee2016}, the simulation undergoes reconnection which is fully 3D, self-generated, and self-sustaining, and exhibits characteristics of both plasmoid-mediated and turbulent reconnection. Consistent with earlier studies of SGTR, which have predominantly considered analogues of magnetic annihilation \citep{Beresnyak2017, OishiEA2015, StrianiEA2016, KowalEA2017, KowalEA2020}, we find that 3D reconnection at $S \geq 10^4$ is qualitatively different to 2D models. We also find that the global magnetic topology has a major influence on the reconnection process, as is reasonably expected from the experience of laminar reconnection models, and that the SGTR analogue of Sweet-Parker reconnection, studied here and in \citet{HuangBhattacharjee2016}, is distinct in important respects from the current slab models that have been more widely studied.

 In the following subsections, we discuss the dynamic evolution in detail and provide new analysis on the topology and structure of the turbulent reconnection regions for SGTR in a Sweet-Parker-type global magnetic topology.

\subsection{Dynamic phases and pathway to ``fast'' reconnection}\label{sec:pathway}

Our simulation shows three major stages: an initial laminar Sweet-Parker phase of slow reconnection during which 2.5D instabilities develop (\textit{i. Laminar 2.5D phase}), followed by secondary 3D instabilities that seed field line stochasticity (\textit{ii. Transition phase}), ending with a quasi-stationary turbulent phase that is self-sustaining over the long term  (\textit{iii. SGTR phase}). 

\begin{figure} 
\centering
\hspace*{-1.5cm}\includegraphics[width=1.25\columnwidth]{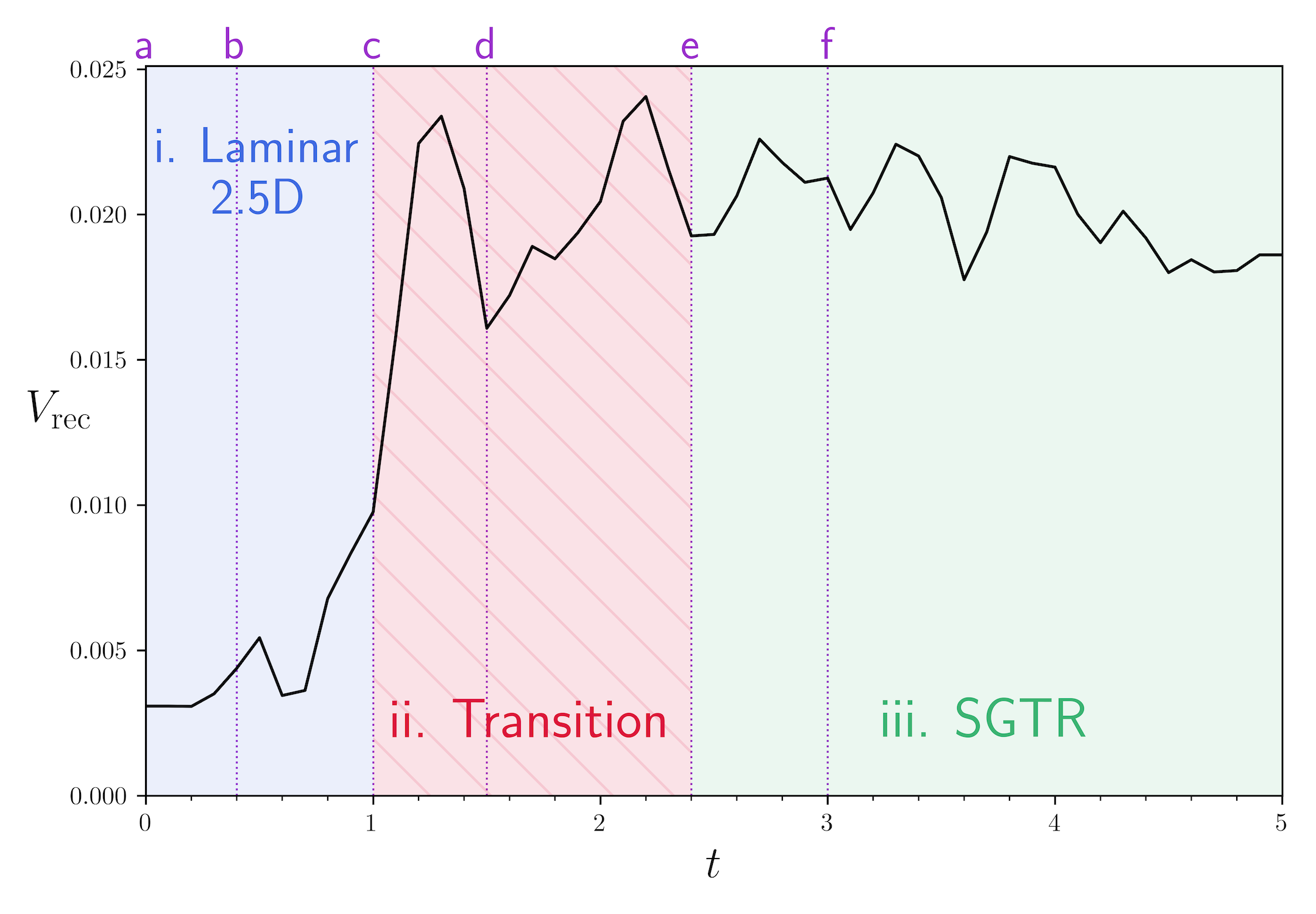}
\caption{Evolution of the reconnection rate $V_{\mathrm{rec}}$ over time. The three general stages are highlighted in different colours: \textit{i. Laminar 2.5D phase} (blue);  \textit{ii. Transition} (red, striped); \textit{iii. SGTR} (green). The vertical dotted lines mark additional times of interest: \textbf{a} 2.5D Sweet-Parker reconnection onset ($t=0.0$); \textbf{b} 2.5D nonlinear tearing instability onset ($t=0.4$); \textbf{c} Kink instability of central flux rope (CFR) and turbulent reconnection onset ($t=1.0$); \textbf{d} CFR breaks down and reconnection layer becomes (almost) fully stochastic ($t=1.5$); \textbf{e} CFR remnant ejected ($t=2.4$); \textbf{f} CFR remnant fully absorbed at outflows and ``pure'' SGTR onset ($t=3.0$). \label{fig:Vrec}}
\end{figure}

Figure~\ref{fig:Vrec} shows the evolution of the global reconnection rate $V_{\mathrm{rec}}$ and the approximate time intervals for each of the three main simulation stages. The labelled vertical dashed lines indicate finer steps in the simulation that will be elaborated on later, some of which are specific to this particular simulation. Here we have calculated the global reconnection rate as
\begin{linenomath*}\begin{equation*}
V_{\mathrm{rec}} = \frac{d}{dt}  \max_{x \in [-0.5,0.5]}  \int_{-0.5}^x \int_{-0.5}^{0.5} \left. B_y \right\vert_{y=0} \ dz' \ dx',
\end{equation*}\end{linenomath*}
using the same definition as \citet{HuangBhattacharjee2016}, which is the time derivative of one particular approximation of the reconnected flux. See Appendix~\ref{sec:vrec} for a discussion on the evaluation of the reconnection rate and a comparison between alternative definitions of the reconnected flux. The most important outcome of Figure~\ref{fig:Vrec} is the ``switch-on'' nature of fast reconnection; we clearly observe a sudden increase in the global reconnection rate $V_{\mathrm{rec}}$, by a factor of $6\mbox{--}8$ times from the initial value, after the onset of turbulent reconnection. 

\begin{figure*} 
\vspace*{-0.3cm}
\centering
\hspace*{-1.2cm} \includegraphics[width=1.14\textwidth]{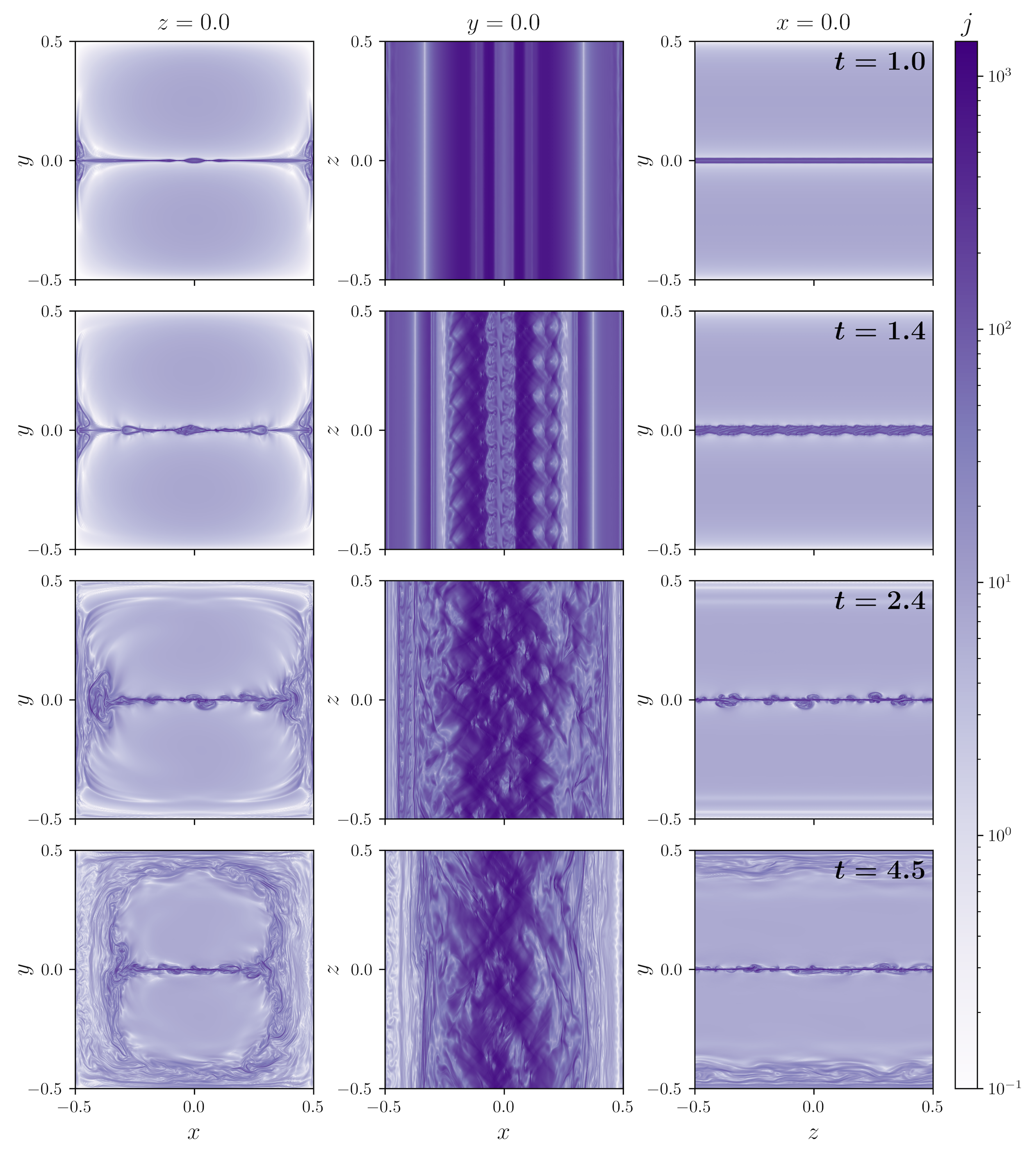}
\caption{Current density strength $j = \Vert \boldsymbol{\nabla \times B}\Vert$ over time. The left column is a cross-section at $z=0.0$ through the two large flux ropes; the middle column is a slice at $y=0.0$ inside the reconnection layer; the right column is a slice at $x=0.0$ through the current sheet along the guide-field direction. The top row ($t=1.0$) shows three flux ropes on $y=0.0$ formed by the tearing instability of the initial current layer; the second row ($t=1.4$) is snapshot during the transition phase while the 3D kink instability of the CFR is developing; the third row ($t=2.4$) shows the highly nonuniform reconnection layer during the SGTR phase while the CFR reaches the left outflow; the bottom row ($t=4.5$) shows how far the pair of initial large scale flux ropes have merged at an advanced stage of the simulation during ``pure'' SGTR. \label{fig:j} } 
\end{figure*}

To facilitate our description of the simulation phases in the following subsections, visualisations of the current density strength $j = \Vert \boldsymbol{\nabla \times B}\Vert$ are provided in Figure~\ref{fig:j} at four particular times. The left and right columns show cross-sections across the reconnection layer at midplanes $z = 0.0$ and $x = 0.0$, respectively. The central column gives a slice at $y = 0.0$ within the reconnection layer.

\subsubsection{i. Laminar 2.5D phase: $t=0.0\mbox{--}1.0$} 

The simulation begins with the relaxation of the two large flux ropes, which immediately begin to merge, causing thinning of the current sheet at the reconnection interface. This first stage observed, during $t=0.0\mbox{--}0.4$ (\textbf{a}--\textbf{b} in Figure~\ref{fig:Vrec}), is \textit{laminar 2.5D Sweet-Parker} reconnection, during which the dynamics are invariant in the guide-field direction and the evolution on each $z$-slice is consistent with the standard laminar Sweet-Parker slow reconnection model. Clear inflow velocity regions form above and below the current sheet, and long symmetric outflow velocity regions develop along $y=0.0$. During this stage a modest stable reconnection rate $V_{\mathrm{rec}} \approx 0.003\mbox{--}0.0035$ is measured,  in agreement with Sweet-Parker and the rate for early times reported by \citet{HuangBhattacharjee2016}. 

This initial stage is soon interrupted by the formation of a chain of islands within the current layer by the \textit{2.5D tearing instability} over $t=0.4\mbox{--}0.7$ (after \textbf{b} in Figure~\ref{fig:Vrec}). Three plasmoids become particularly prominent and are visible from $z$-slices of the current density $j$ (top row of Figure~\ref{fig:j}) and other variables, the 3D structures of which take the form of long straight flux ropes extending over the whole $z$ direction. These prominent flux ropes slowly grow over $t = 0.7\mbox{--}1.0$ (up to \textbf{c} in Figure~\ref{fig:Vrec}), which we interpret as the development of the \textit{nonlinear tearing mode}, and the leftmost and rightmost flux ropes gradually move outwards. Other plasmoids with significantly smaller length scales are also observed in the outflows, their growth being limited by their expulsion from the reconnection region, while new plasmoids are generated in their place. However, the largest structure, which we will refer to as the \textit{central flux rope} (CFR), persists at the centre ($x=0.0$) and has a major influence on the subsequent evolution. Over $t = 0.7\mbox{--}1.0$, $V_{\mathrm{rec}}$ (Figure~\ref{fig:Vrec}) gradually increases as the CFR slowly enlarges with a ``cat-eye'' cross-section. 

The development of a large CFR was not reported in \citet{HuangBhattacharjee2016}, nor is one evident to us in their figures. The reason for this difference between our simulations and theirs is unclear, but the CFR was also found to form for 2.5D simulations on the \textit{Lare2d} code for grid resolutions up to $10000^2$, matching the minimum grid size $\Delta y = 10^{-4}$ employed in \citet{HuangBhattacharjee2016}, with or without initial velocity noise. This implies that the CFR formation in our simulation is not a numerical artifact due to lower grid resolution. Further, using 2.5D simulations, a visible CFR was found to develop for effective Lundquist number values above $S = 10^4$, consistent with the widely quoted critical value $S_{\mathrm{crit}} \approx 10^4$ for the Sweet-Parker current layer to be self-unstable in the absence of sufficiently large perturbations \citep{LapentaLazarian2012}, and below $S = 10^7$, beyond which the rapid formation of a thin chain of similarly sized plasmoids was more prominent. These observations reinforce the conclusion that the CFR is a robust feature of our \textit{Lare} simulations for the Lundquist number we have applied.
 
In a broader context, the early dominance of parallel tearing modes in our 3D simulation is consistent with the results of \citet{OishiEA2015}, but simulations also exist in which oblique tearing modes form initially \citep{DaughtonEA2011, LiuEA2013, HuangBhattacharjee2016, Beresnyak2017, StanierEA2019}. Studies have shown that the fastest growing tearing modes can be parallel or oblique, with the properties of the dominant mode depending on critical parameters such as the length of the current layer and the magnetic shear angle across the current layer \citep[e.g.,][]{BaalrudEA2012, LeakeEA2020}. For our initial magnetic field, we evaluated the predicted linear growth rate for a range of tearing modes characterised by integers $(m,n)$ corresponding with wavenumbers $k_x = 2\pi m$ and $k_z = 2\pi n$, where parallel modes possess $n=0$. Here we followed the approach in \citet{LeakeEA2020} who applied linear theory derived from reduced MHD by \citet{BaalrudEA2012}. The fastest growing tearing mode was found to be parallel with $(m,n) = (1,0)$, which supports the formation and domination of the CFR that we observed over $t = 0.4\mbox{--}1.0$ (\textbf{b}--\textbf{c} in Figure~\ref{fig:Vrec}). It is possible that subdominant oblique tearing modes are not sufficiently resolved, causing their nonlinear interaction and growth to be dampened; however, the pathway to SGTR following from parallel tearing modes is an important topic that we analyse in the next section.

\subsubsection{ii. Transition phase: $t = 1.0\mbox{--}2.4$}\label{sec:transitionphase} 

In the second major phase over $t = 1.0\mbox{--}2.4$ (\textbf{c}--\textbf{e} in Figure~\ref{fig:Vrec}), the dynamics undergo a fundamental change in character: the 2.5D symmetry breaks, 3D instabilities seed the developing stochasticity of field lines, the system exhibits self-generated turbulence, and the reconnection rate rapidly increases to reach a global maximum of $V_{\mathrm{rec}} \approx 0.024$ (at $t=2.2$), approximately 7.8 times the rate of the initial Sweet-Parker phase. This ``switch-on'' behaviour is qualitatively and quantitatively different to the evolution observed by \citet{HuangBhattacharjee2016}, in whose simulation the reconnection rate gradually increased over the simulation runtime and only reached a maximum of $V_{\mathrm{rec}} \approx 0.009$. The reason for this difference appears to be that our simulation and theirs capture different pathways to SGTR, as we elaborate on below. There are in fact strong grounds to expect that SGTR can be reached by a variety of routes that depend on specific circumstances, e.g., depending on whether the dominant modes of the tearing instability are parallel or oblique. An interesting observation from comparing our simulation to \citet{HuangBhattacharjee2016} is that different pathways to SGTR may affect the switch-on and reconnection rate properties. This may help to explain why some reconnection events rise rapidly whereas others rise slowly, e.g., impulsive versus gradual solar flares \citep{FletcherEA2011}.
 
One of the significant driving mechanisms of this dynamic stage is the onset of a \textit{3D helical kink instability} of the CFR at $t = 1.0$ (\textbf{c} in Figure~\ref{fig:Vrec}) that makes its axis increasingly coiled. Concurrently, the CFR dislodges from $x=0.0$ and gradually moves to the left, and the two flanking flux ropes accelerate in opposite directions towards the edges of the reconnection layer (see second row of Figure~\ref{fig:j}). The secondary instability of a flux rope generated from a parallel mode of the preceding tearing instability is consistent with the simulation findings of \citet{LapentaBettarini2011} and \citet{OishiEA2015}. To detect and investigate this instability, we identify the magnetic field line that is the axis of the CFR, then trace its evolution. 

We carry out the following procedure. In our simulation, which has $B_z > 0$ everywhere, we parameterise field lines by $\boldsymbol{x}_B(z) = (x_B(z),y_B(z),z)$ with initial point  $\boldsymbol{x}_B(z_0) = \boldsymbol{x}_0$, governed by
\begin{equation}
\frac{d\boldsymbol{x}_B}{dz} = \frac{\boldsymbol{B}\left(\boldsymbol{x}_B(z)\right)}{B_z\left(\boldsymbol{x}_B(z)\right)}. \label{eqn:odesys}
\end{equation}
Since the domain is periodic in $z$, we allow $z \in \mathbb{R}$. The 2D field line mapping from an initial point at $z=z_0$ to $z=z_0+n$ is then
\begin{equation}
F_{z_0}^{n}(x_0,y_0) = \bigl(x_B(z_0+n),y_B(z_0+n)\bigr) .  \label{eqn:Fmap}
\end{equation}
For our simulation, $F_{z_0}^{n}$ denotes the field line mapping over $n$ cycles for domain length $Z=1$, where we have $n \in \mathbb{R}$, e.g.,\ forward traces for $n>0$ and backward traces for $n<0$. To analyse the topological properties of the field line mapping, we consider the \textit{colour map} \citep{PolymilisEA2003, YeatesEA2010, YeatesHornig2011a, YeatesHornig2011b, YeatesEA2015} defined by
\begin{equation}
C_{z_0}^{n}(x_0,y_0) = \left\{\arraycolsep=5pt\def\arraystretch{1.2}\begin{array}{lr}
\text{Blue (B)}, & [F_{z_0}^{n}]^{x} > x_0 \wedge [F_{z_0}^{n}]^{y} > y_0; \\ 
\text{Green (G)}, & [F_{z_0}^{n}]^{x} < x_0 \wedge [F_{z_0}^{n}]^{y} > y_0; \\ 
\text{Red (R)}, & [F_{z_0}^{n}]^{x} < x_0 \wedge [F_{z_0}^{n}]^{y} < y_0; \\ 
\text{Yellow (Y)}, & [F_{z_0}^{n}]^{x} > x_0 \wedge [F_{z_0}^{n}]^{y} < y_0; 
\end{array}\right. \label{eqn:colmap}
\end{equation}
whose colour space corresponds to the displacement vector between initial point $(x_0,y_0)$ and $F_{z_0}^{n}(x_0,y_0)$. The colour map $C_{z_0}^{n}$ over a complete number of cycles $n\in \mathbb{N}$ can then be used to identify periodic points of degree $n$. Boundaries of R-B or G-Y interfaces are positions of general periodic orbits, with integer twist about some axis.  Points where the four different colours meet are isolated periodic orbits. By topological degree theory, if the anticlockwise sequence of colours is B-G-R-Y, it is an \textit{elliptic} periodic point; if the sequence is Y-R-G-B, it is a \textit{hyperbolic} periodic point. The locations of isolated periodic orbits of interest were identified using the \textit{characteristic bisection method} \citep{Vrahatis1995, PolymilisEA2003}. Elliptic periodic points should coincide with axes of flux ropes, the curves of which form \textit{closed loops}. 

\begin{figure*}[h] 
\centering
\hspace*{-1.8cm}\includegraphics[width=1.2\textwidth]{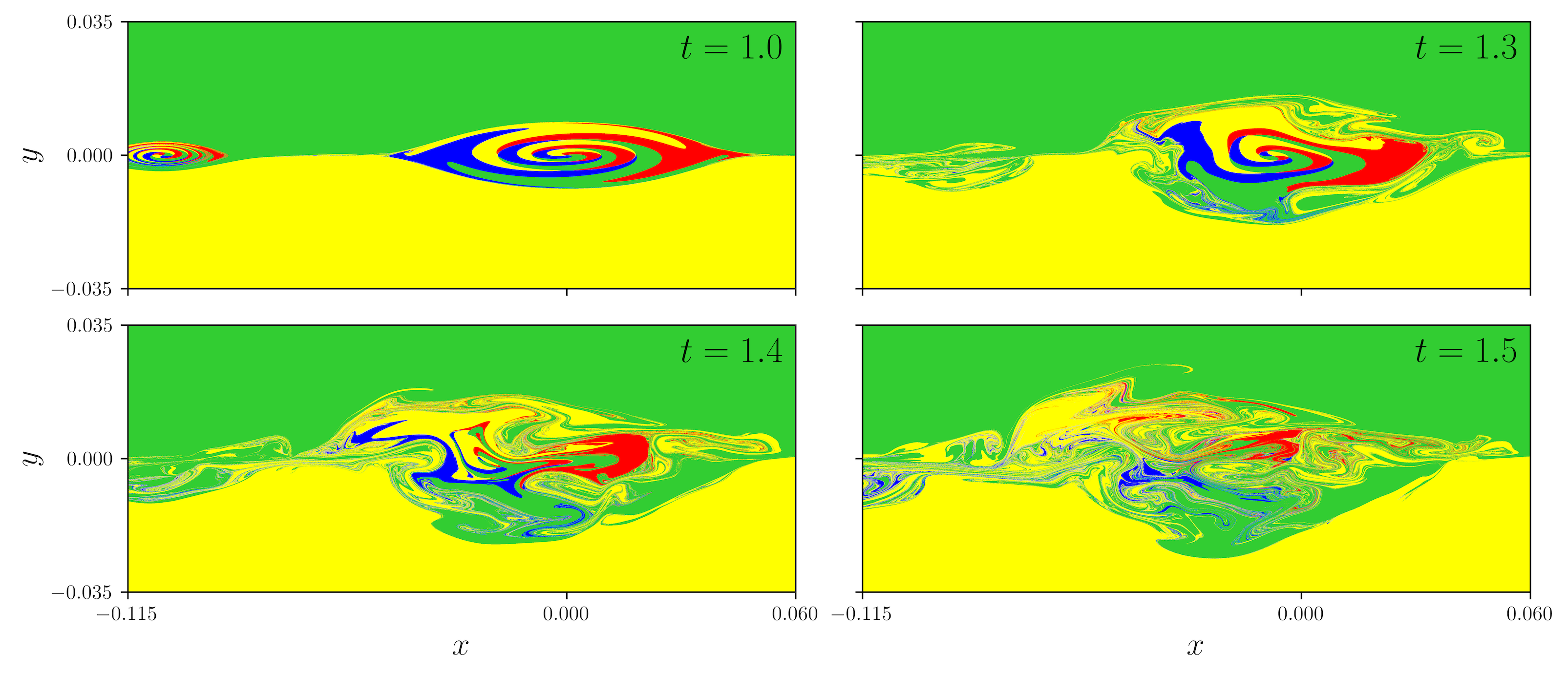}
\caption{Closeups of the colour map $C_{z_0}^{1}$ at the bottom boundary $z_0=-0.5$ during the 3D kink instability of the central flux rope (CFR). Prior to the turbulent reconnection onset $t = 1.0$ (\textbf{c} in Figure~\ref{fig:Vrec}), the CFR is stable and has a cat-eye cross-section with an elliptical core. The top left panel at $t=1.0$ also picks up the smaller flux rope to the left of the CFR, which accelerates to the left and soon exits the closeup window. After $t = 1.2$, the CFR axis moves to the left, and the cross-section becomes increasingly deformed with a growing outer structure. By $t= 1.5$ (\textbf{d} in Figure~\ref{fig:Vrec}), the CFR breaks down; the axis either disappears within the stochastic field or it can no longer be detected with the colour map.} \label{fig:colmap}
\end{figure*}

\begin{figure*}[h] 
\centering
\hspace*{-1.8cm}\includegraphics[width=1.2\textwidth]{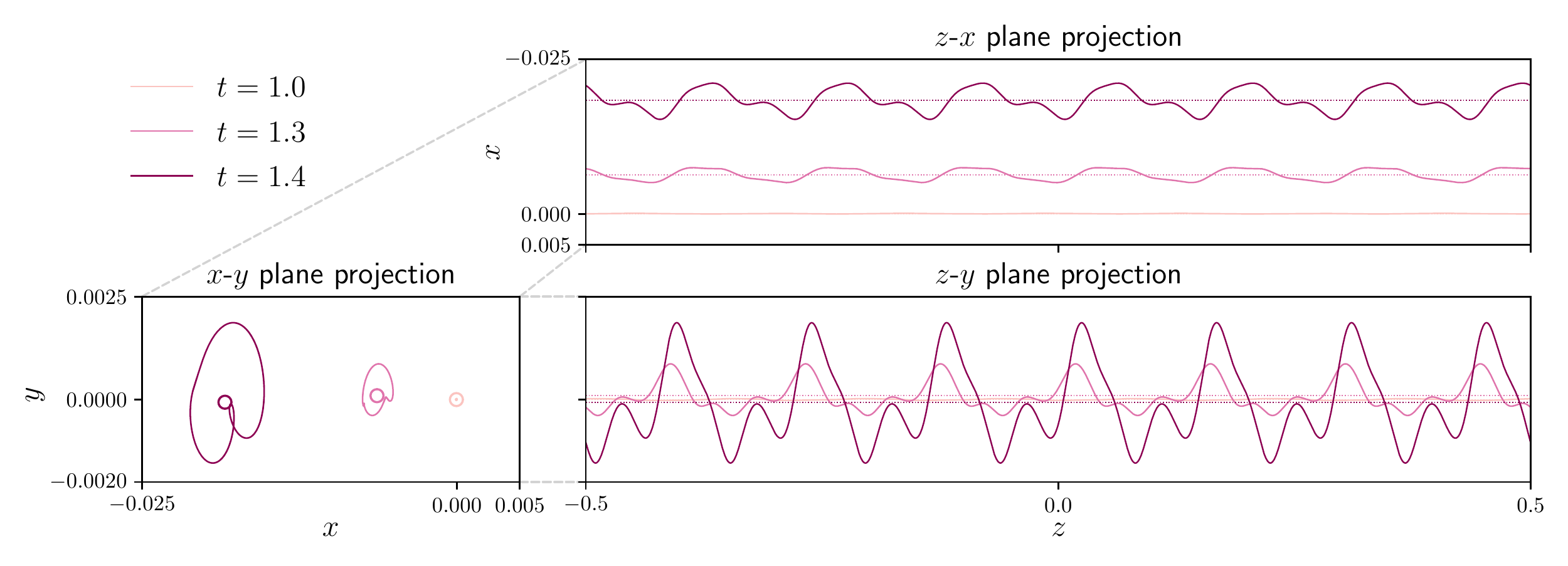}
\caption{Evolution of the central flux rope (CFR) axis $\boldsymbol{x}_{\mathrm{axis}}$ on various plane projections over $t = 1.0\mbox{--}1.4$. The 3D kink instability occurs at $t = 1.0$ (\textbf{c} in Figure~\ref{fig:Vrec}). Solid curves represent the CFR axis and the dashed horizontal lines or circles denote the centre of mass $\boldsymbol{x}_{\mathrm{center}}$ of the CFR axis. The CFR evolves from a straight flux rope prior to $t = 1.0$ to a fully 3D helical kink which is asymmetric and weakly nonlinear. The helix rapidly grows in amplitude after $t = 1.2$ with dominant wavenumber $k_z/(2\pi) = 7$ while moving to the left on the $x$-$y$ plane. At $t = 1.5$ onwards (\textbf{d} in Figure~\ref{fig:Vrec}), the axis cannot be detected. \label{fig:3Dkink}}
\end{figure*}

Figure~\ref{fig:colmap} shows the colour map $C_{z_0}^{1}$ for the bottom boundary $z_0=-0.5$, illustrating the topology of the field line mapping and periodic surfaces within the CFR during the kink instability (\textbf{c}--\textbf{d} in Figure~\ref{fig:Vrec}). We identify the elliptic period-1 orbit that intersects $(x_0,y_0)\approx(0,0)$ as the axis of the CFR. During the development of the nonlinear tearing stage over $t = 0.7\mbox{--}1.0$ (up to \textbf{c} in Figure~\ref{fig:Vrec}), the CFR rapidly grows with a ``cat-eye'' structure containing a highly elliptical core possessing an exceptional level of internal twist. The ellipticity of this expanding core quickly decreases due to magnetic tension $ (\boldsymbol{B \cdot \nabla})\boldsymbol{B}$ from the high curvature of field lines within the plasmoid. 

The time evolution of the CFR axis on various plane projections is displayed in Figure~\ref{fig:3Dkink}; this is denoted $\boldsymbol{x}_{\mathrm{axis}}(z)$ and parameterised by $z\in [-0.5, 0.5)$. For comparison, the corresponding centre of mass $\boldsymbol{x}_{\mathrm{center}}$, i.e., the mean coordinate $\bigl<\bigl(x_{\mathrm{axis}}(z),y_{\mathrm{axis}}(z)\bigr)\bigr>$ over $z\in[-0.5,0.5)$, is also provided. Prior to the kink instability at $t = 1.0$ (\textbf{c} in Figure~\ref{fig:Vrec}), the CFR axis is initially straight since the system is still 2.5D. The CFR axis then begins moving to the left, particularly after a dynamic eruption at $t = 1.2$, and becomes increasingly kinked.  This deformation is fully 3D with dominant wavenumber $k_z/(2\pi) = 7$; the same periodicity can also be observed in other measures such as the current density strength $j$ (see second row of Figure~\ref{fig:j}). Fourier analysis of $\boldsymbol{x}_{\mathrm{axis}}$ reveals that the kink is a superposition of multiple modes with $k_z/(2\pi) = 7n$ for $n\in \mathbb{N}$, which significantly decrease in amplitude for increasing $n>1$, making the instability weakly nonlinear and asymmetric. 

In time, as seen from Figure~\ref{fig:colmap}, the cross-sectional geometry becomes increasingly deformed, with some interesting topological substructure evident by $t = 1.4$. By $t = 1.5$ (\textbf{d} in Figure~\ref{fig:Vrec}), the laminar core of the CFR breaks down and the colour map no longer detects an axis. Section~\ref{sec:FRS} discusses the topological complexity of ``frayed'' flux ropes in the reconnection layer. Further analyses on this kink instability can be found in Appendix~\ref{sec:kink}.

Accompanying the breaking of 2.5D symmetry by the helical kink instability, we observe the formation and fast broadening of a \textit{stochastic layer}, which are the mixing regions about $y=0.0$ where field lines are no longer laminar but instead ``wander''. From the perspective of magnetic topology, these distinct stochastic and laminar regions can be approximated using various tools. Since our simulation is periodic in $z$, a \textit{Poincar\'e section} \citep[e.g.,][]{BorgognoEA2008, BorgognoEA2011a, BorgognoEA2011b, BorgognoEA2015, RubinoEA2015, FalessiEA2015, VerandaEA2020a, BorgognoEA2017, DiGiannataleEA2017b, DiGiannataleEA2018, DiGiannataleEA2021} is an effective initial approach to illustrate this transitional stage \citep{DaughtonEA2014, DahlinEA2017, StanierEA2019}. For a collection of seed points $(x_0,y_0)$ on a fixed $z_0$-slice, we generate a scatter plot of the orbit points where the field lines repeatedly intersect the $x$-$y$ plane as they cycle around the toroidal space, i.e.,\ the mapping $F_{z_0}^{n}(x_0,y_0)$ [definition (\ref{eqn:Fmap})] for all iterations $n \in \mathbb{N}$ up to some limit $n \leq n_{\max}$. In theory, the field lines seeded within a stochastic region will randomly  fill in that particular space statistically as $n\to\infty$ due to ergodicity, whereas field lines seeded within laminar regions will trace contours of flux surfaces. 

\begin{figure*} 
\centering
\hspace*{-1.9cm} \includegraphics[width=1.19\textwidth]{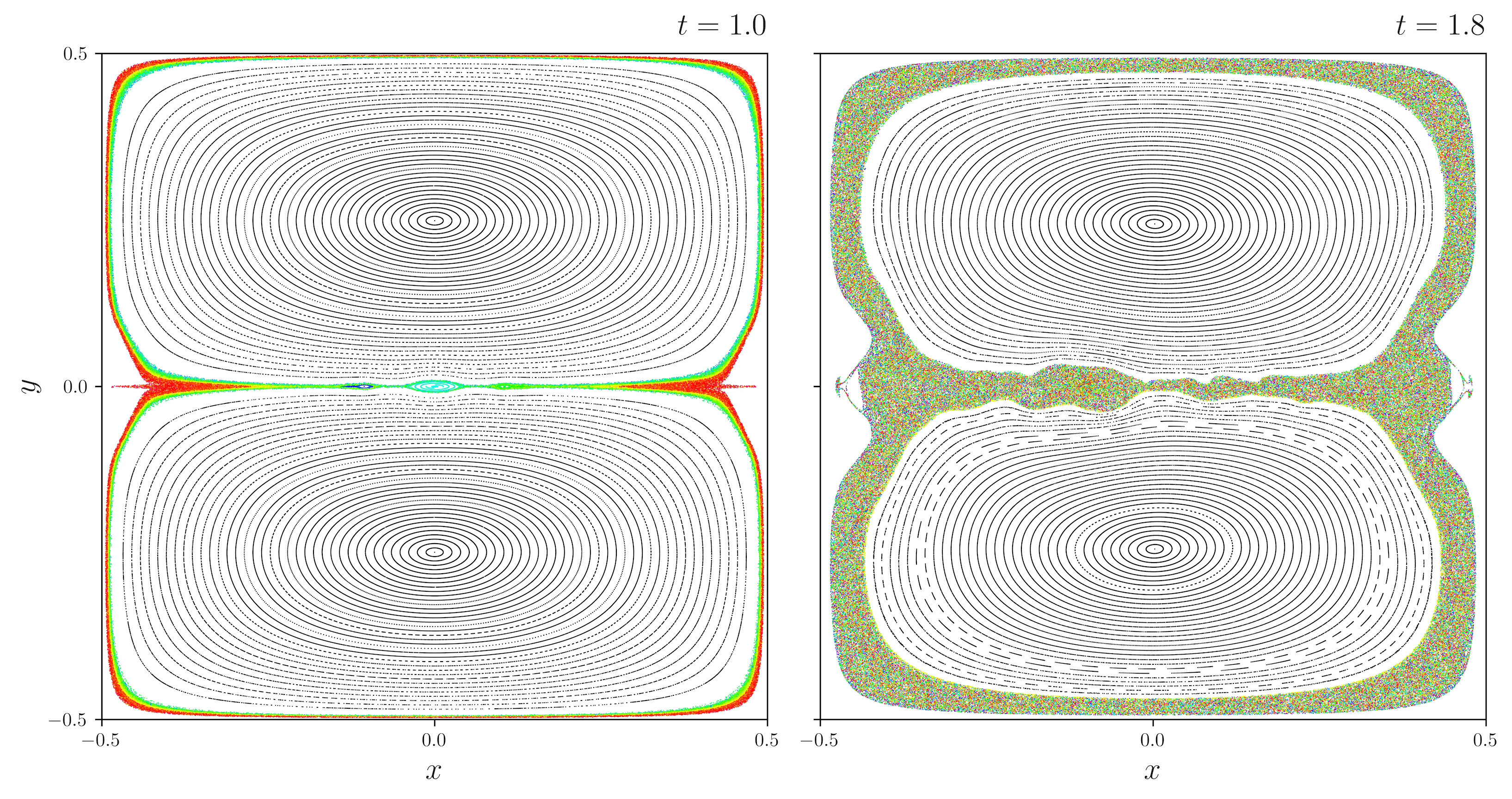} 
\caption{Poincar\'e sections of the field line mapping $F_{z_0}^{n}$ for all iterations up to $n = 2000$ at the bottom boundary $z_0=-0.5$ for two selected times during the dynamic transition phase $t= 1.0\mbox{--}2.4$ (\textbf{c}--\textbf{e} in Figure~\ref{fig:Vrec}).  The multicoloured dots, seeded over $x \in [-0.42,0.42]$ at $y=0.0$, spread over the reconnection-connected volume (RCV) that is topologically identified with the reconnection layer; points with the same colour correspond to the same field line. This RCV rapidly expands after the turbulent reconnection onset $t= 1.0$ (left panel) and becomes (almost) fully stochastic by $t = 1.8$ (right panel). The black dots, seeded within the upper and lower flux ropes, lie on simple contours of laminar flux surfaces.} \label{fig:poincare}
\end{figure*}

Figure~\ref{fig:poincare} shows the Poincar\'e sections at the bottom boundary $z_0=-0.5$ up to $n = 2000$ iterations at $t = 1.0$ and $t = 1.8$ during the dynamic transition phase over $t= 1.0\mbox{--}2.4$ (\textbf{c}--\textbf{e} in Figure~\ref{fig:Vrec}). We identify four general regions that are topologically separate, up to at least $n=2000$. Firstly, we have two laminar regions containing the large flux ropes, i.e., the ``upper'' flux rope (in the $y>0$ half of the domain) and ``lower'' flux rope (in the $y<0$ half of the domain); these are illustrated with black dots using seed lines extending from their respective axes (the axes having been detected using the colour map). The laminar flux ropes are enclosed by a \textit{reconnection-connected volume} (RCV), with a figure-of-eight cross-section, that is topologically associated with the reconnection layer; this domain is highlighted by multicoloured dots corresponding to distinct field lines seeded along the midplane $x \in [-0.42,0.42]$ at $y=0.0$. Since field lines that permeate the reconnection layer become stochastic, the RCV is an effective proxy for observing the development and spread of stochasticity generated from the reconnection process. Section~\ref{sec:outerlayer} provides a deeper treatment of alternative tools, discussion on determining whether a field line is stochastic, and the effectiveness of the Poincar\'e section approach. Lastly, we have an outer region that extends to the $x$ and $y$ boundaries. The evolution of the RCV is demonstrated in Section~\ref{sec:outertopology}. 

During the laminar 2.5D phase $t= 0.0\mbox{--}1.0$ (\textbf{a}--\textbf{c} in Figure~\ref{fig:Vrec}), the RCV forms a very thin band around the large flux ropes (left panel of Figure~\ref{fig:poincare}). However, at $t = 1.0$ (\textbf{c} in Figure~\ref{fig:Vrec}), coincident with the 3D kink instability of the CFR, the RCV rapidly expands and the large laminar flux ropes begin to shrink (right panel of Figure~\ref{fig:poincare}). During this transition, stochastic field lines proliferate throughout the RCV and surround the flux ropes nested within the reconnection layer. It is currently difficult to determine whether the kink instability generates a substantial proportion of stochasticity, or if the development of a stochastic environment around or within the CFR makes it kink-unstable. 

We refer to $t = 1.0$ as the \textit{turbulent reconnection onset}, since this is when self-generated turbulence begins to develop. After this, the system is increasingly non-laminar. Further, the current sheet becomes fragmented, and the reconnection layer becomes threaded with numerous small scale structures resembling oblique twisted flux ropes and turbulent eddies (see second row of Figure~\ref{fig:j}). When the CFR breaks down at $t = 1.5$ (\textbf{d} in Figure~\ref{fig:Vrec}), the whole reconnection layer appears to be made up almost entirely of stochastic field lines, with the exception of the left and right plasmoid-type flux ropes. Once these flux ropes exit the reconnection layer at $t = 1.8$ (see right panel of Figure~\ref{fig:poincare}), we consider the reconnection layer and the RCV to be fully stochastic. 

For the remainder of the transition phase, the dynamics inside the reconnection layer are dominated by the production, merging and expulsion of flux rope structures that are 3D analogues of plasmoids, but which lack a detectable axis field line. The stochastic layer continues to broaden, the (stochastic) RCV expands towards the $x$ and $y$ boundaries, and the reconnection rate remains at an enhanced level. The large structure produced from the previously laminar CFR continues to grow while accelerating to the left, before being expelled from the left outflow at $t = 2.4$ (\textbf{e} in Figure~\ref{fig:Vrec}; third row of Figure~\ref{fig:j}).

\subsubsection{\textit{iii. SGTR phase:} $t = 2.4\mbox{--}5.0$} 

In this third and final major phase over $t = 2.4\mbox{--}5.0$ (after \textbf{e} in Figure~\ref{fig:Vrec}), the system exhibits \textit{self-generated turbulent reconnection} that is fully 3D and globally quasi-stationary.

Within this part of the simulation, the CFR structure is expelled from the reconnection layer at $t = 2.4$ (third row of Figure~\ref{fig:j}) and it is absorbed at the termination of the outflow jet by $t = 3.0$ (\textbf{e}--\textbf{f} in Figure~\ref{fig:Vrec}). The reconnection layer appears to be fully turbulent, evidenced by the highly irregular flow pattern and fragmented current density strength (middle column of Figure 3). The properties of the turbulence have previously been investigated in depth by \citet{HuangBhattacharjee2016} and \citet{KowalEA2017}, and applying their techniques to our simulation confirms these previous results. On the global scale, the configuration has a dominant inflow $v_y$ towards the reconnection layer and outflows $v_x$ toward the $x$ boundaries, consistent with the expected reconnection flow pattern for the global magnetic topology. By $t = 3.0$ (after \textbf{f} in Figure~\ref{fig:Vrec}), the expanding stochastic RCV (see Figure~\ref{fig:poincare}) engulfs the outer topological region (see Section~\ref{sec:outertopology}). Further, the reconnection rate approximately plateaus to a large typical value of $V_{\mathrm{rec}} \approx 0.020$ for the remainder of the simulation, approximately 6.4 times the rate of the initial Sweet-Parker phase.

We therefore consider that from $t = 3.0$ onwards (after \textbf{f} in Figure~\ref{fig:Vrec}) we observe \textit{``pure'' SGTR} in which the system has settled into quasi-stationary global dynamics. A true steady state is not strictly achieved since reconnection consumes the merging laminar flux ropes (see bottom row of Figure~\ref{fig:j}). If the simulation runtime was extended, we anticipate the large laminar flux ropes would fully reconnect and the reconnection rate would decay to zero. Nonetheless, the secular merging of the laminar flux ropes is sufficiently slow compared to the dynamics inside the reconnection layer that a quasi-stationary conceptualisation is instructive, if treated with due care.

\subsection{Reconnection layer thickness scales}\label{sec:thickness}

Turbulent reconnection is a dynamic 3D multiscale process involving the interaction of numerous spatial and temporal scales. Due to the mass conservation argument introduced by \citet{Parker1957}, the thickness $\delta$ of the Sweet-Parker layer is closely related to the reconnection rate $V_{\mathrm{rec}}$; details of this are discussed in Section~\ref{sec:SPscalings}. Hence, the characteristic thickness scales of the reconnection layer are important for understanding the global dynamics, due to the Sweet-Parker-type global magnetic topology of our simulation. The main aim of this section is to describe the various thicknesses that characterise the reconnection layer and compare their evolution over time. We also make important remarks on the properties of the magnetic topology inside the SGTR layer in Sections~\ref{sec:innertopology} and \ref{sec:outertopology}. 

We detect two major characteristic thickness scales: \textit{(a)} An \textit{inner thickness scale} associated with current and vorticity densities, reconnection outflows, and turbulent fluctuations; and \textit{(b)} An \textit{outer thickness scale} corresponding to the wider stochastic layer from a magnetic topology standpoint. The terms \textit{inner} and \textit{outer} scale are restricted to the description of the averaged thicknesses and properties only. We also identify flux rope structures inside the SGTR layer that are linked with both of these thickness scales. 

In this section, we determine thickness scales by averaging many individual measurements of the layer thickness $\delta y$ using a variety of quantities of interest. Later in Section~\ref{sec:meanprofiles}, we compare the mean profiles of variables during the SGTR phase when the global dynamics become quasi-stationary. The two approaches are complementary and determine consistent thickness scales; the method in this section has the advantage of displaying the time evolution over the whole simulation, while the method in Section~\ref{sec:meanprofiles} provides additional insights into the shapes of the mean profiles during SGTR.  It will be shown in Section~\ref{sec:SPscalings} that the observed reconnection rate connects best with the inner thickness scale; this implies that it is preferable to interpret SGTR as being controlled by the thickness of the effective reconnection electric field produced by turbulent fluctuations, rather than the larger thickness of the stochastic layer which includes additional regions referred to as the SGTR wings. 

\subsubsection{General approach}\label{sec:temporalevol}

\begin{figure*} 
\centering
\hspace*{-1.8cm}\includegraphics[width=1.2\textwidth]{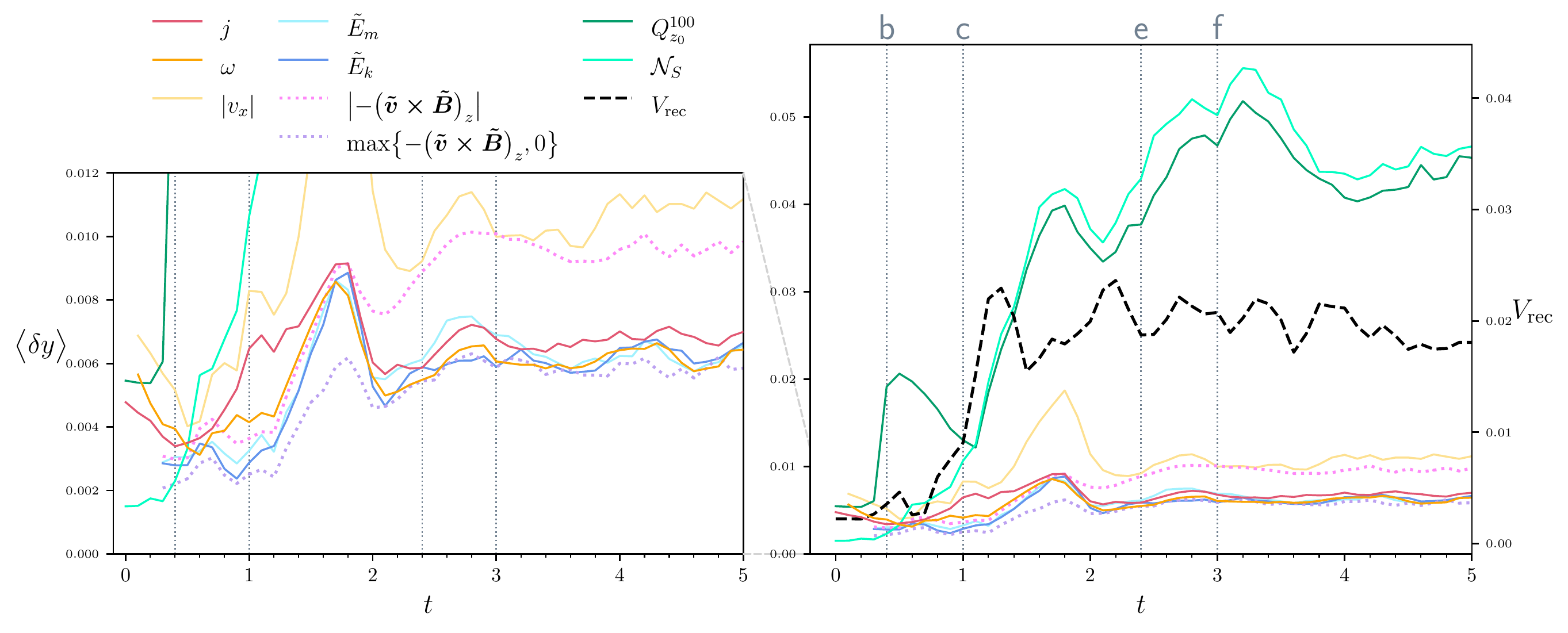}
\caption{Comparison of different measures of the characteristic reconnection layer thickness $\bigl< \delta y \bigr>$. The right panel compares $\bigl< \delta y \bigr>$ (left axis) with the global reconnection rate $V_{\mathrm{rec}}$ (right axis) over time. The left panel is a closeup of the $\bigl< \delta y \bigr>$ curves in the right panel over $\bigl< \delta y \bigr> \in [0, 0.012]$. Dark red: current density strength $j$; Orange: vorticity strength $\omega$; Yellow: outflow velocity $\vert v_x \vert$; Light blue: energy of the magnetic field fluctuations $\tilde{E}_m$; Dark blue: energy of the kinetic fluctuations $\tilde{E}_k$; Pink (dotted): unsigned turbulent EMF $\bigl\vert- \bigl(\boldsymbol{\tilde{v} \times \tilde{B}}\bigr)_z\bigr\vert$; Purple (dotted): positive turbulent EMF $\max\bigl\{-\bigl(\boldsymbol{\tilde{v}} \boldsymbol{\times} \boldsymbol{\tilde{B}}\bigr)_z,0\bigr\}$; Dark green: squashing factor $Q_{z_0}^{100}$ for $z_0 = -0.5$; Turquoise: Poincar\'e section count $\mathcal{N}_S$ for $n=2000$ iterations;  Black (dashed): global reconnection rate $V_{\mathrm{rec}}$. The vertical dotted lines mark times of interest: \textbf{b} 2.5D nonlinear tearing instability onset ($t= 0.4$); \textbf{c} Turbulent reconnection onset ($t= 1.0$); \textbf{e} SGTR onset ($t = 2.4$); \textbf{f} ``Pure'' SGTR  onset ($t = 3.0$).  \label{fig:delta}}
\end{figure*}

The general motivation and procedure of the time-dependent scale analysis is as follows. The thickness of the reconnection layer can be measured from many different quantities that all reflect complementary aspects of the reconnection dynamics. Common to all of these, the reconnection layer is periodic in the $z$ direction, localised away from the $x=\pm 0.5$ boundaries, and situated approximately about the $x$-$z$ plane at $y=0.0$. Therefore, we focus on measuring the layer thickness in the $y$ direction, denoted $\delta y$. From the onset of the 2.5D tearing instability at $t = 0.4$ (\textbf{b} in Figure~\ref{fig:Vrec}), the surfaces of the reconnection layer become convoluted and nonuniform and the layer slowly shortens in the $x$ direction. Furthermore, we are most interested in the thickness at the centre of the reconnection region, rather than the thickness of the outflow jets. Hence, we approximate $\delta y$ over a local region $(x,z) \in [-0.15,0.15] \times [-0.5,0.5)$, then take the arithmetic mean to obtain a single statistical measure of the layer thickness for each quantity of interest, denoted $\bigl< \delta y \bigr>$. Shorter intervals than $x \in [-0.15, 0.15]$ yield similar results, but are more sensitive to the expulsion of the large CFR structure over $t = 1.5\mbox{--}2.0$. 

For a variable $f(\boldsymbol{x})$, the thickness $\delta y$ at fixed coordinate $(x,z)$ is quantified using one of two methods:

\paragraph{i. Contour thickness} If $f(\boldsymbol{x})$ exhibits a well-defined natural boundary that can be estimated using contours of $f(\boldsymbol{x})$ for a suitable threshold $f_0$, then the \textit{contour thickness} can be found, i.e., the maximum $y$-distance between intersections of the major $f(\boldsymbol{x}) = f_0$ contour surface with each line of constant $(x,z)$.

\paragraph{ii. Effective FWHM} Otherwise, for $f(\boldsymbol{x})$ with a more complicated distribution and/or a less sharp boundary, $\delta y$ can be robustly quantified as the \textit{full-width at half-maximum} (FWHM) of an appropriate bell-curve model $g(y)$. For nonnegative variable $f(\boldsymbol{x}) \geq 0$, at fixed $(x,z)$, we evaluate the FWHM of $g(y)$ satisfying
\begin{linenomath*}\begin{equation*}
\int_{-a}^{a} f(\boldsymbol{x}) \ dy = \int_{-a}^{a} g(y) \ dy, \quad \max_{y\in[-a,a]} g(y)   = \max_{y\in[-a,a]} f(\boldsymbol{x})    
\end{equation*}\end{linenomath*} 
over a sufficiently wide window $y \in [-a,a]$. We chose $a=0.1$. This approach does not require fitting the chosen bell curve to the $f(\boldsymbol{x})$ profile; instead, it redistributes area under the curve into a single peak while preserving the maximum value, to estimate the spread of the $f(\boldsymbol{x})$ profile. The resultant \textit{``effective'' FWHM} is most productive when the dominant fluctuations of $f(\boldsymbol{x})$ are concentrated about $y=0.0$ and not highly granular. While the effective FWHM is quantitively dependent on the chosen bell-curve, the evolution of $\bigl< \delta y \bigr>$ will be qualitatively consistent for any Gaussian-like $g(y)$. Using suitable bell-curve 
\begin{linenomath*}\begin{equation*}
g(y) = \frac{g_{\max}}{\bigl(1+(y/\sigma)^2\bigr)^{3/2}}, \label{eqn:bell}
\end{equation*}\end{linenomath*}
with free parameters $g_{\max}>0 $ and $\sigma >0$, we obtain the following closed-form solution:
\begin{linenomath*}\begin{equation*}
\delta y (x,z) = 2a \sqrt{\frac{2^{2/3}-1}{\kappa^2 - 1}}, \quad \kappa = 2a \frac{\displaystyle \max_{y\in [-a,a]} f(\boldsymbol{x})}{\displaystyle\int_{-a}^{a} f(\boldsymbol{x}) \ dy}. 
\end{equation*}\end{linenomath*}

If we have $f(\boldsymbol{x}) < 0$ for some $\boldsymbol{x}$, then the effective FWHM can be evaluated after additional processing. For example, to filter and measure the positive fluctuations only we used $\max\left\{f(\boldsymbol{x}), 0\right\}$, whereas to quantify the unsigned fluctuations we used $\vert f(\boldsymbol{x}) \vert$. 

In practice, the $\delta y$ measurements employing the effective FWHM were found to be very robust and reliable. Consistent results for $\bigl< \delta y \bigr>$ were also found using the FWHM of a best-fit Gaussian distribution to the profile of $f(\boldsymbol{x})$. However, this alternative technique was prone to erroneous $\delta y$ values, mainly due to errors with fitting a Gaussian distribution to profiles that exhibit substantial numerical noise or multiple peaks. 

The temporal evolution of all the characteristic thicknesses that we will consider in the proceeding subsections is shown in Figure~\ref{fig:delta}. For comparison, the global reconnection rate $V_{\mathrm{rec}}$ from Figure~\ref{fig:Vrec} has been superimposed.

\subsubsection{Inner thickness scale} \label{sec:innerlayer}

Firstly, we identify an \textit{inner thickness scale} corresponding with the mean properties of the current, vorticity, outflows, and MHD turbulent fluctuations. Due to the fragmented distribution of the physical variables, we employ the effective FWHM method, which was found to be suitable for all the examined quantities from $t = 0.3$ onwards; these are plotted in Figure~\ref{fig:delta}.

\paragraph{Current and vorticity densities} The mean thicknesses $\bigl< \delta y \bigr>$  of the current density strength $j = \Vert \boldsymbol{\nabla \times B} \Vert$ (dark red) and the vorticity strength $\omega = \Vert \boldsymbol{\nabla \times v}\Vert$ (orange) are mostly qualitatively and quantitively consistent. We observe some thinning prior to the 2.5D tearing onset at $t = 0.4$ (\textbf{b} in Figure~\ref{fig:delta}), followed by rapid broadening around the turbulent reconnection onset $t = 1.0$ (\textbf{c} in Figure~\ref{fig:delta}); this correlates with the global reconnection rate $V_{\mathrm{rec}}$. The $\omega$ thickness does not grow substantially until after $t = 1.0$  since the reconnection layer is initially laminar before becoming increasingly turbulent after the kink instability. A global maximum is reached around $t= 1.8$ due to the CFR remnant, then a significant dip occurs before $t= 2.4$ (\textbf{e} in Figure~\ref{fig:delta}) while the CFR remnant exits the local averaging region. The thicknesses reach a mean quasi-stationary value after $t= 3.0$ (\textbf{f} in Figure~\ref{fig:delta}) once pure SGTR sets in: $\bigl< \delta y \bigr> \approx 0.0067$ for $j$ and $\bigl< \delta y \bigr> \approx 0.0061$ for $\omega$. A discussion on the current and vorticity coherent structures that produce these averages can be found in Section~\ref{sec:FRS}.

\paragraph{Outflow jets} The reconnection process also forms distinctive outflow jets in $v_x$ that remain robust and roughly antisymmetric about $x=0.0$ during the simulation. To assign a mean thickness to the central section of these jets, we apply the effective FWHM to the absolute value $\vert v_x \vert$ (yellow). The evolution of this thickness is qualitatively similar to the current density $j$ and vorticity $\omega$, although the typical value is slightly larger: after $t= 3.0$, the mean quasi-stationary value is  $\bigl< \delta y \bigr> \approx 0.011$. The use of $\vert v_x \vert$ returns a thickness that is greater than the interior core of the jets, i.e., approximate maximum ridges of $v_x <0$ for $x <0$ and $v_x>0$ for $x >0$, but it was found to be the most robust approach to measure structures within the outflows. 

\paragraph{MHD turbulent fluctuations} One of the most important features of SGTR is that the effective reconnection electric field is provided by MHD turbulence generated by the reconnection process. To begin with, we can quantify the thicknesses of the turbulence using \textit{energies of the magnetic and kinetic fluctuations} defined as
\begin{linenomath*}\begin{equation*}
 \tilde{E}_m =\frac{ \Vert \boldsymbol{\tilde{B}} \Vert^2}{2}, \quad \tilde{E}_k =\frac{ \Vert \boldsymbol{\tilde{w}} \Vert^2}{2}, 
\end{equation*}\end{linenomath*}
respectively, where $\boldsymbol{w} = \sqrt{\rho} \boldsymbol{v}$  \citep{KidaOrszag1992}. Here, we follow \citet{HuangBhattacharjee2016}, who defined the fluctuating component of variable $f(\boldsymbol{x})$ to be $\tilde{f} = f - \bar{f}$, where $\bar{f}$ denotes the mean or background component. We take the mean over $z \in [-0.5,0.5)$ as a proxy for $\bar{f}$ since the simulation has approximate translational symmetry over the $z$ direction. Therefore, the primary contributions to $\tilde{f}$ correspond to 3D dynamics within the reconnection layer. The mean thicknesses of $\tilde{E}_m$ (light blue) and  $\tilde{E}_k$ (dark blue) are found to closely track each other and the mean thickness associated with the vorticity strength $\omega$. The mean quasi-stationary values after $t= 3.0$ are $\bigl< \delta y \bigr> \approx 0.0063$ for $\tilde{E}_m$ and $\bigl< \delta y \bigr> \approx 0.0062$ for $\tilde{E}_k$.

Next, we consider the electric field component $E_z$; this is a major driver of turbulent reconnection process for our simulation, especially during the SGTR phase. In the quasi-stationary state, several (although not all) of the core principles of Sweet-Parker reconnection can be applied, including that the $E_z$ averaged over time and $z$ is (almost) constant across the reconnection layer, as a consequence of Faraday's law. From Ohm's Law, we have
\begin{eqnarray}
E_z & = & -\bigl(\boldsymbol{v \times B}\bigr)_z + \eta j_z ,\nonumber \\
& = & -\bigl(\boldsymbol{\bar{v}} \boldsymbol{\times} \boldsymbol{\bar{B}}\bigr)_z-\bigl(\boldsymbol{\bar{v}} \boldsymbol{\times} \boldsymbol{\tilde{B}}\bigr)_z  -\bigl(\boldsymbol{\tilde{v}} \boldsymbol{\times} \boldsymbol{\bar{B}}\bigr)_z \nonumber \\
&& -\bigl(\boldsymbol{\tilde{v}} \boldsymbol{\times} \boldsymbol{\tilde{B}}\bigr)_z + \eta j_z, \label{eqn:Ez}
\end{eqnarray}
after decomposing $\boldsymbol{v}$ and $\boldsymbol{B}$ in terms of their mean and fluctuating components. The important terms in equation (\ref{eqn:Ez}) are the background electromotive force (EMF) $-\bigl(\boldsymbol{\bar{v}} \boldsymbol{\times} \boldsymbol{\bar{B}}\bigr)_z$, turbulent EMF $-\bigl(\boldsymbol{\tilde{v}} \boldsymbol{\times} \boldsymbol{\tilde{B}}\bigr)_z$ and resistive EMF $\eta j_z$ \citep[see][]{HuangBhattacharjee2016}. The \textit{turbulent EMF} can also be used to quantify the turbulent layer, by measuring the thickness of the unsigned perturbations $\bigl\vert-\bigl(\boldsymbol{\tilde{v}} \boldsymbol{\times} \boldsymbol{\tilde{B}}\bigr)_z\bigr\vert$ (dotted pink) or positive perturbations $\max\bigl\{-\bigl(\boldsymbol{\tilde{v}} \boldsymbol{\times} \boldsymbol{\tilde{B}}\bigr)_z,0\bigr\}$ (dotted purple). The mean thickness for $\bigl\vert-\bigl(\boldsymbol{\tilde{v}} \boldsymbol{\times} \boldsymbol{\tilde{B}}\bigr)_z\bigr\vert$ closely tracks the inner thicknesses prior to $t = 1.8$, especially $\omega$, $\tilde{E}_m$ and $\tilde{E}_k$ which probe the MHD turbulence region. It continues to broaden until levelling after $t = 2.4$ (\textbf{e} in Figure~\ref{fig:delta}), where it tracks $\vert v_x\vert$ and reaches a mean quasi-stationary value $\bigl< \delta y \bigr> \approx 0.0096$ after $t= 3.0$. The mean thickness for $\max\bigl\{-\bigl(\boldsymbol{\tilde{v}} \boldsymbol{\times} \boldsymbol{\tilde{B}}\bigr)_z,0\bigr\}$ has a similar evolution to $\bigl\vert-\bigl(\boldsymbol{\tilde{v}} \boldsymbol{\times} \boldsymbol{\tilde{B}}\bigr)_z\bigr\vert$ but is smaller and agrees with $j$, $\omega$, $\tilde{E}_m$ and $\tilde{E}_k$ after $t = 2.4$; its mean quasi-stationary value is $\bigl< \delta y \bigr> \approx 0.0058$ after $t= 3.0$. Section~\ref{sec:meanprofiles} further discusses the $E_z$ decomposition in equation (\ref{eqn:Ez}) and demonstrates that the turbulent EMF dominates over the resistive EMF in our simulation.

\subsubsection{Magnetic topology inside the SGTR layer}\label{sec:innertopology}

\begin{figure*} 
\centering
\hspace*{-1.8cm}\includegraphics[width=1.19\textwidth]{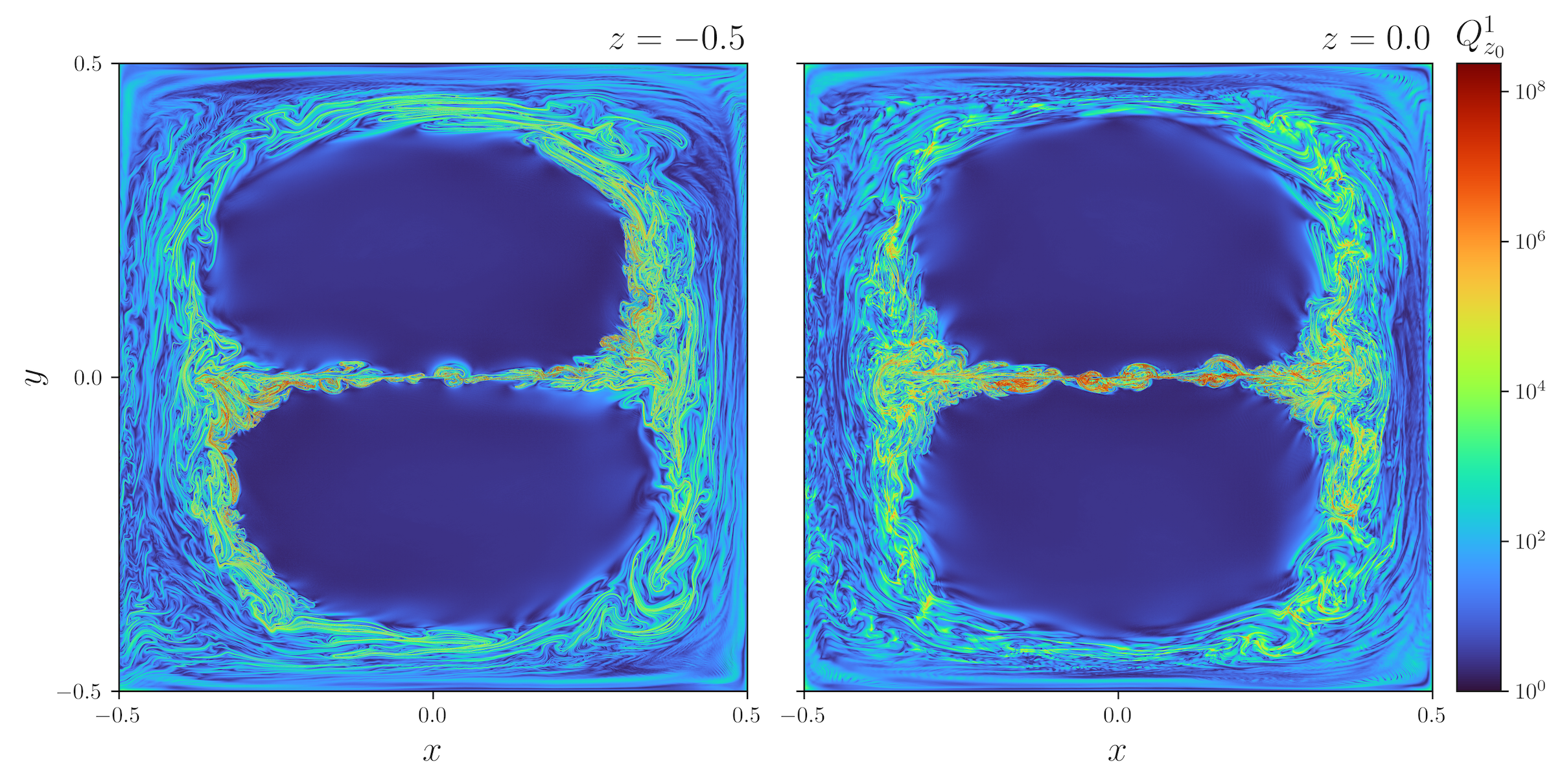} 
\caption{The squashing factor $Q_{z_0}^{1}$ with $z_0 = -0.5$ at $t = 4.0$. The left panel is plotted at the bottom boundary $z=-0.5$. The right panel is plotted at midplane $z=0.0$, assuming each field line connecting $(x_0,y_0)$ to $F_{z_0}^{1}(x_0,y_0)$ is identified with value $Q_{z_0}^{1}$. \label{fig:Q}} 
\end{figure*}

A complementary perspective on the interior of the SGTR layer comes from magnetic topology. In this approach, flux ropes and other regions of topological complexity can be identified using tools such as the \textit{squashing factor} \citep{TitovEA2002,PontinHornig2015,PontinEA2016,PontinEA2017,ScottEA2017}
\begin{equation}
Q_{z_0}^{n}(x_0,y_0) = \frac{\Vert DF_{z_0}^{n} \Vert_{\mathrm{F}}^2}{\vert \mathrm{det}(DF_{z_0}^{n})\vert}, \label{eqn:Q}
\end{equation}
where $DF_{z_0}^{n}$ is the Jacobian of the field line mapping $F_{z_0}^{n}$ [definition (\ref{eqn:Fmap})] and $\Vert\cdot\Vert_{\mathrm{F}}$ denotes the Frobenius norm. This is a metric that quantifies the degree of deformation of the field lines between two planes, which is useful since reconnection preferentially occurs at regions where $F_{z_0}^{n}$ possesses large gradients, through the formation of intense current layers \citep{PontinEA2016}. It also indicates regions with substantial turbulence and field line mixing. In particular, $\Vert DF_{z_0}^{n} \Vert_{\mathrm{F}}^2$ quantifies the rate of divergence of field lines emerging from infinitesimally close foot points at $\boldsymbol{x}_0$, while $\vert \mathrm{det}(DF_{z_0}^{n})\vert$ quantifies the dilation factor of an infinitesimal area element at $\boldsymbol{x}_0$ under the field line mapping. The standard practice is to assign the value $Q_{z_0}^{n}(x_0,y_0)$ to the whole field line from $(x_0,y_0)$ to $F_{z_0}^{n}(x_0,y_0)$, so that $Q_{z_0}^{n}$ can be plotted on any slice between $z = z_0$ and $z = z_0+n$ \citep[e.g.,][]{PontinHornig2015}. For our simulation, the determinant term simplifies to the ratio of the normal field $B_z>0$ evaluated at foot point $\boldsymbol{x}_0$ and mapped point $\bigl(F_{z_0}^{n}(x_0,y_0), z_0+n\bigr)$ \citep{ScottEA2017}:
\begin{linenomath*}\begin{equation*}
\bigl\vert \mathrm{det}\bigl(DF_{z_0}^{n}(x_0,y_0)\bigr)\bigr\vert = \frac{B_z(x_0,y_0,z_0)}{B_z\bigl(F_{z_0}^{n}(x_0,y_0), z_0+n\bigr)}.
\end{equation*}\end{linenomath*}
The partial derivatives within $DF_{z_0}^{n}$  are approximated by integrating four neighbouring field lines about the foot point $\boldsymbol{x}_0$ of the main field line then taking central differences
\begin{eqnarray*}
\frac{\partial [F_{z_0}^{n}]^i}{\partial x}(x_0,y_0) & \approx & \frac{[F_{z_0}^{n}]^i(x_0 + \varepsilon_x,y_0) - [F_{z_0}^{n}]^i(x_0 - \varepsilon_x,y_0)}{2\varepsilon_x}, \\
\frac{\partial [F_{z_0}^{n}]^i}{\partial y}(x_0,y_0) & \approx & \frac{[F_{z_0}^{n}]^i(x_0,y_0 + \varepsilon_y) - [F_{z_0}^{n}]^i(x_0,y_0 - \varepsilon_y)}{2\varepsilon_y},
\end{eqnarray*}
for sufficiently small step sizes $\varepsilon_x$ and $\varepsilon_y$. 

We consider $Q_{z_0}^{1}$ for field lines seeded from the bottom boundary $z_0=-0.5$, to investigate the distortion of field lines during a single passage through simulation domain. The left panel of Figure~\ref{fig:Q} shows $Q_{z_0}^{1}$ at $z = -0.5$ over the cross-sectional plane at $t=4.0$, revealing an internal region of strong gradients that highlight flux rope structures and thin reconnection layers. Here, the distribution of $Q_{z_0}^{1}$ is skewed since the pair of laminar flux ropes that are merging have a left-handed orientation, i.e., field lines that possess $Q_{z_0}^{1} \gg 1$ in this figure include some that rotate and enter the reconnection layer as they are traced upwards from the bottom boundary. To obtain a symmetrical profile, we plot the same $Q_{z_0}^{1}$ sampled at the midplane $z=0.0$. The ``frayed'' flux ropes structures that the squashing factor picks out inside the reconnection layer show strong variation along the $z$ direction (also seen in locally-defined quantities including $j$ and $\rho$), and the midplane diagram reveals a superposition pattern of ridges of $Q_{z_0}^{1}$. 

While cross-sections of the flux rope structures resemble plasmoids in 2D systems, the field lines they consist of are fully stochastic and do not form flux surfaces. This is consistent with many related studies that have explored similarly complex magnetic topologies, such as filamentary flux rope structures ``hidden'' within stochastic field line regions that are otherwise undetected using Poincare sections \citep[e.g.,][]{BorgognoEA2011a, BorgognoEA2015, RubinoEA2015, FalessiEA2015, BorgognoEA2017, DiGiannataleEA2017b, SistiEA2019}. We also find that cross-sections of the flux rope structures agree very well with structures observed in cross-sections of physical variables in Section~\ref{sec:innerlayer}, e.g., for the current density strength $j$ compare left column, bottom two rows of Figure~\ref{fig:j} with Figure~\ref{fig:Q}. These important structures generate the mean properties and govern the underlying layer dynamics, e.g., shaping the SGTR layer and driving the turbulent EMF; they will be touched upon later in Section~\ref{sec:outertopology}, and a detailed examination and a discussion on their possible characterisation is left until Section~\ref{sec:FRS}. 

The squashing factor is very similar to the \textit{(maximal) finite-time Lyapunov exponent} (FTLE) of field lines in the \textit{$z$ direction} at fixed time $t$. Under the field line mapping $F_{z_0}^{n}$ [definition (\ref{eqn:Fmap})], the FTLE is defined as \citep{Haller2015}
\begin{equation}
\Lambda_{z_0}^{n}(x_0,y_0) = \frac{1}{n} \ln\left(\lambda_{\max}(G_{z_0}^{n})^{1/2}\right), \label{eqn:FTLE}
\end{equation}
where $\lambda_{\max}(G_{z_0}^{n})$ denotes the maximum eigenvalue of  the displacement (or left Cauchy-Green strain) tensor $G_{z_0}^{n}(x_0,y_0) = (DF_{z_0}^{n})(DF_{z_0}^{n})^T$. The FTLE measures the average rate of exponential divergence of field lines over an arbitrary distance \citep{KantzSchreiber2003, YeatesEA2012}, i.e., \ for two seed points $\boldsymbol{x}_0,\boldsymbol{x}_1$ at $z=z_0$ with initial distance $\delta_0 = \Vert \boldsymbol{x}_0 - \boldsymbol{x}_1 \Vert \ll 1$, the separation $\delta_n = \Vert F_{z_0}^{n}(\boldsymbol{x}_0) - F_{z_0}^{n}(\boldsymbol{x}_1) \Vert$ at $z=z_0+n$ satisfies $\delta_n \approx \delta_0 \exp\left(n\Lambda_{z_0}^{n}(\boldsymbol{x}_0)\right)$ for $n \gg 1$. 

The squashing factor and FTLE definitions are primarily characterised by different matrix norms of the Jacobian \citep{YeatesEA2012}: $Q_{z_0}^{n}$ uses the Frobenius norm $\Vert DF_{z_0}^{n}\Vert_{\mathrm{F}} = \mathrm{tr}\bigl(G_{z_0}^{n}\bigr)^{1/2}$, whereas $\Lambda_{z_0}^{n}$ uses the $l_2$-norm (spectral norm) $\Vert DF_{z_0}^{n}\Vert_2 = \lambda_{\max}(G_{z_0}^{n})^{1/2}$. In fact, we have the equivalence $\Lambda \approx \ln\left(Q^{1/2}\right)/n$ in the limit as $\Lambda \gg 1$ if $Q_{z_0}^{n} \approx \Vert DF_{z_0}^{n}\Vert_{\mathrm{F}}^2$ \citep{YeatesEA2012,HuangEA2014,PontinHornig2015}; in our case, this agreement is particularly strong. Another equivalent approach to the FTLE is the \textit{exponentiation factor} $\sigma$ or related variations \citep[see][etc.]{Boozer2012, HuangEA2014, DaughtonEA2014, LeEA2018, StanierEA2019, LiEA2019}; these also give similar results to the squashing factor.

\subsubsection{Outer thickness scale}\label{sec:outerlayer}

Now, we recognise an \textit{outer thickness scale} corresponding to the stochastic layer that develops during the turbulent reconnection process, highlighted earlier in Figure~\ref{fig:poincare}. We provide two different topological tools to quantitively measure the characteristic thickness of the stochastic layer, using the local separation rate of field lines, or a 3D Poincar\'e section. 

\paragraph{Local separation rate of field lines} One approach to measure the stochastic layer is to evaluate a metric that indicates where field lines are stochastic. By utilising the simulation's periodicity in $z$, the \textit{squashing factor} $Q_{z_0}^{n}$ in definition (\ref{eqn:Q}) for a \textit{large number of iterations} $n \in \mathbb{N}$ is one suitable candidate. Since field lines that penetrate the stochastic regions are ergodic and eventually experience strong local separation as they cycle around the toroidal space, the \textit{order of magnitude} of $Q_{z_0}^{n}$ for $n\gg 1$ reveals laminar [$Q_{z_0}^{n} = \mathcal{O}(1)$] and stochastic [$Q_{z_0}^{n} \gg 1$] regions of the system.  A similar method, employing the exponentiation factor $\sigma$, was used in \citet{DaughtonEA2014}.\footnote{To identify stochastic separatrices (if they exist) throughout non-periodic domains, the squashing factor $Q_{z_0}^{n}$ and Poincar\'e section count $\mathcal{N}_S$ approaches are not applicable. \citet{DaughtonEA2014} provided a fast method to detect stochastic regions, applicable to both periodic and non-periodic domains, using particle mixing as a proxy within the context of kinetic simulations. The results were found to be consistent with a measurement of the local separation rate of field lines (see their Figure 4). The particle mixing technique was successfully employed or similarly adapted in later kinetic simulations \citep{DahlinEA2017, LeEA2018, StanierEA2019} and an MHD simulation by \citet{YangEA2020}; however, it has been shown to not be robust in general \citep[e.g.,][]{BorgognoEA2017}. } Informal justification for this class of methods comes from the definition of the \textit{maximal Lyapunov exponent} $\Lambda_{\max} = \limsup_{n \to \infty} \Lambda_{z_0}^{n}$ \citep[e.g.,][]{Temam1988, BorgognoEA2011a, RubinoEA2015}, where we have $\Lambda_{\max} = \infty$ for random noise, i.e., stochastic field lines. The numeric value $Q_{z_0}^{n}$ in high-$Q$ regions is not a quantitative measure of the stochasticity, but it does indicate the degree to which the field line mapping is sensitive to the foot points.  

After generating samples of $Q_{z_0}^{n}$ over regular 2D grids, it was found that $n \gtrsim 10$ iterations was sufficient to illuminate the main structures making up the stochastic layer; these regions continue to fill and appear to converge, with the distribution of $Q_{z_0}^{n}$ remaining qualitatively consistent up to at least $n=100$. By assuming that $Q_{z_0}^{n}(x_0,y_0)$ can be assigned to the entire field line connecting $(x_0,y_0)$ to $F_{z_0}^{n}(x_0,y_0)$, we can efficiently approximate $Q_{z_0}^{n}$ throughout the 3D space by exploiting periodicity in $z$ and employing an irregular grid approach. After evaluating $Q_{z_0}^{n}(x_0,y_0)$, we store the respective field line over many grid positions in the $z$ direction. Repeating this for many seed points $(x_0,y_0)$, the final dataset is a collection of points, each weighted by the corresponding $Q_{z_0}^{n}$ value, on every $z$-slice. Once this dataset is sufficiently dense for large $n$, $Q_{z_0}^{n}$ values at coordinates that have not been directly sampled can be approximated using interpolation.
 
A main challenge in implementing this approximation method is that suitable seed points need to be chosen to resolve the interfaces between the stochastic region and large laminar flux ropes, referred to as \textit{stochastic separatrices}  \citep{ParnellEA2010, Pontin2011, DaughtonEA2014}. The stochastic separatrices are convoluted and difficult to approximate, possibly because they form fractals. Further, within the stochastic layer before $t = 1.8$, there are some sizeable regions corresponding to insular orbits that require a dense grid of seed points to ensure sufficient sampling. While these issues were not major in practice, they limited the resolution that could be realistically obtained without excessive computational effort. 

The approximation method also operates under the assumption that the $Q_{z_0}^{n}$ value associated with a particular field line is at most weakly dependent on $z_0$ for a sufficiently large number of iterations $n \gg 1$. This holds well for our simulation, especially for the purposes of an order of magnitude comparison. From comparisons of the 3D approximation with direct samples of $Q_{z_0}^{n}$ over regular 2D grids, we found that they were in very close agreement for $n \gtrsim 50$.

\begin{figure*} 
\centering
\gridline{\hspace*{-1.8cm}\fig{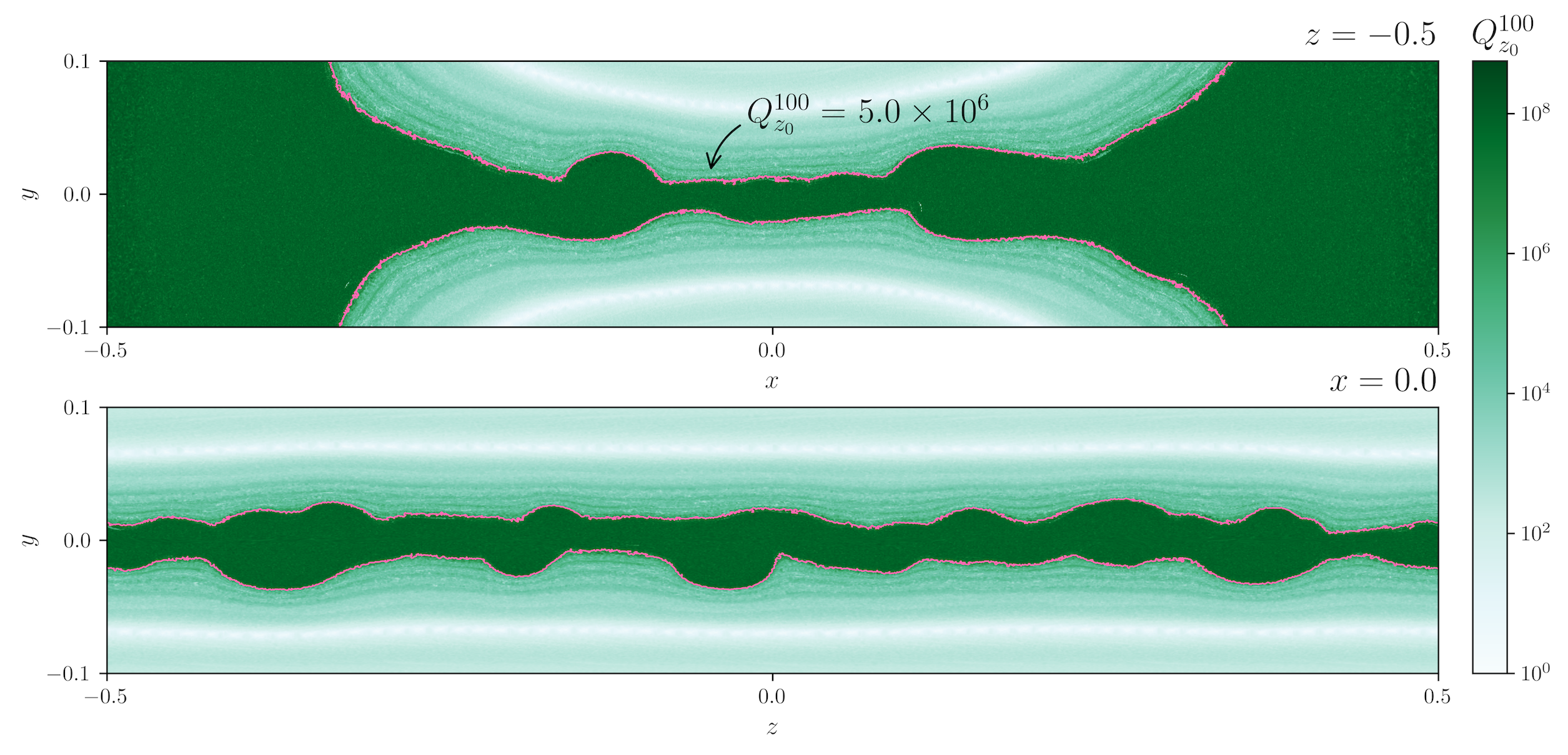}{1.19\textwidth}{(a)}}
\vspace*{-0.5cm}\gridline{\hspace*{-1.8cm}\fig{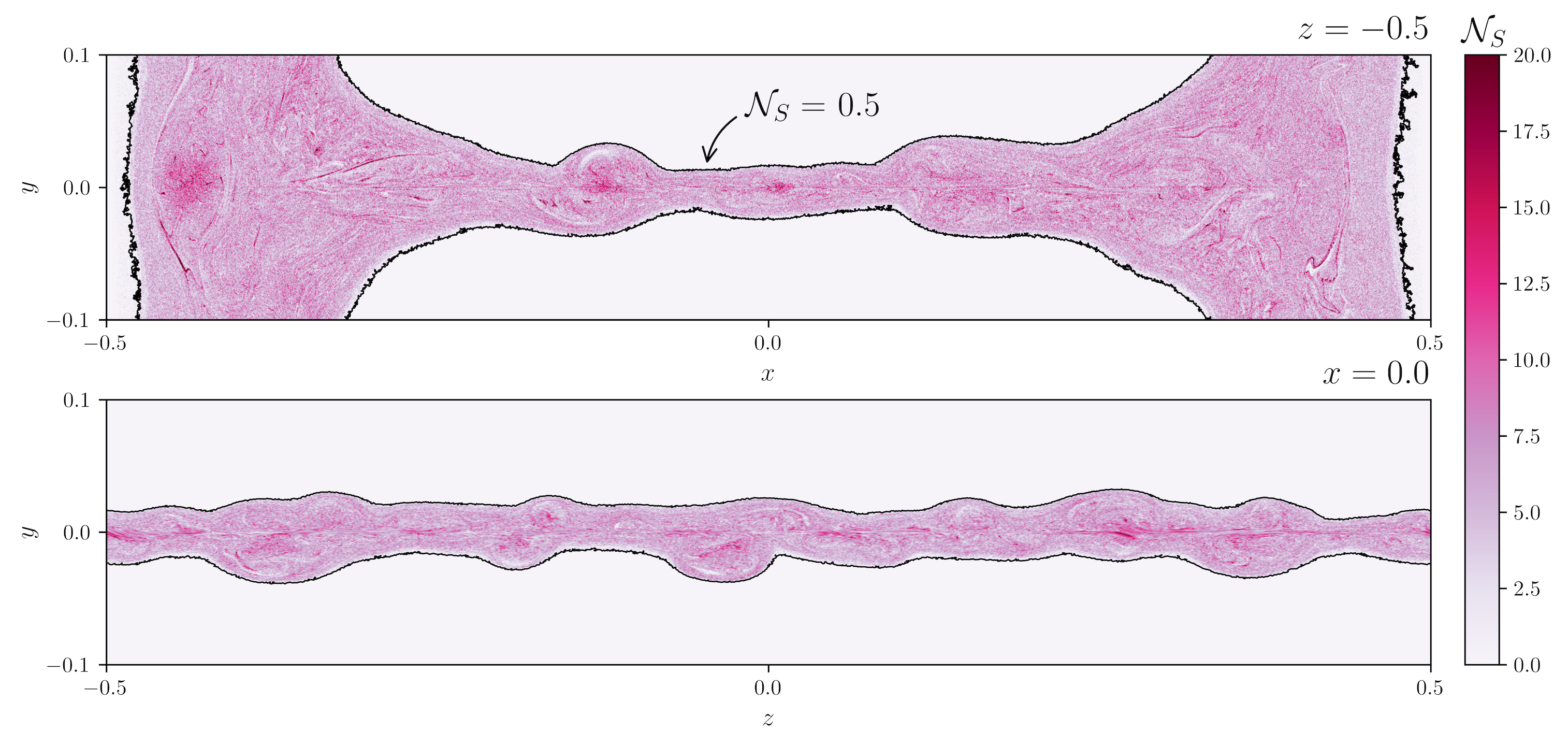}{1.19\textwidth}{(b)}}
\caption{2D slices of variables highlighting the stochastic layer, corresponding with the outer thickness scale, at $t= 2.5$. The top and bottom panels of each subfigure are slices at $z = -0.5$ and $x = 0.0$, respectively. Subfigure (a): squashing factor $Q_{z_0}^{100}$ with $z_0 = -0.5$ approximated over 3D space; the stochastic separatrices are estimated using contours $Q_{z_0}^{100} = 5 \times 10^6$ (pink). Subfigure (b): Poincar\'e section count $\mathcal{N}_S$ for $n=2000$ iterations; contours of $\mathcal{N}_S = 0.5$ (black) approximate the boundary of the reconnection-connected volume (RCV).  \label{fig:QnNS}} 
\end{figure*}

Figure~\ref{fig:QnNS}(a) shows the 3D $Q_{z_0}^{n}$ at $t = 2.5$ using $z_0=-0.5$ over $n=100$ iterations. A $z$-slice (top panel) and $x$-slice (bottom panel) through the reconnection layer are provided, demonstrating that the stochastic layer becomes highly varied in its thickness and shape. We find that the stochastic separatrices are well defined and can be effectively approximated by contours of $Q_{z_0}^{100} = 5 \times 10^6$, shown in pink. 

Figure~\ref{fig:delta} shows the evolution of mean contour thickness $\bigl< \delta y \bigr>$ of $Q_{z_0}^{100}$ (dark green) for the same parameters $z_0=-0.5$ over $n=100$. Consistent with our observations in Section~\ref{sec:pathway}, the stochastic layer is initially thin then rapidly expands over the transition phase $t = 1.0\mbox{--}2.4$ (labels \textbf{c}--\textbf{e}). Most importantly, $\bigl< \delta y \bigr>$ is significantly larger in magnitude compared to the inner thicknesses in Section~\ref{sec:innerlayer}, hence an outer thickness scale. After $t= 3.0$ (label \textbf{f}), we obtain the mean value $\bigl< \delta y \bigr> \approx 0.045$, which is broader than the inner thicknesses by a factor of $4.2\mbox{--}7.6$; however, this quasi-stationary value is more variable than the previous thickness measurements. Qualitatively, the thickness of $Q_{z_0}^{100}$ correlates well with the (narrower) thicknesses of $-\bigl(\boldsymbol{\tilde{v}} \boldsymbol{\times} \boldsymbol{\tilde{B}}\bigr)_z$ (dotted pink and dotted purple) and the global reconnection rate $V_{\mathrm{rec}}$; these results suggest that the stochasticity is an essential part of the global turbulent reconnection properties. The minor bump between $t = 0.4\mbox{--}1.0$ (labels \textbf{b}--\textbf{c}) is also observed in the thicknesses for the fluctuations $\tilde{E}_m$ (light blue), $\tilde{E}_k$ (dark blue), and $-\bigl(\boldsymbol{\tilde{v}} \boldsymbol{\times} \boldsymbol{\tilde{B}}\bigr)_z$ (dotted pink and dotted purple). This bump corresponds to the short-lived occurrence and disappearance of a band of moderately large $Q_{z_0}^{100}$ around the early reconnection layer, coincident with the 2.5D tearing onset (label \textbf{b}); this brief secondary layer may be linked to the formation of the initial flux ropes and the spread of stochasticity.

\paragraph{3D Poincar\'e section} An alternative approach to detect the stochastic layer is by approximating the reconnection-connected volume (RCV), previously observed in the 2D Poincar\'e plots (Figure~\ref{fig:poincare}). The main idea is to construct a \textit{3D Poincar\'e section} by directly filling the ergodic RCV with field lines that occupy it. This is complementary to the squashing factor $Q_{z_0}^{100}$ results, and is considerably less computationally expensive, and considerably less space intensive, since we only need to integrate field lines $\boldsymbol{x}_B$ [equation (\ref{eqn:odesys})]. Using seed points $\boldsymbol{x}_0$ along $y=0.0$ over $x \in [-0.42,0.42]$ at the bottom boundary $z_0=-0.5$, we trace field lines over a large number of iterations of the mapping $F_{z_0}^{n}$ [definition (\ref{eqn:Fmap})] then evaluate their distribution over a regular 3D grid to obtain a \textit{Poincar\'e section count}, denoted $\mathcal{N}_S$. Points inside the RCV are distinguished by $\mathcal{N}_S > 0$. The boundary surfaces between the RCV and the laminar flux ropes are approximated using contours of $\mathcal{N}_S = \mathcal{N}_{S0} < 1$, which in turn approximate the stochastic separatrices, assuming that the ergodic regions are sufficiently filled and the boundaries are well-resolved at the chosen grid scale. 

Figure~\ref{fig:QnNS}(b) displays the Poincar\'e section count $\mathcal{N}_S$ at $t = 2.5$ using $n=2000$ iterations of the field line mapping; for comparison with $Q_{z_0}^{100}$ in Figure~\ref{fig:QnNS}(a), the same 2D slices have been given. We observe that the RCV closely matches the stochastic layer highlighted by $Q_{z_0}^{100}$, and its boundary can be effectively estimated by contours of $\mathcal{N}_S = 0.5$, shown in black.  The thicknesses $\delta y$ are slightly wider than the $Q_{z_0}^{100}$ thicknesses (by about $6\%$ after $t = 3.0$), presumably because the computationally lighter $\mathcal{N}_S$ could be evaluated for a higher number of iterations and therefore better fills the true topological region. 

In principle, regions where $\mathcal{N}_S>0$ may not strictly be stochastic: the initial seed line may pick laminar islands that may exist within the otherwise stochastic RCV, and we have evidence in rare cases of minor numerical leakage into the laminar regions which might exaggerate the spread of the RCV. These issues are not major, and the squashing factor results $Q_{z_0}^{100}$ in Figure~\ref{fig:QnNS}(a) support that the reconnection layer becomes almost entirely stochastic after $t = 1.8$. We also comment that the boundary detected between the RCV and the outer topological region [black lines near the $x=\pm0.5$ boundaries in Figure~\ref{fig:QnNS}(b)] is not strictly a stochastic separatrix: it is only a topological boundary, up to $n=2000$ iterations for our collection of seed points. The outer topological region quickly becomes high-$Q$ after $t = 0.8$ and is nearly indistinguishable from the stochastic RCV by $t = 2.2$ using $Q_{z_0}^{n}$ for $n \gtrsim 50$. The general agreement is clear by comparing the top panels of Figures~\ref{fig:QnNS}(a) and \ref{fig:QnNS}(b). 

The evolution of the mean contour thickness $\bigl< \delta y \bigr>$ of $\mathcal{N}_S$ for $n=2000$ is shown in Figure~\ref{fig:delta} (turquoise). This agrees very closely with the mean thickness for $Q_{z_0}^{100}$ (dark green), excluding the minor bump between $t = 0.4\mbox{--}1.0$, indicating that the RCV is for practical purposes almost equivalent to the stochastic layer. The quasi-stationary mean thickness for $\mathcal{N}_S$ after $t= 3.0$ (label \textbf{f}) is $\bigl< \delta y \bigr> \approx 0.047$, which is $4.5\mbox{--}8.1$ times greater than the inner thicknesses.

\subsubsection{Local magnetic topology versus stochasticity}\label{sec:outertopology}

\begin{figure*} 
\centering
\hspace*{-1.8cm}\includegraphics[width=1.19\textwidth]{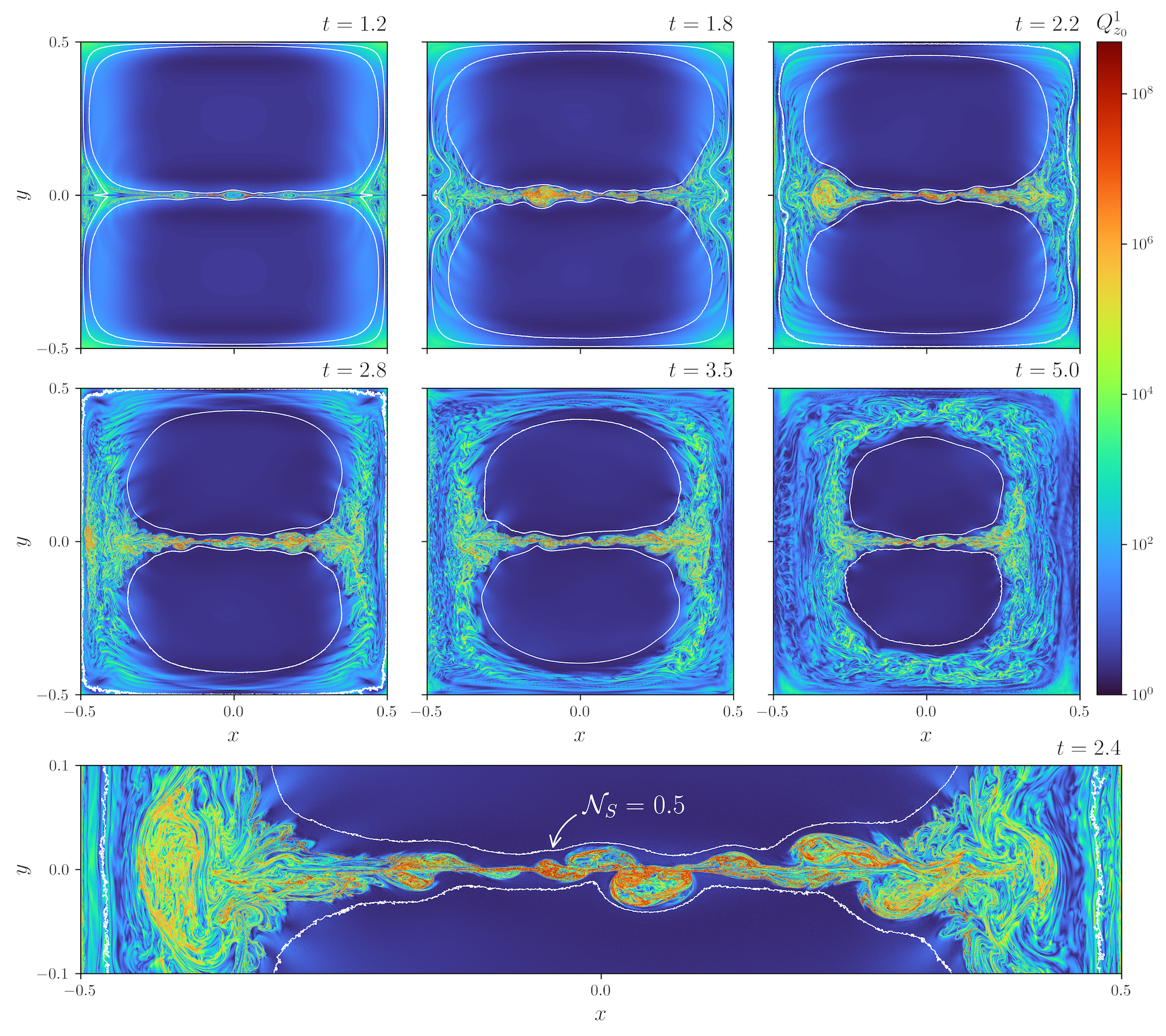} 
\caption{Comparison plots of the squashing factor $Q_{z_0}^{1}$ with $z_0 = -0.5$ and the reconnection-connected volume (RCV) boundaries at $z = 0.0$ over time. The bottom panel is a closeup of the reconnection layer at $t = 2.4$. The RCV boundaries are approximated by contours of the Poincar\'e section count $\mathcal{N}_S = 0.5$ for $n=2000$ iterations (white). The interior contours of the RCV coincide with the stochastic separatrices or laminar flux rope boundaries, whereas $Q_{z_0}^{1}$ highlights the flux rope structures within the reconnection layer. \label{fig:QNScomp}} 
\end{figure*}

Finally, we refer back to the flux rope structures detected in Section~\ref{sec:innertopology} and compare them to the larger stochastic layer in which they are embedded. Firstly, the dual usage of $Q_{z_0}^{n}$ for highlighting the flux rope structures (Section~\ref{sec:innertopology}, $n=1$) and characterising the outer stochastic layer (Section~\ref{sec:outerlayer}, $n=100$) demonstrates that both constructions are connected and threaded by the same field lines. Figure~\ref{fig:QNScomp} overlays the squashing factor $Q_{z_0}^{1}$ with $z_0=-0.5$ at $z=0.0$ (from Figure~\ref{fig:Q}) with contours of $\mathcal{N}_S = 0.5$ for $n=2000$ (white). The upper panels are samples over time, whereas the bottom panel is a close up of the reconnection layer at the SGTR phase onset $t = 2.4$ (\textbf{e} in Figure~\ref{fig:Vrec}). Consistent with Section~\ref{sec:pathway},  Figure~\ref{fig:QNScomp} illustrates that the large laminar flux ropes shrink over time and the outer topological region is engulfed by the RCV by $t = 3.0$ (\textbf{f} in Figure~\ref{fig:Vrec}). We observe that the RCV contains the flux rope structures, evidencing that there exist internal features of the reconnection layer that are smaller than the stochastic thickness, and detectable using only topological properties of the magnetic field. Further, the lack of structural features illuminated by $Q_{z_0}^{1} \gg 1$ within the intermediate blue regions inside the stochastic layer is consistent with previous observations \citep[e.g.,][]{RubinoEA2015, BorgognoEA2017, StanierEA2019} and suggests that these intermediate regions have distinct properties from the SGTR core; Section~\ref{sec:meanprofiles} discusses these SGTR wings in detail from a mean field perspective.

From 2D slices at $z=0.0$, flux rope structures cover approximately $50\%$ of the total area of the stochastic layer during SGTR; this area ratio is greater than the ratio $12\mbox{--}24\%$ of the inner and outer thickness scales, indicating that the flux ropes structures are not assigned to the mean inner or outer characteristic scales, but instead drive the properties of both. In principle, a mean thickness $\bigl< \delta y \bigr>$ could be measured for $Q_{z_0}^{1}$ in a similar style to Sections~\ref{sec:innerlayer} and \ref{sec:outerlayer}. However, attempting to do so would be expensive as squashing factor calculations would have to be carried out for every $z$-slice. Section~\ref{sec:FRS} provides further analysis and discussion of the flux rope structures.

\subsection{Mean profiles} \label{sec:meanprofiles}

Section~\ref{sec:thickness} investigated the layer thickness scales by taking measurements $\delta y$ in the $y$ direction, then averaging these over the central region of the reconnection layer to obtain a mean thickness $\bigl<\delta y \bigr>$  for different variables $f(\boldsymbol{x})$ at each simulation snapshot. A complementary approach is to evaluate the \textit{mean profile} of $f(\boldsymbol{x})$ across the reconnection layer in the $y$ direction, using spatial and time averaging to maximise signal to noise. While this second method is applicable only to the quasi-stationary SGTR phase of the simulation, it delivers additional insights into the shapes of the underlying profiles of $f(\boldsymbol{x})$.

\begin{figure} 
\centering
\hspace*{-0.8cm}\includegraphics[width=1.25\columnwidth]{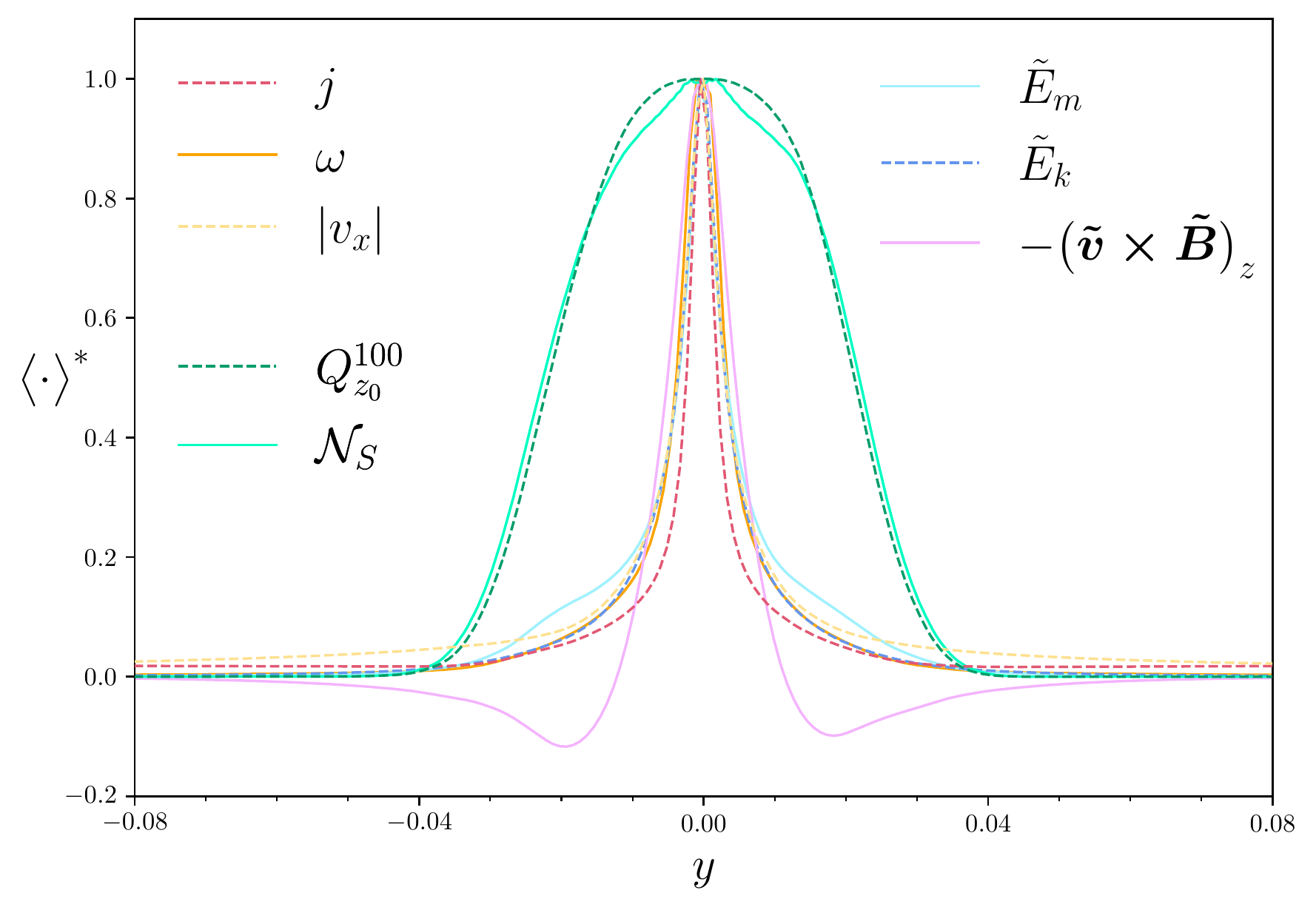} 
\caption{Normalised mean profiles in the $y$ direction across the reconnection layer, over the SGTR phase $t \in [2.4, 5.0]$ (after \textbf{e} in Figure~\ref{fig:Vrec}). Dark red (dashed): current density strength $j$; Orange: vorticity strength $\omega$; Yellow (dashed): outflow velocity absolute value $\vert v_x \vert$; Light blue: energy of the magnetic field fluctuations $\tilde{E}_m$; Dark blue (dashed): energy of the kinetic fluctuations $\tilde{E}_k$; Pink: turbulent EMF $- \bigl(\boldsymbol{\tilde{v} \times \tilde{B}}\bigr)_z$; Dark green (dashed): squashing factor $Q_{z_0}^{100}$ for $z_0 = -0.5$; Turquoise: Poincar\'e section count $\mathcal{N}_S$ for $n=2000$ iterations. \label{fig:meanprof}} 
\end{figure}

Figure~\ref{fig:meanprof} shows the \textit{normalised} mean profiles of the variables considered in Section~\ref{sec:thickness} over $y \in [-0.08, 0.08]$. For variable $f(\boldsymbol{x})$, we take the spatial average over the same local region $(x,z) \in [-0.15,0.15] \times [-0.5,0.5)$ used for $\bigl< \delta y \bigr>$. To further reduce noise, we also take a time average over the SGTR phase $t \in [2.4,5.0]$ (after \textbf{e} in Figure~\ref{fig:delta}); changing the lower bound for the time from $t \geq 2.0$ to $t \geq 4.0$ did not significantly change the resulting profiles. To aid the comparison, the final mean profiles $\bigl< f \bigr>$ are normalised by their local maximum over $y \in [-0.08, 0.08]$, denoted $\bigl< f \bigr>^* = \bigl< f \bigr>/\max \bigl< f \bigr>$.

\paragraph{(a) Inner core} The mean profiles for the current density strength $j$ (dashed dark red), vorticity strength $\omega$ (orange), energy of the magnetic field fluctuations $\tilde{E}_m$ (light blue), and energy of the kinetic fluctuations $\tilde{E}_k$ (dashed dark blue) are very consistent in the interior region and form a narrow peak, in agreement with Figure~\ref{fig:delta}. To obtain a mean profile for the outflow jets, the absolute value $\vert v_x \vert$ (dashed yellow) is taken prior to averaging (similar to Section~\ref{sec:innerlayer}) to avoid cancellation due to approximate antisymmetry about $x=0.0$. The mean profile of the turbulent EMF $- \bigl(\boldsymbol{\tilde{v} \times \tilde{B}}\bigr)_z$ (light pink) has also been provided; a FWHM estimation of the thickness of the positive central peak is very consistent with the quasi-stationary mean thickness [$\bigl< \delta y \bigr> \approx 0.0096$] for $\bigl\vert-\bigl(\boldsymbol{\tilde{v}} \boldsymbol{\times} \boldsymbol{\tilde{B}}\bigr)_z\bigr\vert$ measured in Section~\ref{sec:innerlayer}. While the mean profiles for $j$, $\omega$, $\tilde{E}_m$, $\tilde{E}_k$, $\vert v_x \vert$, and $- \bigl(\boldsymbol{\tilde{v} \times \tilde{B}}\bigr)_z$ are not identical, the inner thicknesses obtained in Section~\ref{sec:innerlayer} adequately describes the cores for all of these quantities, further supporting the existence of an inner scale for SGTR dynamics.

\paragraph{(b) Outer scale} For the squashing factor $Q_{z_0}^{100}$ (for $z_0 = -0.5$) [dashed dark green] and Poincar\'e section count $\mathcal{N}_S$ (for $n=2000$) [turquoise], the mean profiles show close agreement with each other, forming a broad indicator-like distribution. This further confirms the existence of a thicker layer in which the magnetic field is stochastic and connected to the interior parts of the reconnection layer; Sections~\ref{sec:FRS} and \ref{sec:dichotomy} discuss this property in detail. Estimating the outer and inner thickness from the FWHM of the mean profiles shown in Figure~\ref{fig:meanprof} finds that the outer scale is a factor $4.6\mbox{--}9.5$ 
broader than the inner scale, which overlaps with the scale ratio of $4.2\mbox{--}8.1$ determined from $\bigl< \delta y \bigr>$ measurements in Section~\ref{sec:outerlayer}.

\begin{figure} 
\centering
\hspace*{-1.6cm}\includegraphics[width=1.27\columnwidth]{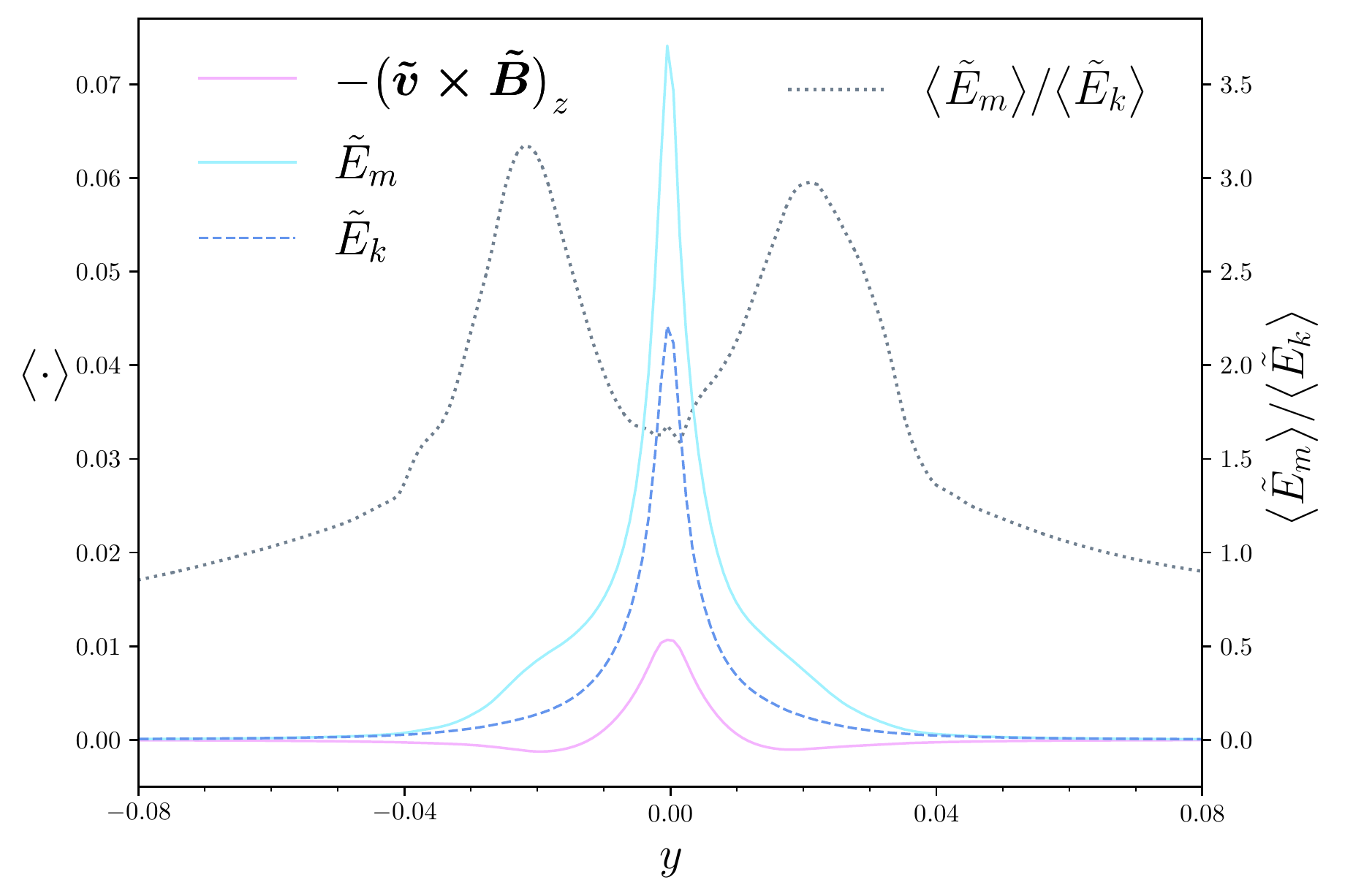} 
\caption{Mean profiles of the fluctuation variables in the $y$ direction across the reconnection layer, over the SGTR phase $t \in [2.4, 5.0]$ (after \textbf{e} in Figure~\ref{fig:Vrec}). Pink: turbulent EMF $- \bigl(\boldsymbol{\tilde{v} \times \tilde{B}}\bigr)_z$; Light blue: energy of the magnetic field fluctuations $\tilde{E}_m$; Dark blue (dashed): energy of the kinetic fluctuations $\tilde{E}_k$;  Grey (dotted): turbulent energy fluctuation ratio $\bigl< \tilde{E}_m\bigr>/\bigl< \tilde{E}_k\bigr>$ (right axis). \label{fig:Eflucratio}} 
\end{figure}

\paragraph{(c) SGTR wings} We refer to the regions between the inner core and the stochastic separatrices as the \textit{SGTR wings}. The amplitude of MHD turbulence (quantified by $\tilde{E}_m$ and $\tilde{E}_k$) is much lower in the wings than in the core, but while the wings are less turbulent, they nonetheless are part of the wider stochastic layer that is magnetically connected to the core. We also see evidence that fluctuations in the wings have a different nature to fluctuations in the core: Figure~\ref{fig:Eflucratio} shows the fluctuation variables and the turbulent energy fluctuation ratio $\bigl< \tilde{E}_m\bigr>/\bigl< \tilde{E}_k\bigr>$ (dotted grey). In the innermost region we have a roughly constant ratio of $\bigl< \tilde{E}_m\bigr>/\bigl< \tilde{E}_k\bigr> \approx 1.7$ [previously found by \citet{HuangBhattacharjee2016} (see their Figure 3) to be approximately $1.9$]; in the wings, however, a significantly greater excess of $\tilde{E}_m$ to $\tilde{E}_k$ is found, up to $\bigl< \tilde{E}_m\bigr>/\bigl< \tilde{E}_k\bigr> \approx 3.2$. Further, the mean profile of $- \bigl(\boldsymbol{\tilde{v} \times \tilde{B}}\bigr)_z$ has reversals on either side of the central peak, that span the wings. These empirical results suggest that the SGTR wings have their own particular dynamics, which could potentially be important to a fuller understanding of SGTR. This mean property is also consistent with Section~\ref{sec:outertopology} where we observed sizeable voids flanking the core of the reconnection layer that do not contain flux rope structures highlighted by $Q_{z_0}^{1} \gg 1$. 

\begin{figure}
\centering
\hspace*{-0.5cm}\includegraphics[width=1.25\columnwidth]{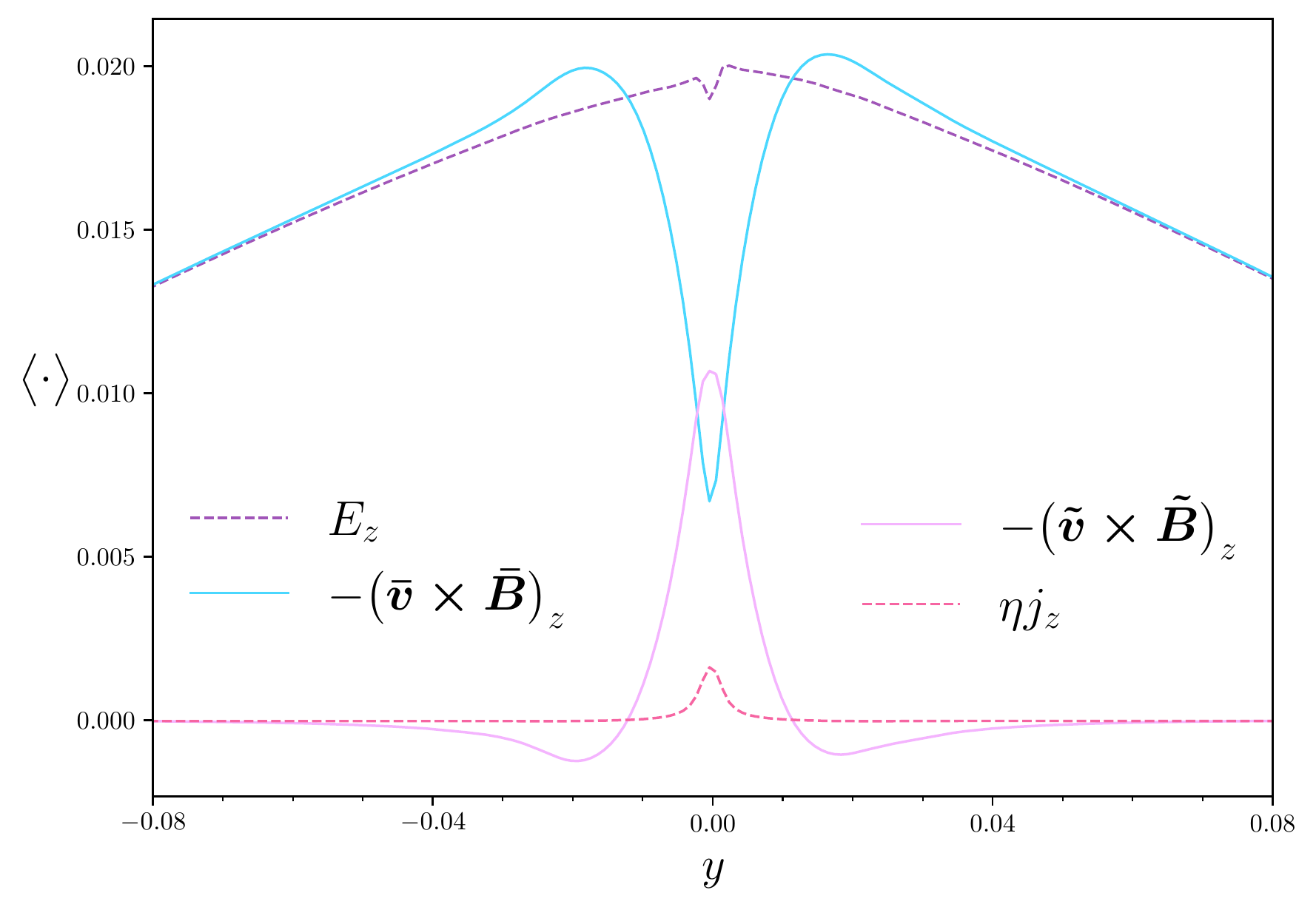} 
\caption{Mean profiles in the $y$ direction across the reconnection layer, over the SGTR phase $t \in [2.4, 5.0]$ (after \textbf{e} in Figure~\ref{fig:Vrec}), using terms from the decomposition of the electric field component $E_z$. Purple (dashed): $E_z$; Blue: background EMF $- \bigl(\boldsymbol{\bar{v} \times \bar{B}}\bigr)_z$; Light pink: turbulent EMF $- \bigl(\boldsymbol{\tilde{v} \times \tilde{B}}\bigr)_z$ (c.f., Figure~\ref{fig:meanprof}); Dark pink (dashed): resistive EMF $\eta j_z$. \label{fig:Ezdecomp}} 
\end{figure} 

\paragraph{Reconnection electric field} To obtain additional insight into the SGTR layer structure, we also consider remaining terms from the decomposition of the electric field component $E_z$ shown in equation (\ref{eqn:Ez}). Similar analysis was carried out by \citet{HuangBhattacharjee2016} and in the context of kinetic simulations by \citet{LeEA2018, LiuEA2011, LiuEA2013}. By taking the spatial average over $(x,z) \in [-0.15,0.15] \times [-0.5,0.5)$ and time average over $t \in [2.4,5.0]$, we obtain
\begin{equation}
\bigl<E_z \bigr> = - \bigl<\bigl(\boldsymbol{\bar{v} \times \bar{B}}\bigr)_z\bigr> - \bigl<\bigl(\boldsymbol{\tilde{v} \times \tilde{B}}\bigr)_z\bigr> + \eta \bigl< j_z\bigr>, \label{eqn:Ezmean}
\end{equation}
i.e., an equation for the mean profiles of $E_z$, the background EMF, the turbulent EMF, and the resistive EMF. The reduction from equation (\ref{eqn:Ez}) to equation (\ref{eqn:Ezmean}) is only due to the mean over $z \in [-0.5,0.5)$ by the definition of the mean $\bar{f}$ and fluctuating $\tilde{f}$ components. 

The decomposition in equation (\ref{eqn:Ezmean}) is shown in Figure~\ref{fig:Ezdecomp}. The results resemble those found in \citet{HuangBhattacharjee2016} (see their Figure 6), although the time average over the SGTR phase that we have employed produces a cleaner plot than inspecting the EMF terms at a single snapshot, which allows for a more confident interpretation and detection of new features. The ratio between $\bigl<E_z \bigr>$ (dashed purple) about the midplane $y = 0.0$ and $V_{\mathrm{rec}}$ is approximately one, which is also consistent with \citet{HuangBhattacharjee2016} (see their Figures 2 and 6). In an exact steady state,  $\bigl<E_z \bigr>$  would be constant in $y$, but in this simulation, it falls off slightly for increasing $|y|$, as a consequence of a gradual decay in $|B_x|$ that occurs as the laminar flux ropes that merge to drive the SGTR are consumed on a secular timescale. This departure from a constant $\bigl<E_z \bigr>$ model is minor, and our focus is on the changing dominant contributions to $\bigl<E_z \bigr>$. In the outer regions, $\bigl<E_z \bigr>$ is strongly dominated by the background EMF (blue) produced by the inflow of magnetic flux towards the reconnection region. Near to the midplane, the background EMF drops greatly as the inflows stop; the blue curve does not reduce all the way to zero at $y=0.0$ because of the contribution of reconnection outflows within the region we have averaged over. The difference required to maintain near constancy of $\bigl<E_z \bigr>$ across the reconnection region is provided by the turbulent EMF (light pink, c.f., Figures~\ref{fig:meanprof} and \ref{fig:Eflucratio}). By comparison, the resistive EMF (dashed dark pink) is much smaller, which is an important test that the reconnection in the SGTR phase of our simulation is indeed attributable to turbulence, not resistivity. There is also no significant missing contribution to $\bigl<E_z \bigr>$ at the $y=0.0$, which evidences that the reconnection is not attributable to numerical resistivity. Moreover, the results of Figure~\ref{fig:Ezdecomp} demonstrate that during the quasi-stationary SGTR phase, from a mean field perspective, our simulation behaves in a manner reconcilable with a Sweet-Parker type of model, but with the reconnection electric field provided by turbulence that is generated and sustained within the reconnection region, instead of by resistivity. At the same time, the mean profile of the turbulent EMF also reinforces that the full picture of SGTR is more complicated than simply invoking a ``turbulent resistivity'', due to the reversal of the turbulent EMF in the SGTR wings.

\section{Discussion}\label{sec:discussion}

\subsection{Sweet-Parker scalings}\label{sec:SPscalings}

The results of Section~\ref{sec:thickness} have shown that the SGTR layer displays more than one characteristic thickness $\bigl<\delta y \bigr>$, most notably an inner scale associated with current density, vorticity, outflows, and turbulence, and an outer scale associated with stochastic magnetic field line mixing. Both layer thickness scales correlate with the global reconnection rate $V_{\mathrm{rec}}$ (Figure~\ref{fig:delta}), in a sense that they rapidly grow at the turbulent reconnection onset and approximately plateau during the SGTR stage. This leads to an important question: what is the ``correct'' thickness to predict the global reconnection rate of \textit{quasi-stationary SGTR}, using a Sweet-Parker-like estimation of the reconnection layer aspect ratio?

The classical Sweet-Parker model \citep{Sweet1958, Parker1957} involves a simple laminar 2D configuration containing a thin extended current sheet of length $L$ and thickness $\delta$ with $\delta \ll L$. The model assumes that a steady state is obtained where the inflow velocity of plasma $v_{\mathrm{in}}$ balances the outward diffusion rate of the field lines \citep{PriestForbes2000, KowalEA2012}, which yields the critical approximation $v_{\mathrm{in}} \approx 2\eta/\delta$. Two of the Sweet-Parker model assumptions do not apply to SGTR, namely that the magnetic field is laminar and that the reconnection electric field is resistive. However, many component parts of the Sweet-Parker analysis are independent of those assumptions, and others are readily modified, making it possible to update Sweet-Parker insights for quasi-stationary SGRT.

Under mass conservation, the total mass flux that enters the reconnection layer from both sides along length $L$ must equal the total mass flux that leaves both edges of the reconnection layer of thickness $\delta$,  yielding the approximation $v_{\mathrm{in}} L \approx  v_{\mathrm{out}} \delta$. Further, the outflow is found to satisfy $v_{\mathrm{out}} \approx v_A$ where $v_A$ is the Alfv\'en speed with respect to the reconnecting inflow field. The inflow speed is then taken as the reconnection rate $V_{\mathrm{rec}} = v_{\mathrm{in}}$.  Hence, when the densities of the inflow and outflow are similar, we have the aspect ratio scaling
\begin{equation}
V_{\mathrm{rec}} = v_{\mathrm{in}} \approx  v_A\frac{\delta}{L}, \label{eqn:deltaL}
\end{equation}
which does not depend on the processes (e.g., resistive versus turbulent) inside the reconnection region.

Since the magnetic field in our simulation possesses a Sweet-Parker-type global magnetic topology, it is insightful to test the Sweet-Parker scaling (\ref{eqn:deltaL}) over time. Under normalisation, we have $v_A \approx 1$. The outflow velocity is consistent with the Sweet-Parker approximation, with $\vert v_x \vert \approx 1$. While our simulation is compressible, the mass density $\rho$ does not widely fluctuate, with the global range remaining within $\rho \approx 0.73\mbox{--}1.18$. The steady state assumption is only approximately satisfied for the  initial 2.5D laminar stage (\textbf{a}--\textbf{c} in Figure~\ref{fig:Vrec}) and the SGTR  stage (after \textbf{e} in Figure~\ref{fig:Vrec}) where we have the SGTR analogue of Sweet-Parker reconnection, especially during the quasi-stationary phase (after \textbf{f} in Figure~\ref{fig:Vrec}). Hence, we do not expect the scaling (\ref{eqn:deltaL}) to be a satisfactory approximation during the intermediate stages of the simulation (\textbf{c}--\textbf{e} in Figure~\ref{fig:Vrec}). 

\begin{figure*} 
\centering
\hspace*{-1.8cm}\includegraphics[width=1.2\textwidth]{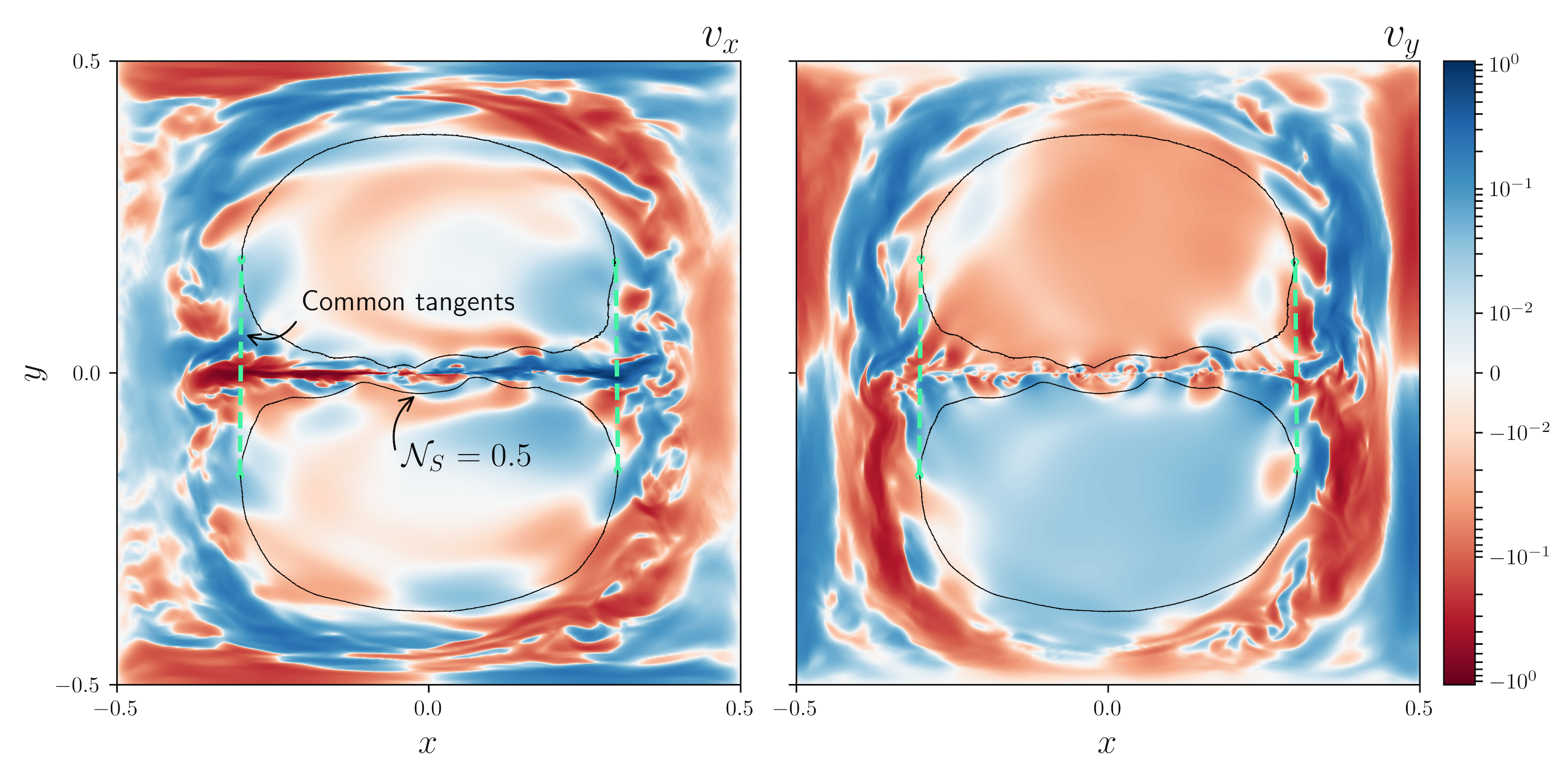}
\caption{Overlay plots of the $v_x$ (left panel) and $v_y$ (right panel) components with the laminar boundaries of the upper and lower flux ropes (black), i.e., stochastic separatrices, using a sample at $z = 0.0$, $t = 4.0$. The laminar boundaries are approximated using contours of the Poincar\'e section count $\mathcal{N}_S = 0.5$ (for $n=2000$ iterations). The mean inflow velocity $\bigl< v_{\mathrm{in}} \bigr>$ is approximated within the interior of the laminar flux ropes. The common tangents of the laminar boundaries (dashed green) are used to approximate the mean length $\bigl< L\bigr>$ of the reconnection layer. \label{fig:vin}}
\end{figure*}

The inflow velocity $v_{\mathrm{in}}$ can be approximated by taking an appropriate mean of the component $v_y$ towards the reconnection layer within the large laminar flux ropes. The upper $V_{\mathrm{upper}}$ and lower $V_{\mathrm{lower}}$ flux rope volumes are defined as the regions enclosed by the corresponding stochastic separatrices; we approximate the laminar boundaries $\partial V_{\mathrm{upper}}$ and $\partial V_{\mathrm{lower}}$ using contours $\mathcal{N}_S = 0.5$ of the Poincar\'e section count for $n=2000$ iterations (see Section~\ref{sec:outerlayer}). Figure~\ref{fig:vin} shows the laminar boundaries overlaid on the $v_x$ and $v_y$ components at $z=0.0$, $t=4.0$. We then evaluate the absolute value of the mean $\bigl\vert\bigl< v_y \bigr>\bigr\vert$ over each flux rope volume $V_{\mathrm{upper}}$ and $V_{\mathrm{lower}}$, then take the average of these to approximate the mean inflow velocity $\bigl< v_{\mathrm{in}} \bigr>$. Other averages of $v_y$ were tested and found to be consistent, e.g., taking various $\bigl\vert\bigl< v_y \bigr>\bigr\vert$ after filtering for $v_y < 0$ in $V_{\mathrm{upper}}$ and $v_y > 0$ in $V_{\mathrm{lower}}$. 

The layer length $L$ was estimated as follows. Using the laminar boundaries $\partial V_{\mathrm{upper}}$ and $\partial V_{\mathrm{lower}}$ on every $z$-slice, we found the common tangents on the left and right (dashed green lines in Figure~\ref{fig:vin}). We then approximated $L$  by measuring the distance between the intersection points of each common tangent with the midplane $y=0.0$. The average was then taken over the whole domain $z \in [-0.5,0.5)$ to obtain a mean layer length $\bigl< L\bigr>$, the standard error of which was very small. The mean decreases monotonically over time from $\bigl< L\bigr> \approx 1.0$ at $t = 0.0$ to $\bigl< L\bigr> \approx 0.51$ at $t = 5.0$ as the laminar flux ropes reconnect. The evolution is approximately piecewise linear with distinct changes in the gradient during each general stage of the simulation (\textit{i.}, \textit{ii.} and \textit{iii.} in Figure~\ref{fig:Vrec}); the steepest gradient occurs during the transition phase. The common tangents generally align with the outflow termini, at which the Alfv\'enic reconnection outflows begin to brake significantly in $v_x$ and are redirected, creating a reversal in $v_y$ between the inflow of the laminar flux ropes and the return flow around them (see Figure~\ref{fig:vin}). The mean length $\bigl< L\bigr>$ can also be approximated using contours of the outflow $v_x$ and inflow $v_y$ velocities directly; these yield consistent results with the common tangents method but are less stable due to turbulence.

\begin{figure} 
\centering 
\hspace*{-1.5cm}\includegraphics[width=1.27\columnwidth]{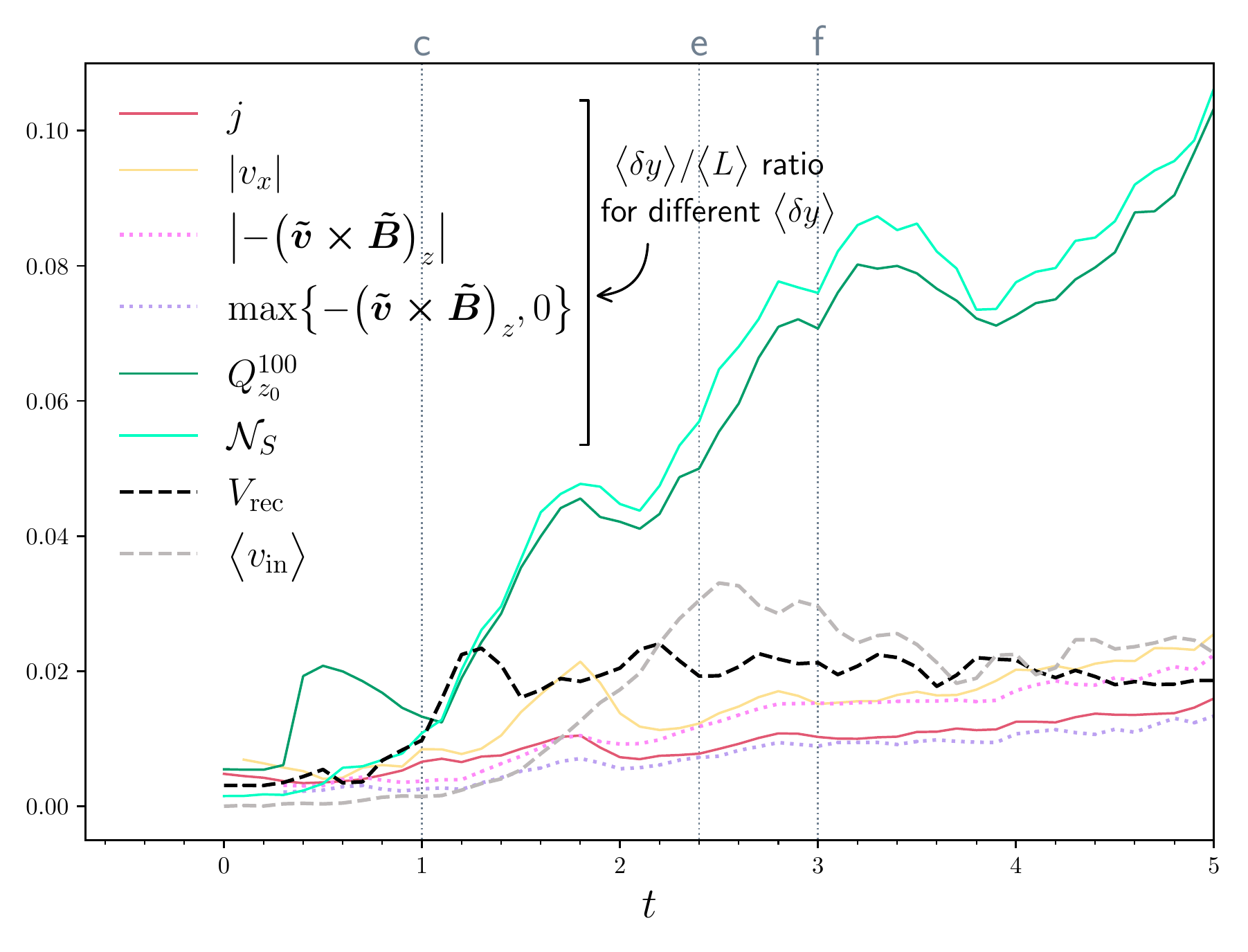}
\caption{Comparison of the aspect ratios $\bigl<\delta y \bigr>/\bigl<L \bigr>$ for various characteristic thicknesses $\bigl<\delta y \bigr>$ with the global reconnection rate $V_{\mathrm{rec}}$ and mean inflow velocity $\bigl<v_{\mathrm{in}} \bigr>$ over time. Dark red: current density strength $j$;  Yellow: outflow velocity $\vert v_x \vert$; Pink (dotted): unsigned turbulent EMF $\bigl\vert- \bigl(\boldsymbol{\tilde{v} \times \tilde{B}}\bigr)_z\bigr\vert$; Purple (dotted): positive turbulent EMF $\max\bigl\{-\bigl(\boldsymbol{\tilde{v}} \boldsymbol{\times} \boldsymbol{\tilde{B}}\bigr)_z,0\bigr\}$; Dark green: squashing factor $Q_{z_0}^{100}$ for $z_0 = -0.5$; Turquoise: Poincar\'e section count $\mathcal{N}_S$ for $n=2000$ iterations;  Black (dashed): global reconnection rate $V_{\mathrm{rec}}$; Grey (dashed): mean inflow velocity $\bigl< v_{\mathrm{in}} \bigr>$. The vertical dotted lines mark times of interest: \textbf{c} Turbulent reconnection onset ($t= 1.0$); \textbf{e} SGTR onset ($t = 2.4$); \textbf{f} ``Pure'' SGTR  onset ($t = 3.0$).  \label{fig:deltaLcomp}}
\end{figure}

Figure~\ref{fig:deltaLcomp} compares the aspect ratios $\bigl<\delta y \bigr>/\bigl<L \bigr>$ for selected characteristic thicknesses $\bigl<\delta y \bigr>$ evaluated earlier in Section~\ref{sec:thickness} (see Figure~\ref{fig:delta}). The $\bigl<\delta y \bigr>/\bigl<L \bigr>$ approximations increase over time, which is not unexpected since the mean thicknesses $\bigl<\delta y \bigr>$ reach quasi-stationary values while the length $\bigl<L \bigr>$ decreases. To test the Sweet-Parker scaling (\ref{eqn:deltaL}), we also plot the global reconnection rate $V_{\mathrm{rec}}$ (dashed black) and the mean inflow velocity $\bigl< v_{\mathrm{in}} \bigr>$ (dashed grey).  The $\bigl< v_{\mathrm{in}} \bigr>$ is comparable in magnitude with $V_{\mathrm{rec}}$ before the turbulent reconnection onset at $t = 1.0$ (label \textbf{c}) and during the pure SGTR phase $t = 3.0\mbox{--}5.0$ (after label \textbf{f}), as anticipated. During the transition phase $t=1.0\mbox{--}2.4$ (labels \textbf{c}--\textbf{e}), the increase in $\bigl< v_{\mathrm{in}} \bigr>$ lags behind the increase in $V_{\mathrm{rec}}$. In this part of the simulation, reconnection is not quasi-stationary and the laminar flux ropes are eroded by the expansion of the reconnection layer, such that $V_{\mathrm{rec}}$ can exceed $\bigl< v_{\mathrm{in}} \bigr>$. There is then evidence of $\bigl< v_{\mathrm{in}} \bigr>$ overshooting $V_{\mathrm{rec}}$ before the system settles into quasi-stationary SGTR;  for this reason, we focus on the \textit{mean values} after $t = 3.0$ (label \textbf{f}). During quasi-stationary SGTR we observe $V_{\mathrm{rec}} \approx 0.020$ and $\bigl< v_{\mathrm{in}} \bigr>\approx 0.023$. 

The aspect ratio for the Poincar\'e section count $\mathcal{N}_S$ (for $n=2000$) [turquoise] agrees well with $V_{\mathrm{rec}}$ prior to $t = 1.3$, but afterwards the aspect ratios for both $\mathcal{N}_S$ and $Q_{z_0}^{100}$ (for $z_0 = -0.5$) [dark green] exceed $V_{\mathrm{rec}}$ and continue to grow. After $t = 3.0$, the aspect ratios using the outer thicknesses are $4.0\mbox{--}4.2$ times $V_{\mathrm{rec}}$ and $3.2\mbox{--}3.4$ times $\bigl< v_{\mathrm{in}} \bigr>$. The aspect ratios for the inner thicknesses, e.g., current density strength $j$ (dark red), turbulent EMF $- \bigl(\boldsymbol{\tilde{v} \times \tilde{B}}\bigr)_z$ (dotted pink and dotted purple), and outflow velocity $\vert v_x\vert$ (yellow), also agree well with $V_{\mathrm{rec}}$ and $\bigl< v_{\mathrm{in}} \bigr>$ prior to $t = 1.0$. For $j$ and $\max\bigl\{-\bigl(\boldsymbol{\tilde{v}} \boldsymbol{\times} \boldsymbol{\tilde{B}}\bigr)_z,0\bigr\}$, the measured aspect ratios after $t = 3.0$ are $0.53\mbox{--}0.61$ times $V_{\mathrm{rec}}$ and $0.42\mbox{--}0.49$ times $\bigl< v_{\mathrm{in}} \bigr>$. The best agreement is obtained for $\vert v_x\vert$, which is closely tracked by $\bigl\vert- \bigl(\boldsymbol{\tilde{v} \times \tilde{B}}\bigr)_z\bigr\vert$; the measured aspect ratio for these after $t = 3.0$ are $0.87\mbox{--}0.96$ times $V_{\mathrm{rec}}$ and $0.69\mbox{--}0.77$ times $\bigl< v_{\mathrm{in}} \bigr>$. Finally, we comment that since the Sweet-Parker scaling (\ref{eqn:deltaL}) is derived from mass flux considerations, it is well founded that $V_{\mathrm{rec}}$ and $\bigl< v_{\mathrm{in}} \bigr>$ should be closely related to the mean thickness and profile of the outflows $\vert v_x\vert$. Similarly, the induction equation ensures that the characteristic thickness of the outflows should be closely related to the mean thickness and profile of the dominant source of the reconnection EMF, which for SGTR is the turbulent contribution, $-\bigl(\boldsymbol{\tilde{v}} \boldsymbol{\times} \boldsymbol{\tilde{B}}\bigr)_z$. Our results in Figure~\ref{fig:deltaLcomp} confirm that during the pure SGTR phase there is indeed consistency between the reconnection rate, inflow speed, outflow thickness, and turbulent EMF thickness.  

From the Sweet-Parker scaling (\ref{eqn:deltaL}), the large differences between $\bigl<\delta y \bigr>/\bigl<L \bigr>$ for the outer thickness scale and $V_{\mathrm{rec}}$ or $\bigl< v_{\mathrm{in}} \bigr>$ imply that the stochastic layer thickness \textit{is not} the ``effective'' thickness to predict the global reconnection rate. The reconnection rate is instead in agreement with the inner scale associated with the thicknesses of the current density, vorticity density, reconnection outflows, and turbulent fluctuations. 
 
There have previously been two major conceptual models of how 3D reconnection may be enhanced under turbulence by the broadening of the effective thickness $\delta$ of the reconnection layer. One of these models, which is found in field line cartoons similar to Figure 2 of \citet{LazarianVishniac1999}, has invoked broadening of the reconnection layer through field line wandering, leading various studies to focus on identifying the stochastic or mixing region \citep[e.g.,][]{DaughtonEA2014}. The alternative, which is less convenient to represent in cartoon form, invokes the broadening of the reconnection layer by replacing the resistive EMF with the turbulent EMF [see equation (\ref{eqn:Ezmean})]. Until now, it may have been widely assumed that the stochastic and turbulent EMF thicknesses would be equivalent; however, the results of this paper show that these can differ significantly (by a factor of roughly six in our simulation), and it is the inner scale associated with the turbulent EMF, i.e., effective reconnection electric field produced by turbulent fluctuations, that sets the reconnection rate of quasi-stationary SGTR, not the larger stochastic thickness. 

The significance of the inner scale for the reconnection rate can be understood by observing that, in mean models, fluctuations of relatively large amplitude are located in the SGTR core, and since they provide the reconnection EMF, the width of this inner region sets the global reconnection rate. Meanwhile, fluctuations are also present in the SGTR wings, but at drastically lower amplitudes than in the core (Figure~\ref{fig:meanprof} and \ref{fig:Eflucratio}). The relatively weak turbulence in the wings scatters magnetic field lines, making the field stochastic there as well, magnetically connecting the core and wings and promoting mixing of plasma from the two sides of the reconnection layer; but due to low amplitudes, it does not contribute significantly to the global reconnection rate. This does not necessarily mean that the wings are unimportant: the reversed turbulent EMF and greater $\bigl< \tilde{E}_m\bigr>/\bigl< \tilde{E}_k\bigr>$ ratio in the wings compared to the core (Figures~\ref{fig:meanprof}, \ref{fig:Eflucratio} and \ref{fig:Ezdecomp}) are evidence that turbulence in the wings is potentially of a qualitatively different nature to turbulence in the core. The existence of the SGTR wings is an interesting and unanticipated finding of this study, and their role in SGTR will be a compelling topic for future research.

\subsection{Flux rope structures}\label{sec:FRS}

To work towards a complete understanding of SGTR, it is important that mean models for the global dynamics (e.g., Sections~\ref{sec:thickness}, \ref{sec:meanprofiles} and \ref{sec:SPscalings}) are complemented with knowledge of the dynamics within the reconnection layer. For example, one would like to understand the nature of the turbulence in the core, especially the source of the non-alignment and dominant handedness of $\boldsymbol{\tilde{v}}$ and $\boldsymbol{\tilde{B}}$ required to produce the turbulent EMF, and why these properties change in the SGTR wings, where the $\bigl< \tilde{E}_m\bigr>/\bigl< \tilde{E}_k\bigr>$ ratio is larger and the turbulent EMF reverses. Individual snapshots (e.g., Figures~\ref{fig:j}, \ref{fig:Q}, and \ref{fig:QNScomp}) show that dynamics in the SGTR layer are intimately related to the generation and interaction of fully 3D structures have similarities with plasmoids; however, they have a highly fragmented internal structure and are ``frayed'' in the sense that field lines traced from inside each flux rope are stochastic and ultimately exit it into the wider stochastic layer. In this section, we further explore the properties and magnetic topology of these flux rope structures.

\begin{figure*} 
\centering
\hspace*{-1.8cm}\includegraphics[width=1.19\textwidth]{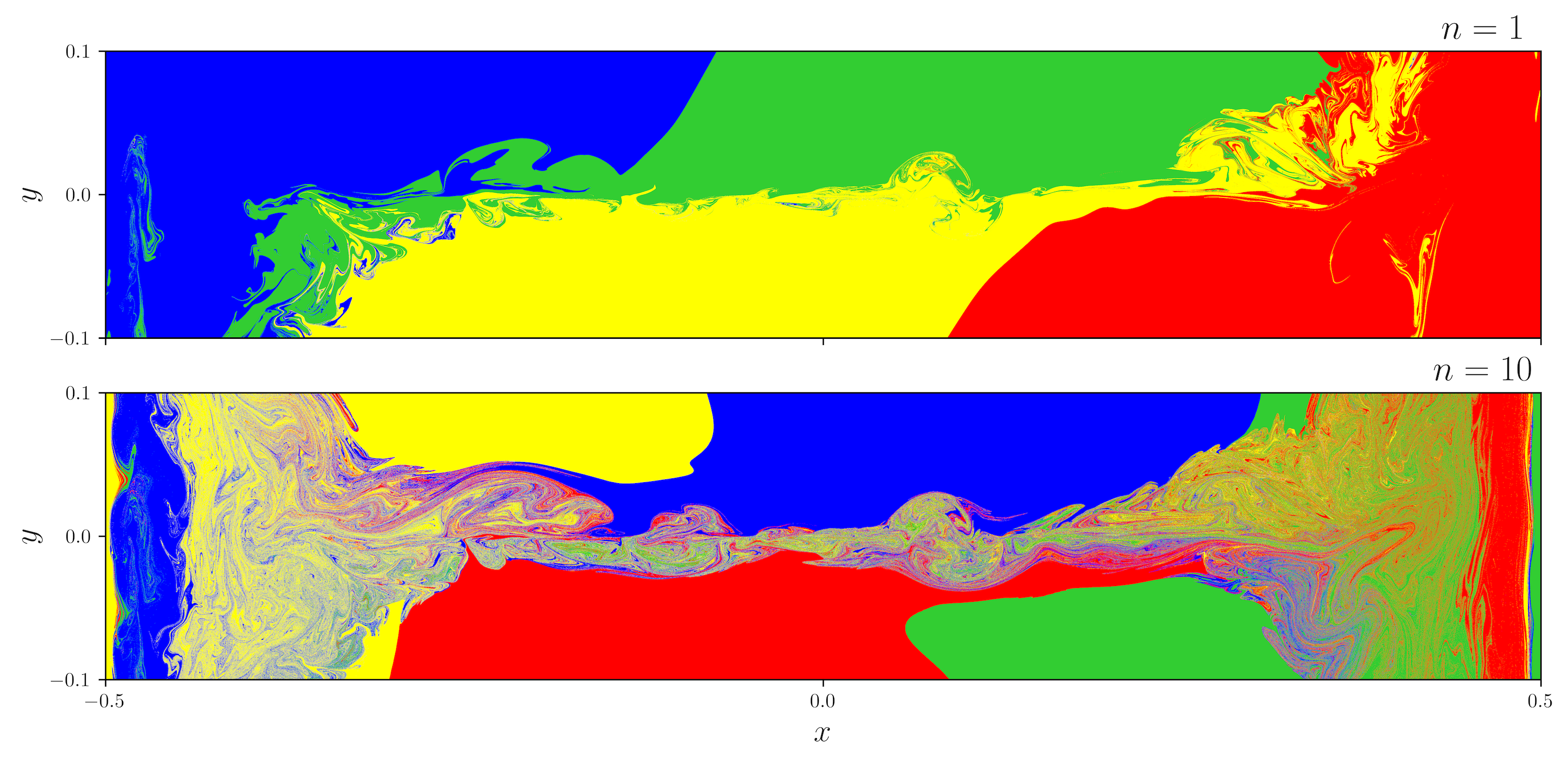}
\caption{Comparison of colour maps $C_{z_0}^{n}$ at $t = 3.0$. The top and bottom panels show $n=1$ and $n=10$ at $z_0 = -0.5$, respectively. Although we can identify flux rope structures within the stochastic field, the field line mapping is too topologically complicated to allow the detection of distinct flux ropes using colour maps, presumably because they are too small or not structured around a periodic point. \label{fig:colmapextra}} 
\end{figure*}

Over $t = 0.0\mbox{--}1.8$, we were able to successfully identify and trace the axes of some of the initial ropes that develop within the reconnection layer with the use of a colour map $C_{z_0}^{n}$ in definition (\ref{eqn:colmap}).  Examples include the CFR and the two neighbouring prominent flux ropes highlighted in Section~\ref{sec:transitionphase}. However, the colour map method is ineffective in general, because: \textit{i.} It can only detect flux ropes whose axes form closed loops; \textit{ii.} It fails to identify the flux ropes at the smallest observable length scales; \textit{iii.} Once stochasticity becomes widespread within the reconnection layer,  periodic points are rarely discernible. The colour map is only adequate primarily during the early stages of the simulation while the magnetic field is still weakly 2.5D, especially for identifying flux ropes developed from parallel tearing modes. Figure~\ref{fig:colmapextra} shows the colour map $C_{z_0}^{n}$ at $z_0 = -0.5$ over the reconnection region at $t = 3.0$ for $n = 1$ and $n=10$ iterations of the field line mapping $F_{z_0}^{n}$ [definition (\ref{eqn:Fmap})]. Although large structures in the magnetic field can be observed, especially for $n=10$ (or greater), the field line mapping is too topologically complicated to allow the identification of any distinct elliptic points. The failure of the colour map method is possibly in part due to limitations in numerically integrating stochastic field lines over the whole interval $z \in [-0.5,0.5)$ to high precision, or resolving intricate details of $F_{z_0}^{n}$ such as elliptic points at length scales comparable to the underlying spatial grid of the \textit{Lare3d} simulation. However, another consideration is that flux ropes may predominantly be oblique with axes that form \textit{open curves}, i.e., do not coincide with elliptic periodic points. 

Alternatively, it is also very likely that the flux ropes in SGTR are \textit{not structured around well-defined axes} in the first place, similar to a braided magnetic field \citep{Wilmot-SmithEA2010, YeatesEA2010, PontinEA2011, YeatesHornig2011a, YeatesHornig2013, PontinEA2013, WyperHesse2015, PontinHornig2015, YeatesEA2015, PontinEA2016, PontinEA2017, PriorYeates2018}. In this case, they are better described as loose stochastic flux bundles \citep{HuangBhattacharjee2016} in the form of structures that only attain coherence \textit{locally} in $z$ direction. These crucial properties raise the important problem of how best to characterise, trace and analyse these flux rope structures using topological tools. It also necessitates the local analysis of the flux rope structures, using the field line mapping over smaller distances $\vert n \vert \in (0,1)$. Research into this will also be illuminating for simulations in nonperiodic domains exhibiting SGTR.  

From the perspective of fluid mechanics, the flux rope structures we observe in the magnetic field closely resemble \textit{coherent structures} that are widely studied to assess patterns in unsteady velocity fields. These structures are defined as influential material surfaces that exhibit considerable temporal coherence and are independent of the reference frame. A large collection of tools to rigorously characterise and analyse these coherent structures exist in the literature. \textit{Lagrangian Coherent Structures} (LCSs) are defined by Lagrangian particle trajectories and correspond to regions in the flow displaying the strongest shearing, attraction or repulsion \cite[see review by][and references therein]{Haller2015}, whereas \textit{Objective Eulerian Coherent Structures} (OECSs) are defined as the most dominant material surfaces using properties of the instantaneous velocity field \citep[e.g.,][]{SerraHaller2016,NolanEA2020}.
 
Coherent structure tools have been adapted in previous magnetic reconnection studies and shown to be effective in illuminating important structures embedded within stochastic magnetic fields and turbulent plasmas. At fixed snapshots, Lagrangian tools have frequently been used to reveal LCS along magnetic field lines, i.e., in ``field line time''  \citep{BorgognoEA2011a, BorgognoEA2011b, RempelEA2013, RubinoEA2015, FalessiEA2015, BorgognoEA2017, DiGiannataleEA2018, DiGiannataleEA2017a, DiGiannataleEA2017b, PegoraroEA2019, VerandaEA2020a, VerandaEA2020b, DiGiannataleEA2021}. The finite-time Lyapunov exponent (FTLE) [e.g., definition (\ref{eqn:FTLE})] is a common method, equivalent to our earlier implementation of the squashing factor [equation (\ref{eqn:Q})]. Further, barriers formed by these coherent structures that temporarily confine the magnetic field \citep[e.g.,][]{Burrell1997, TalaEA2006, ConnorEA2004} have been identified in field line time by evaluating \textit{maximal repulsion-attraction material lines} from \citet{Haller2011} \citep[e.g.,][]{OnuEA2015, FalessiEA2015, DiGiannataleEA2018, DiGiannataleEA2017a, DiGiannataleEA2017b, DiGiannataleEA2021} or \textit{second-derivative ridges} of the FTLE from \citet{ShaddenEA2005} \citep[e.g.,][]{BorgognoEA2011a, RubinoEA2015, PegoraroEA2019, VerandaEA2020a, VerandaEA2020b}. LCSs have also been identified using the FTLE along Lagrangian particle trajectories in 2D for observational photospheric data \citep{YeatesEA2012, ChianEA2014} and in 3D simulations \citep{RempelEA2013}. Eulerian tools, such as the \textit{Q-criterion} \citep[e.g.,][]{Haller2005} and related metrics, have additionally been applied to fixed snapshots of the velocity and/or magnetic fields \citep{RempelEA2013, ChianEA2014}. 

Identification of alternative surfaces that are closely linked with LCSs has also been useful in characterising magnetic field barriers. Examples of these include: \textit{invariant Kolmogorov-Arnold-Moser} (KAM) flux surfaces \citep[e.g.,][]{LichtenbergLieberman1992, DiGiannataleEA2021}; \textit{Cantori} (broken KAM flux surfaces) \citep{MacKayEA1984, RubinoEA2015, BorgognoEA2017, DiGiannataleEA2021} or their approximation through \textit{ghost surfaces} \citep{HudsonBreslau2008, HudsonSuzuki2014}; and \textit{invariant stable/unstable manifolds} corresponding with \textit{distinguished hyperbolic trajectories} (DHTs) \citep{Haller2000, MentinkEA2005, BorgognoEA2008, BorgognoEA2011a, BorgognoEA2017}, i.e., the 3D generalisation of hyperbolic lines stemming from X-points in 2D vector fields, or related approximations \citep[e.g.,][]{MadridMancho2009, RempelEA2013}.

Finally, regarding the inner core variables that align with the flux rope structures, the topic of \textit{current and vorticity coherent structures} has also received considerable attention. The detection, formation, evolution, local or internal structure, and statistical properties of these coherent structures and other important shear regions have been investigated in detail within turbulent magnetised plasma in 2D \citep{ServidioEA2010, SistiEA2021b} and 3D \citep{UritskyEA2010, ZhdankinEA2013, KowalEA2020, SistiEA2021a}. Many related studies have also explored the role of the Kelvin-Helmholtz instability in the generation of these structures \citep[e.g.,][]{HenriEA2012, DaughtonEA2014, FaganelloEA2014, BorgognoEA2015, SistiEA2019, KowalEA2020}. The connection between the coherent current structures and the development of a complex magnetic topologies and structures has been strongly emphasised in several papers \citep[e.g.,][]{DaughtonEA2014, BorgognoEA2015, BorgognoEA2017, SistiEA2019}. 

In our simulation, a simple method to characterise the instantaneous flux ropes structures is by the \textit{maximising ridges} of the FTLE defined earlier as $\Lambda_{z_0}^{n}$ [definition (\ref{eqn:FTLE})]; these are locations where $\left\vert \Lambda_{z_0}^{n}\right\vert$ exceeds a specific threshold, e.g.,\ $50\%$ of $\max\left\vert \Lambda_{z_0}^{n}\right\vert$ over a plane for a particular $n$ \citep[e.g.,][]{ChianEA2014, LiuEA2018}. These ridges are an effectively approach to distinguish surfaces within the reconnection regions where field lines exhibit strong local convergence/attraction (backward traces $n<0$) or divergence/repulsion (forward traces $n>0$) about a particular $z$-slice, especially for small $\vert n\vert \in (0,1)$. The maximising ridges for $Q_{z_0}^{n}$ [definition (\ref{eqn:Q})] were found to be almost identical to $\Lambda_{z_0}^{n}$, since both metrics have very close agreement in our simulation (see Section~\ref{sec:innertopology}).

\begin{figure*} 
\centering
\hspace*{-1.8cm}\includegraphics[width=1.19\textwidth]{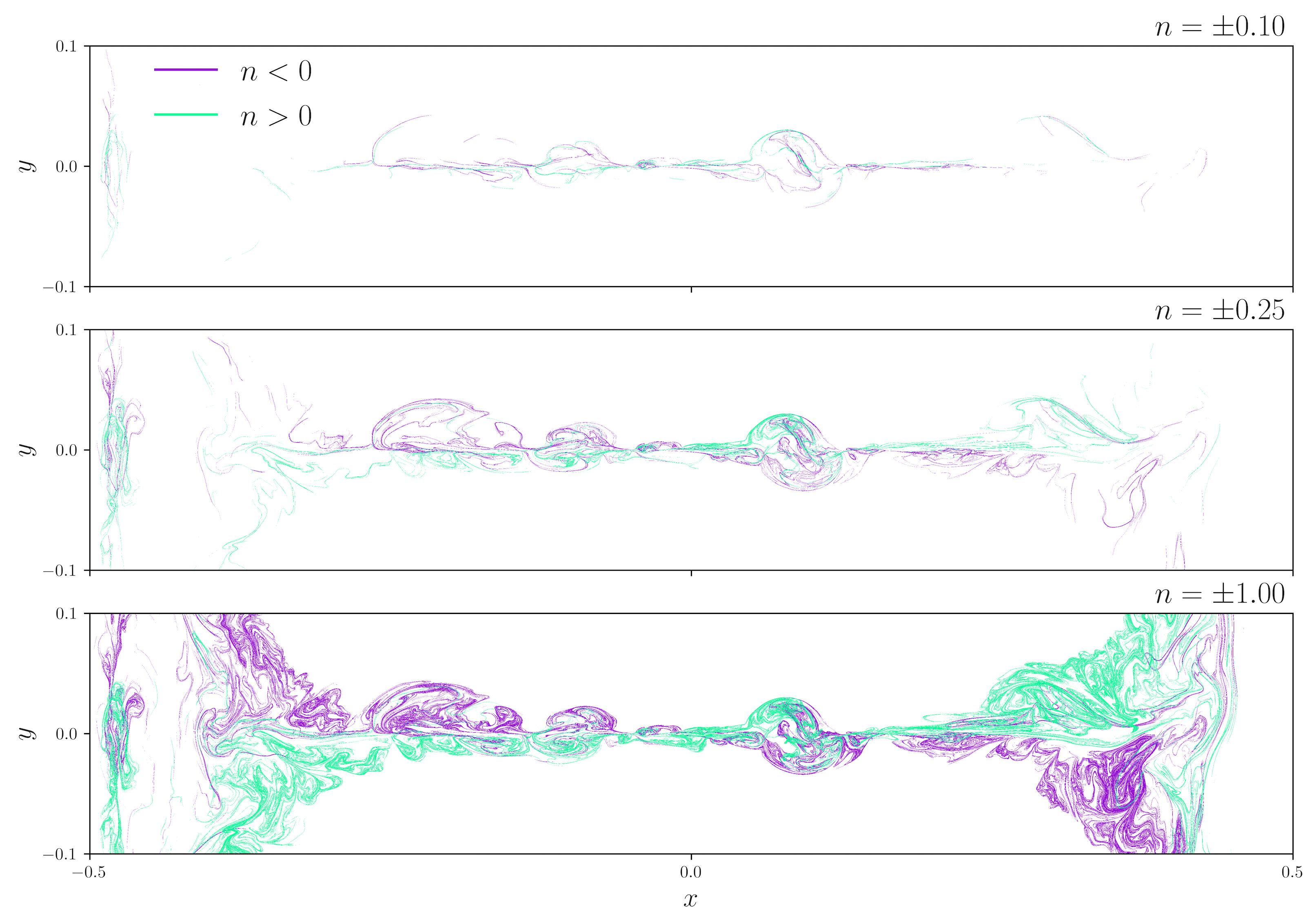} 
\caption{Comparison plots of the maximising ridges of the forward and backward finite-time Lyapunov exponent (FTLE) $\Lambda_{z_0}^{n}$. Samples over the bottom boundary $z_0 = -0.5$ at $t = 3.0$ for different iterations $n$ are shown. The repelling ridges (forward $n>0$ in green) and attracting ridges (backward $n<0$ in purple) are regions where $\left\vert \Lambda_{z_0}^{n}\right\vert$ exceeds 50\% of the maximum over the plane for a particular $n$. Top panel: $n = \pm 0.1$;  Middle panel: $n = \pm 0.25$; Bottom panel: $n = \pm 1.0$.  \label{fig:FTLE}}
\end{figure*}

\begin{figure*} 
\centering
\hspace*{-1.8cm}\includegraphics[width=1.19\textwidth]{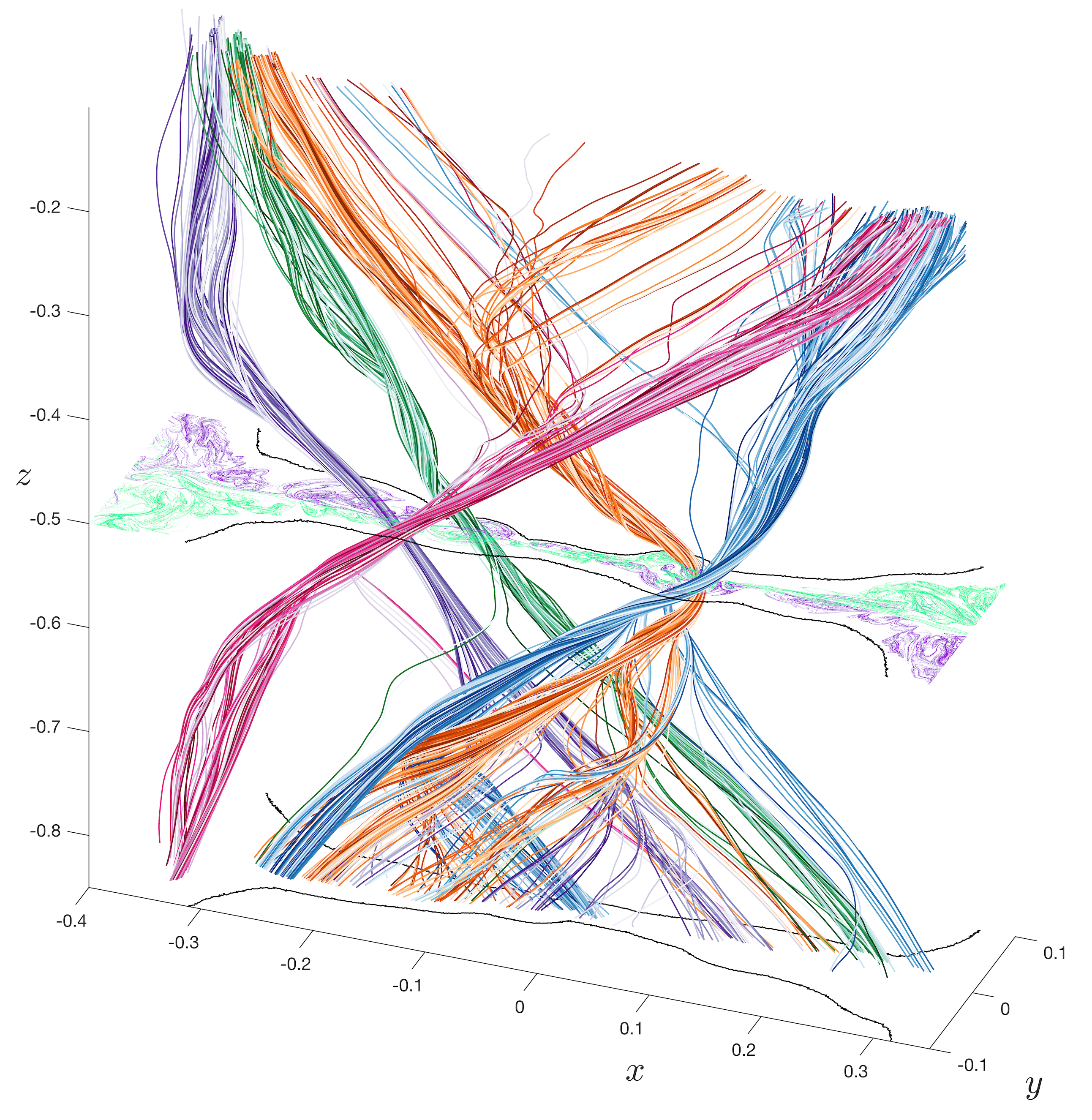}
\caption{Sample flux rope structures in the reconnection layer at $t = 3.0$. The forward (green) and  backward (purple) finite-time Lyapunov exponent (FTLE) $\Lambda_{z}^{\pm1}$ maximising ridges (50\% of the maximum over the plane) are shown at $z=-0.5$ (c.f., Figure~\ref{fig:FTLE}). The stochastic separatrices (black), approximated using contours of the Poincar\'e section count $\mathcal{N}_S = 0.5$ (for $n=2000$), are provided on the $z=-0.5$ and $z=-0.85$ planes. The field line bundles (multicoloured) are traced through selected low-pressure regions within the flux rope structure cross-sections at $z=-0.5$. Here, flux rope structures are made up of field lines that are locally coherent about $z=-0.5$ before diverging at various rates in the positive and/or negative $z$ direction. 
\label{fig:FTLE3DFL}} 
\end{figure*}

Figure~\ref{fig:FTLE} shows the FTLE maximising ridges at $z_0 = -0.5$ for various iterations $n$ at $t = 3.0$, consistent with Figures~\ref{fig:colmapextra}. In each panel, the attracting ($n<0$ in purple) and repelling ($n>0$ in green) ridges are compared for the same $\vert n\vert$; the top, middle and bottom panels show $n = \pm 0.1$, $n = \pm 0.25$, and $n = \pm 1$ iterations, respectively. The arbitrary maximising ridge threshold was chosen to be $50\%$ for robustness, but values $25\mbox{--}75\%$ were also effective for visualising the flux rope structures.  

\begin{figure*} 
\centering
\hspace*{-1.9cm}\includegraphics[width=1.21\textwidth]{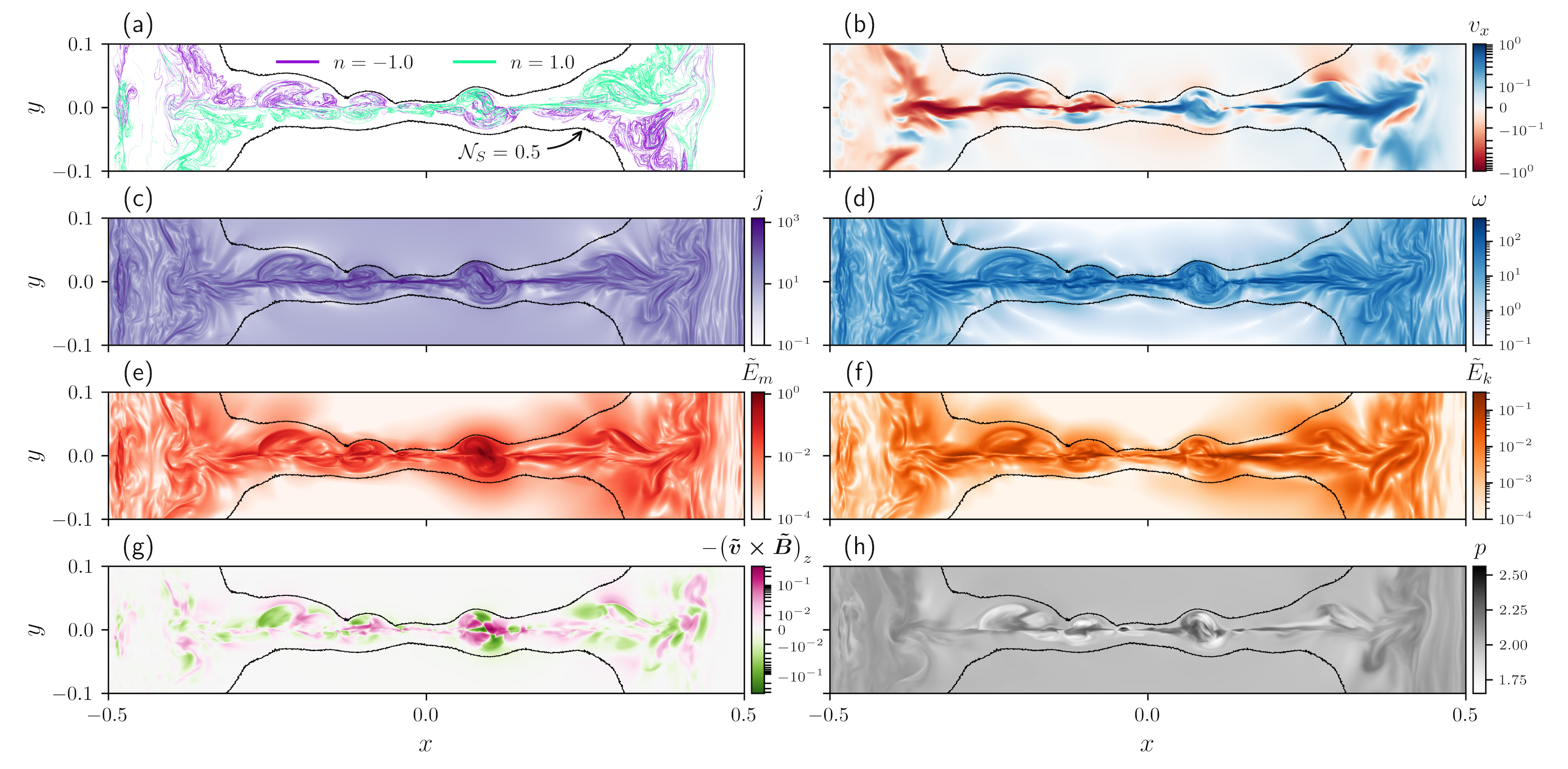} 
\caption{Comparison of physical variables, flux rope structures, and stochastic separatrices within the reconnection layer over $z = -0.5$ at $t = 3.0$. The interior contours (black) of the Poincar\'e section count $\mathcal{N}_S = 0.5$ (for $n=2000$ iterations) are superimposed over each plot. Panel (a): maximising ridges (50\% of the maximum over the plane) of the forward (green) and backward (purple) finite-time Lyapunov exponent (FTLE) $\Lambda_{z}^{\pm1}$ (c.f., Figures~\ref{fig:FTLE} and \ref{fig:FTLE3DFL}); Panel (b): outflow velocity $v_x$; Panel (c): current density strength $j$; Panel (d): vorticity strength $\omega$; Panel (e): energy of the magnetic field fluctuations $\tilde{E}_m$; Panel (f): energy of the kinetic fluctuations $\tilde{E}_k$; Panel (g): turbulent EMF $- \bigl(\boldsymbol{\tilde{v} \times \tilde{B}}\bigr)_z$; Panel (h): pressure $p$. \label{fig:FTLEinnercomp}}
\end{figure*}

A 3D diagram is provided in Figure~\ref{fig:FTLE3DFL} to give a physical understanding of the relationship between the FTLE structures and properties of the field lines. The FTLE maximising ridges are plotted at $z = -0.5$ for $n=\pm 1$ consistent with Figure~\ref{fig:FTLE}, and selected bundles of field lines are traced from $z=-0.5$, placing seed points in regions where the pressure is low $p \in (1.7,1.9)$ (see Figure~\ref{fig:FTLEinnercomp}). The stochastic separatrices (black) are also shown on the $z=-0.5$ and $z=-0.85$ planes, approximated using contours of $\mathcal{N}_S = 0.5$ for $n=2000$ iterations (see Section~\ref{sec:outerlayer}). The key idea is that a field line that passes through a FTLE ridge for a particular $n$ at $z=-0.5$ will deviate substantially from nearby field lines as it traced to $z=-0.5+n$, whereas a field line that passes through a space without FTLE ridges at $z=-0.5$ will (typically) remain relatively close to adjacent field lines over the respective distance in $z$. 

Examining Figure~\ref{fig:FTLE}, the attracting and repelling ridges highlight different substructures within the reconnection layer; a distinction not made when using $Q_{z_0}^{1}$ (Figures~\ref{fig:Q} and \ref{fig:QNScomp}). Firstly, the separation rate is not uniform over the plane: some field lines within the flux rope structures quickly diverge for small distances from $z=-0.5$ in the forward and/or backward directions, while others field lines form coherent braids with neighbouring field lines over longer distances before separating. For example, the orange flux bundle in Figure~\ref{fig:FTLE3DFL} intersects $z=-0.5$ in a region which is mostly empty of FTLE ridges for $n=\pm 0.1$ in Figure~\ref{fig:FTLE} except for some faint attracting ridges, indicating that the field lines are relatively cohesive over $z \in [-0.6,-0.4]$ with some separation in the backwards direction. However, by $n=\pm 0.25$, the traced region in Figure~\ref{fig:FTLE} is significantly covered by overlapping attracting and repelling ridges; in Figure~\ref{fig:FTLE3DFL}, the field lines have separated substantially once they are traced to $z=-0.75$ or $z=-0.25$.

Secondly, the plasmoid-type cross-sectional structures, identified earlier from $j$ or $Q_{z_0}^{1}$, are clearly outlined by a thin skeleton of field lines displaying strong local deformation for small trace distances $n = \pm 0.1$; by considering longer distances $\vert n \vert >1$, we exchange sharpness of the skeleton for greater coverage within the flux ropes structures, due to the topological complexity of the field line mapping and inherent stochasticity. Lastly, as the trace distance $\vert n\vert$ increases, the flux bundles slowly mix; the attracting and repelling ridges eventually start to fill the stochastic regions and coincide in the limit as $\vert n\vert \gg 1$ \citep{BorgognoEA2017}. Field lines are therefore temporarily restricted within the flux rope structures before wandering off to join other structures or the broader stochastic layer \citep{BorgognoEA2011a, StanierEA2019}; this frayed property allows the energetic inner core to be magnetically connected to the rest of the stochastic layer. From the perspective of SGTR, these internal local confinement structures are far more relevant than the asymptotic confinement boundaries represented by the stochastic separatrices \citep[e.g.,][]{BorgognoEA2011a}.  

Finally, we address the connection between the flux rope structures and the inner and outer scales. Figure~\ref{fig:FTLEinnercomp} is an ensemble plot comparing different variables over $z = -0.5$ at $t=3.0$, consistent with Figures~\ref{fig:FTLE} and \ref{fig:FTLE3DFL}, superimposed with the stochastic separatrices (black): panel (a) shows the FTLE maximising ridges for $n=\pm1$ (green and purple), panel (h) shows the pressure $p$, and the remaining panels plot the physical variables described in Section~\ref{sec:innerlayer}. While the plasmoid-type structures on the $z$-slice have regions of both low pressure and high pressure, they do not resemble magnetostatic equilibria. Comparing the panels, the maximising ridges clearly coincide with prominent features of the physical variables inside the SGTR layer and disturb the enclosing stochastic separatrices. This indicates that the flux rope structures detected from the field line mapping control the dynamics throughout the SGTR layer, help to shape the stochastic layer and outflows jets corresponding to $v_x$, and drive the turbulent EMF $- \bigl(\boldsymbol{\tilde{v} \times \tilde{B}}\bigr)_z$. Hence, it is very likely that the dynamics of the flux rope structures produce the mean field properties (Figures~\ref{fig:meanprof}, \ref{fig:Eflucratio}, and \ref{fig:Ezdecomp}), set the inner and outer scales of the reconnection layer, and play a central role in the enhancement of the reconnection rate and enabling SGTR. 
This reflects a recurring theme in complexity science, that the dynamics of ``microscopic'' individuals (in this case, individual flux rope structures) often give rise to seemingly simpler ``macroscopic'' properties (in this case, the mean field properties of SGTR).

\subsection{Dichotomy: Plasmoid-mediated and Lazarian-Vishniac perspectives}\label{sec:dichotomy}
Our simulation and analysis results show that SGTR possesses a variety of rich properties in agreement with prior studies, such as a highly dynamic nonlinear evolution and a topologically complex and stochastic reconnection layer. However, we have an \textit{apparent dichotomy}, since the simulation exhibits aspects of both nonlinear plasmoid reconnection \citep{BaalrudEA2012, EdmondsonEA2010, EdmondsonLynch2017, ComissoEA2017, LingamComisso2018, ComissoEA2018, LeakeEA2020} and the Lazarian-Vishniac model for stochastic reconnection \citep{LazarianVishniac1999}. 

On the one hand, the simulation pathway towards SGTR from Sweet-Parker reconnection is initiated by the onset of the 2.5D tearing instability (\textbf{b} in Figure~\ref{fig:Vrec}). Also, the quasi-stationary SGTR phase (after \textbf{f} in Figure~\ref{fig:Vrec}) is distinguished by the formation and nonlinear interaction of flux rope structures, due to the development of parallel and oblique tearing modes \citep{LeakeEA2020}. This is consistent with the plasmoid-mediated perspective reported in \citet{HuangBhattacharjee2016} and other MHD simulations by \citet{Beresnyak2017},  \citet{OishiEA2015} and \citet{StrianiEA2016}. This was also observed in several kinetic simulations in nonrelativistic \citep{BowersLi2007, DaughtonEA2011, LiuEA2013, NakamuraEA2013, DaughtonEA2014, DahlinEA2015, DahlinEA2017, NakamuraEA2017, LiEA2019, StanierEA2019, AgudeloRuedaEA2021, Zhang2EA2021} and relativistic \citep{LiuEA2011, GuoEA2015, GuoEA2021, ZhangEA2021} regimes.
However, since the magnetic field is ergodic within the reconnection layer, the field lines that constitute the flux rope structures are fully connected into the rest of the stochastic layer. Further, the flux rope structures are extremely variable and topologically complicated with highly deformed cross-sections, in contrast with 2D simulations where field lines lie neatly on nested flux surfaces corresponding with chains of laminar plasmoids. This makes it especially difficult to analyse reconnection locally using the common proxy $\int_{\boldsymbol{x}_B} E_{\parallel} \ ds$ along field lines stemming from periodic points corresponding to flux rope axes \citep[e.g.,][]{PontinEA2011, LiuEA2013, HuangEA2014, GekelmanEA2020}, since most of the axes are nontrivial to detect and may not be well-defined in the first place.

On the other hand, many aspects of the simulation are sensibly interpreted from the perspective of the Lazarian-Vishniac model for stochastic reconnection, consistent with simulations by \citet{KowalEA2017, KowalEA2020}, provided a suitable replacement is made for the applied turbulence originally assumed by \citet{LazarianVishniac1999}. 3D instabilities seed stochasticity throughout the reconnection layer, resulting in the rapid development of a mixing region that is strongly turbulent, along with the enhancement of the reconnection rate, before reaching quasi-stationary global dynamics. The stochastic layer plays an important part in the evolution and its characteristic thickness correlates well with the global reconnection rate. However, the reconnection layer in our simulation displays distinct inner and outer thickness scales that differ by a factor of about six, and the SGTR core and SGTR wings exhibit fundamentally different properties in the direction of the turbulent EMF and $\bigl< \tilde{E}_m\bigr>/\bigl< \tilde{E}_k\bigr>$ ratio; whereas \citet{LazarianVishniac1999} assumed a single layer thickness and no substructure. Moreover, the stochasticity does not appear to enhance the reconnection rate by broadening the effective thickness of the reconnection layer; instead, it is the thickness of the narrower region of strong turbulence in the SGTR core that sets the global reconnection rate. We therefore propose that Lazarian-Vishniac models of SGTR should be modified to account for the ``core and wings'' structure found in this study, and to make the important distinction between the turbulent EMF (which is the primary driver of SGTR) and magnetic field line stochasticity. 

Both perspectives are currently valuable, and they require new development before they can fully describe SGTR dynamics like those analysed in this paper. This leads to the important question: are the simulation's dynamics dominated by turbulent reconnection or plasmoid reconnection, or do we obtain an irreducible dualism that is better described by an existent or new alternative model? \citet{KowalEA2020} and \citet{ChittaLazarian2020} have asserted that self-generated reconnection is dominated by MHD turbulence and the development of tearing modes is subdued, so the plasmoid instability is not a significant driver. However, our results indicate that the problem is not so straightforward to answer, and that the flux rope structures are an important part of the dynamics and should be primary targets for analysis to understand the reconnection within the SGTR layer. Indeed, the dynamics of the flux rope structures may determine the fluctuations that provide the turbulent EMF of the mean field picture of SGTR, making the two perspectives inseparable.

\subsection{Path dependence}\label{sec:pathdependence}

Lastly, we found that the formation and kinking of a CFR was a major component of the turbulent reconnection onset during our simulation, which was not found in \citet{HuangBhattacharjee2016} upon which this study builds. This raises a significant question: are the properties of SGTR dependent on the pathway that sets it up? Further exploration of this, and the role of kink instabilities within the SGTR context, is crucial. As an alternative to magnetically-driven instabilities, \citet{KowalEA2017} and \citet{StrianiEA2016} have presented simulations in which they determined that Kelvin-Helmholtz instabilities play a considerable role. \citet{KowalEA2020} later claimed that the onset and driving mechanism for SGTR is governed by the Kelvin-Helmholtz instability instead of tearing modes. The development of a framework that encompasses all of these possibilities should be an important goal within the magnetic reconnection community.

\section{Conclusions}\label{sec:conclusion}

Throughout this paper, we analysed the dynamics that occur when two laminar flux ropes merge in a $1064^3$ 3D compressive visco-resistive MHD numerical simulation. The resultant process demonstrates ``fast'' turbulent reconnection that is fully 3D, self-generated, and self-sustaining within a global Sweet-Parker-type magnetic topology, consistent with the results of \citet{HuangBhattacharjee2016} from which this study is based on. Following our results in Section~\ref{sec:results} and discussions in Section~\ref{sec:discussion}, the most important outcomes are as follows:

\begin{enumerate}
\item 3D reconnection at $S \geq 10^4$ is qualitatively different to 2D models, with fast SGTR initiating in 3D. This result agrees with earlier studies on SGTR \citep{Beresnyak2017, OishiEA2015, HuangBhattacharjee2016, StrianiEA2016, KowalEA2017, KowalEA2020}. 

\item The reconnection process is punctuated by three main stages: \textit{i. Laminar 2.5D phase}, \textit{ii. Transition phase}, and \textit{iii. SGTR phase}. During stage \textit{i.}, the simulation briefly experiences slow laminar 2.5D Sweet-Parker reconnection followed by the development of the 2.5D nonlinear tearing instability. Stage \textit{ii.} is characterised by the sudden growth of secondary 3D instabilities, development of turbulence, fragmentation of the initial current sheet, and the seeding and spread of stochastic field lines throughout the broadening reconnection layer. In stage \textit{iii.}, the simulation settles into self-sustaining fast turbulent reconnection and gradually reaches a global quasi-stationary state.

\item The global reconnection rate $V_{\mathrm{rec}}$ reveals a distinctive ``switch-on'' property at the turbulent reconnection onset, not seen in similar work by \citet{HuangBhattacharjee2016}. From an initial Sweet-Parker rate of $V_{\mathrm{rec}} \approx 0.003$, $V_{\mathrm{rec}}$ rapidly increases during the transition phase, before reaching a fast quasi-stationary value of $V_{\mathrm{rec}} \approx 0.02$ during the SGTR phase, an enhancement by a factor of approximately $6.4$ from the starting rate.

\item The onset of turbulent reconnection occurs at the same time as a 3D helical kink instability of a large cat-eye flux rope, i.e., the central flux rope (CFR), and the proliferation of a distinct topological region corresponding with stochastic field lines identified with the reconnection layer. The large flux rope structure has a significant influence on the evolution, with its expulsion from the reconnection layer marking the development towards ``pure'' SGTR. 

\item The reconnection layer has two general characteristic thickness $\delta$ scales, evident from the evaluation of the mean layer thicknesses and mean profiles with respect to different variables. The inner thickness scale is governed by an SGTR core associated with the current density, vorticity density, outflow jets, and turbulent fluctuations. The outer thickness scale is determined by the stochastic layer, which is about six times larger than the inner thickness scale for the simulation analysed in this paper. The mean layer thicknesses are found to correlate with the global reconnection rate, i.e., they sharply increase during the transition phase and eventually attain quasi-stationary values during SGTR.

\item Between the SGTR core and stochastic sepatrices exist intermediate regions, i.e., SGTR wings. Within the SGTR wings, from a mean field perspective, the turbulent EMF changes sign and MHD turbulence is different in character compared to the core. The ``core and wings'' structure could potentially be crucial for a complete understanding of SGTR, since the weak turbulence within the SGTR wings generates stochasticity and magnetically connects the SGTR core with the rest of the stochastic layer. 

\item From tests of the Sweet-Parker scaling $V_{\mathrm{rec}} = v_{\mathrm{in}} \approx v_A\delta/L$ during the quasi-stationary SGTR phase, we conclude that the effective thickness of the reconnection layer is not the stochastic layer thickness; it is instead the narrower inner thickness of the turbulent EMF, i.e., effective reconnection electric field produced by turbulent fluctuations. This result implies that the Lazarian-Vishniac model of turbulent reconnection requires modification.

\item The ``frayed'' flux rope structures that dominate the dynamic reconnection layer during SGTR are topologically complicated, have structural properties reminiscent of magnetic braids, and consist entirely of stochastic field lines. These flux bundles may not be structured around well-defined axes, which means that quantifying a local reconnection rate, e.g., evaluating $\int_{\boldsymbol{x}_B} E_{\parallel} \ ds$ along critical curves such as flux rope axes, is nontrivial.

\item The flux rope structures, which are ``hidden'' within the stochastic layer, can be characterised using attracting or repelling ridges of the magnetic field over short distances. Within the SGTR layer, these structures generate the mean field properties (including the core and wings), govern the physical variables, and determine the inner and outer scales. Ultimately, the coherent field line arrangements likely have an important function in facilitating SGTR and enhancing the reconnection to rates that are fast. Hence, future local and interior analyses of these flux rope structures are crucial to understanding the reconnection inside.

\item Both the plasmoid-mediated and Lazarian-Vishniac perspectives of fast reconnection appear valuable for interpreting aspects of SGTR; however, both perspectives require modification to deal with the full complexity of SGTR simulations like the one analysed in this paper

\end{enumerate}

Concerning future research, there are many different potential avenues to be explored. The dependence of the SGTR properties on the path from slow to fast reconnection, the role of different instabilities in sustaining the turbulence, and the relationship between the dynamics of flux rope structures and mean turbulence are still open questions. Other research directions include: the properties of SGTR in different global magnetic topologies, such as a solar flare model; investigating how the layer thicknesses, reconnection rate, and pathway towards SGTR vary with the Lundquist number $S$; analytical study of the layer thickness scales that emerge in SGTR; rigorous analysis of the flux rope structures that dominate the SGTR layer; and the exploration of SGTR from the perspective of the evolution of magnetic helicity.

\paragraph{Funding} This work was supported by the Science and Technology Facilities Council (STFC) studentship ST/T506023/1 and STFC grant ST/S000267/1. 

\paragraph{Acknowledgements}  We thank the anonymous referee for their helpful suggestions. Computations were carried out on Magneto, a small high performance computing (HPC) cluster funded by STFC grant ST/K000993/1 and the University of Dundee. Computations were also performed using the Cambridge Service for Data Driven Discovery (CSD3), part of which is operated by the University of Cambridge Research Computing on behalf of the STFC DiRAC HPC Facility (www.dirac.ac.uk). The DiRAC component of CSD3 was funded by BEIS capital funding via STFC capital grants ST/P002307/1 and ST/R002452/1 and STFC operations grant ST/R00689X/1. DiRAC is part of the National e-Infrastructure. This research has made use of NASA’s Astrophysics Data System (NASA/ADS) Bibliographic Services.

\appendix

\section{Reconnection rate in 3D}\label{sec:vrec}

\begin{figure} 
\centering 
\hspace*{-1.4cm}\includegraphics[width=1.25\columnwidth]{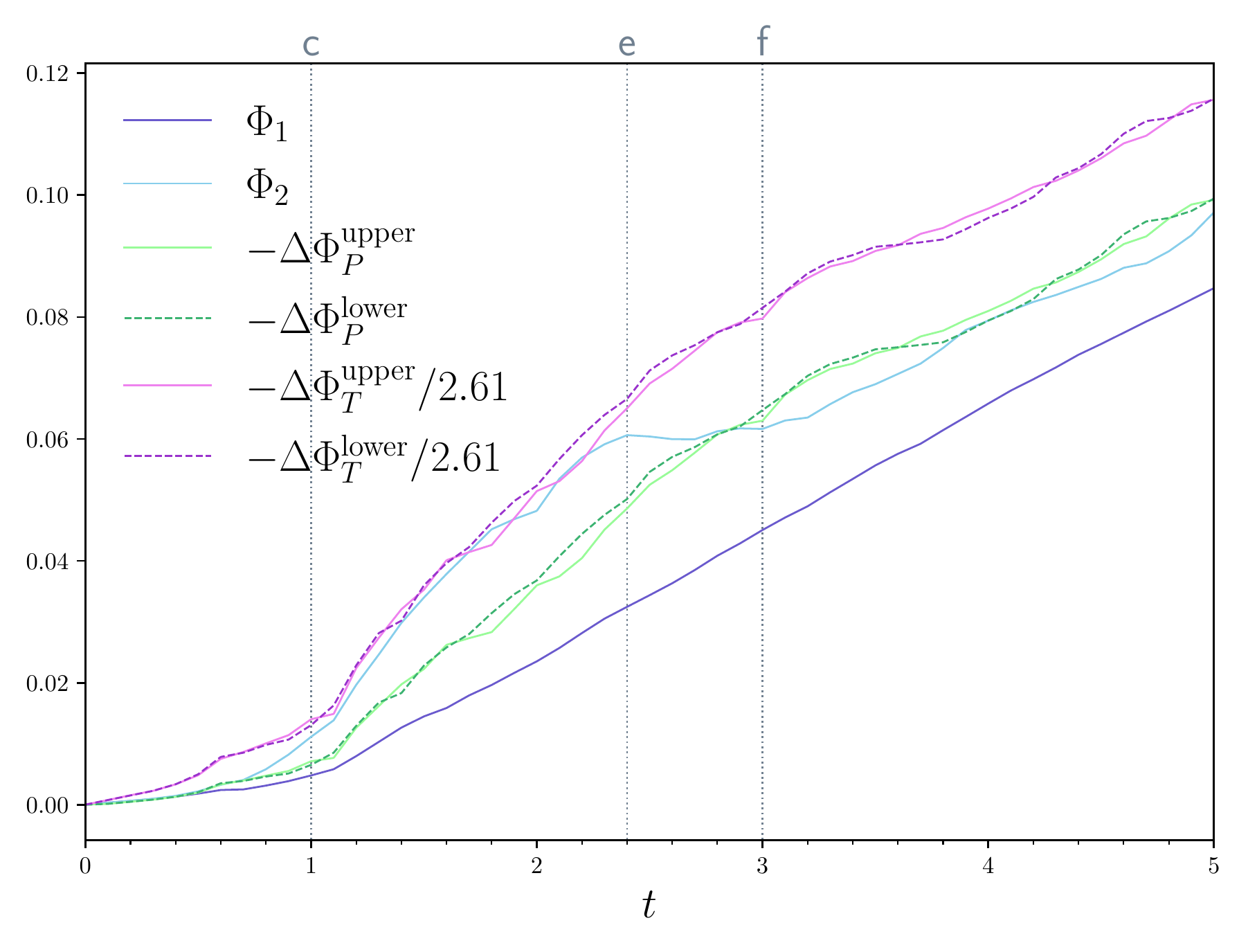} 
\caption{Comparison plot of different flux functions that can be used as a proxy for the reconnection rate. Dark blue: flux $\Phi_1$; Light blue: modified flux $\Phi_2$; Light green and (dashed) dark green: loss of poloidal fluxes $\Phi_P$ from the upper and lower laminar flux ropes; Pink and (dashed) purple: loss of toroidal fluxes $\Phi_T$ from the upper and lower laminar flux ropes, scaled down by the average flux ratio over $t \in [2.4,5.0]$ of $\Phi_T/\Phi_P \approx 2.61$. The vertical dotted lines indicate times of interest: \textbf{c} Turbulent reconnection onset ($t= 1.0$); \textbf{e} SGTR onset ($t = 2.4$); \textbf{f} ``Pure'' SGTR  onset ($t = 3.0$). \label{fig:fluxcomp}}
\end{figure}

While the reconnection rate $V_{\mathrm{rec}}$ in 2D configurations has an established definition, the analogous measure in 3D is not trivial to define and is currently a point of discussion \citep[e.g,][and references therein]{YeatesHornig2011b, DaughtonEA2014, HuangEA2014, WyperPontin2014, WyperHesse2015}. To ensure robust conclusions, we tested a variety of magnetic flux terms, the rate of change of which are suitable proxies for the global reconnection rate, with each flux capturing different nuances of the reconnection process. The evolution of these flux measurements are compared in Figure~\ref{fig:fluxcomp}. 

The dark blue curve represents the flux definition employed in this paper and \citet{HuangBhattacharjee2016}, given by
\begin{linenomath*}\begin{equation*}
\Phi_1(t) = \max_{x \in [-0.5,0.5]}  \int_{-0.5}^x\int_{-0.5}^{0.5} \left.B_y\right\vert_{y=0} \ dz'\ dx', \label{eqn:HB2016flux}
\end{equation*}\end{linenomath*}
which is the maximum net flux through a rectangular plane on $y=0.0$ with one boundary at $x=-0.5$. This is a generalisation of the flux lost from the initial flux ropes in a 2D or 2.5D geometry under the robust assumption that reconnection primarily occurs across $y=0.0$ during the simulation. Over the long term, $\Phi_1$ is dominated by the reconnected flux at large scales, which forms a growing band of magnetic field around the pair of merging twisted flux ropes. Therefore, it is a fair approximation to regard $\Phi_1$ as a measure of the magnetic flux that has finished reconnecting. 

However, while $\Phi_1$ can effectively evaluate the flux associated with a single isolated reconnection event, it fails in general to incorporate the total flux corresponding with many simultaneous reconnection processes due to cancellation of the signed area over the integral of $B_y$. For example, for a chain of same-sized magnetic islands aligned with $y=0.0$ in 2D or 2.5D, $\Phi_1$ will only measure the flux in one island; this argument extends to flux ropes in 3D. A useful modification, denoted by the light blue curve in Figure~\ref{fig:fluxcomp}, is
\begin{linenomath*}\begin{equation*}
\Phi_2(t) = \frac{1}{2} \int_{-0.5}^{0.5} \int_{-0.5}^{0.5} \left\vert \left.B_y\right\vert_{y=0} \right\vert \ dz'\ dx', \label{eqn:fluxmod}
\end{equation*}\end{linenomath*}
which takes the average of the positive and negative magnetic fluxes through the entire $y=0.0$ midplane. Physically, the major difference is that $\Phi_2$ includes all flux ropes that intersect $y=0.0$ as flux that has reconnected but may still be reconnecting, whereas $\Phi_1$ is more focused on global scale flux that has finished reconnecting. The flux $\Phi_2$ displays a significantly larger gradient than $\Phi_1$ over $t = 0.4\mbox{--}2.4$ (before \textbf{e} in Figure~\ref{fig:fluxcomp}) before reducing to approximate linear growth similar to $\Phi_1$ by $t = 3.0$ (after \textbf{f} in Figure~\ref{fig:fluxcomp}) during the phase of quasi-stationary SGTR. Practically, the more globally-focused measure of $\Phi_1$ is smoother, which is desirable when quantifying the reconnection rate.

Since our simulation contains two large merging flux ropes whose interiors remain laminar, it is also sensible to measure the poloidal and toroidal fluxes of these. In this case, the rate of loss of flux from the flux ropes provides an alternative view of the reconnection rate, such that $V_{\mathrm{rec}} = -d \Phi/dt$. The boundary surface of each laminar flux rope, i.e., stochastic separatrix, was identified from interior contours of the Poincar\'e section count using $\mathcal{N}_S = 0.5$ (see Section~\ref{sec:outerlayer}). By Stokes' theorem, the magnetic flux over an oriented surface $R$ can be readily evaluated using
\begin{linenomath*}\begin{equation*}
\Phi = \iint_R \boldsymbol{B} \boldsymbol{\cdot} d\boldsymbol{a} = \oint_{\partial R} \boldsymbol{A} \boldsymbol{\cdot} d \boldsymbol{l}, \quad \boldsymbol{B} = \boldsymbol{\nabla \times A}.
\end{equation*}\end{linenomath*}
We require the vector potential $\boldsymbol{A}$ to be periodic in $z$ and sufficiently smooth, otherwise it is arbitrary; a suitable choice is 
\begin{linenomath*}\begin{equation*}
A_x = - \int_{-0.5}^{y} B_z \ dy', \quad A_y = 0, \quad A_z = \int_{-0.5}^{y} B_x \ dy'.
\end{equation*}\end{linenomath*}
The poloidal flux in each laminar flux rope is calculated as the magnetic flux through a periodic ribbon $R$, such that the boundary $\partial R$ consists of two loops: one lying on the flux rope's boundary surface $L_S$ and another coinciding with the flux rope's axis $L_{\mathrm{axis}}$. The flux rope axes were obtained by identifying the associated elliptic periodic points at $z=-0.5$ using the colour map (see Section~\ref{sec:transitionphase}) then tracing the corresponding field lines. For simplicity, we considered the flux through a ribbon $R(\theta)$ at a fixed local poloidal coordinate value $\theta \in [0,2\pi)$ with respect to each $L_{\mathrm{axis}}$ to ensure that the ribbon was periodic in the toroidal $z$ direction. Hence, we evaluated the poloidal fluxes using
\begin{linenomath*}\begin{equation}
\Phi_P(\theta) = \oint_{L_S(\theta)} \boldsymbol{A} \boldsymbol{\cdot} d\boldsymbol{l} -  \oint_{L_{\mathrm{axis}}} \boldsymbol{A} \boldsymbol{\cdot} d\boldsymbol{l}. \label{eqn:poloidalflux}
\end{equation}\end{linenomath*}
Since the poloidal fluxes should be invariant to the measurement angle for both of the laminar flux ropes, we evaluated the mean $\bigl< \Phi_P(\theta)\bigr>$ over many samples $\theta \in [0,2\pi)$ to estimate the error, which was found to be very small. In Figure~\ref{fig:fluxcomp}, the loss of poloidal fluxes $-\Delta\Phi_P$ from the upper flux rope (light green) and lower flux rope (dashed dark green) agree well with each other, with minor fluctuations due to short-term variations in reconnection between each rope and the turbulent reconnection layer. The consistency between both poloidal fluxes confirms balanced reconnection above and below the reconnection region, as expected due to symmetry of the initial setup about $y=0.0$. The poloidal fluxes removed from the laminar flux ropes approximately track the previous flux measurements, especially $\Phi_2$ during quasi-stationary SGTR (after \textbf{f} in Figure~\ref{fig:fluxcomp}), indicating that the loss rate of the poloidal fluxes correspond well with the global reconnection rate. 

The evaluation of the poloidal fluxes $\Phi_P$ has close similarities with a method used in \citet{DaughtonEA2014} [see also \citet{YangEA2020}], who computed a reconnection rate related to the loss rate of the laminar magnetic flux $\Phi$ in regions above or below their stochastic layer by directly calculating $d\Phi/dt$. In principle, the time derivative of a flux can be evaluated via Leibniz integral rule, Faraday's law, and Stoke's theorem, giving
\begin{linenomath*}\begin{equation*}
\frac{d\Phi}{dt} = -\oint_{\partial R} \bigl(\boldsymbol{E} + \boldsymbol{w} \boldsymbol{\times B}\bigr)\boldsymbol{\cdot} d\boldsymbol{l},
\end{equation*}\end{linenomath*}
which is the electric voltage with an additional contribution from the velocity $\boldsymbol{w}$ of the surface boundary $\partial R$. However, in general, an arbitrary boundary subpath $\partial R_i$ lying on a stochastic separatrix is not comoving with the plasma ($\boldsymbol{w} \neq \boldsymbol{v}$), which leaves the difficult task of approximating $\boldsymbol{w}$, and $\partial R_i$ does not coincide with a single field line, hence the $\boldsymbol{w \times B \cdot} d\boldsymbol{l}$ term does not vanish by the usual arguments. We comment that equation (4) in \citet{DaughtonEA2014}, which uses the electric voltage only, does not hold since the stochastic separatrices in their simulation and ours are time dependent, i.e., the $\boldsymbol{w}$ term cannot be neglected. Further, from comparison tests between the electric voltage and $\Phi_P$ in our simulation, the $\boldsymbol{w}$ term contribution was found to be significant.

The above definitions of $\Phi$ track reconnection of the horizontal components of the magnetic field, i.e., the components that bear strongest analogy to 2D reconnection. For a 3D magnetic field, one can also examine reconnection of the toroidal flux within each laminar flux rope, which we calculate as the magnetic flux through a cross-sectional surface $R$ with closed boundary $\partial R = L_S$ lying on the flux rope boundary surface. We considered the toroidal flux at a fixed $z \in [-0.5,0.5)$ value for convenience, evaluated using
\begin{linenomath*}\begin{equation*}
\Phi_T(z) = \oint_{L_S(z)} \boldsymbol{A} \boldsymbol{\cdot} d\boldsymbol{l}.
\end{equation*}\end{linenomath*}
The mean $\bigl< \Phi_T(z)\bigr>$ was then taken over many samples $z \in [-0.5,0.5)$, which were confirmed to be very consistent. In Figure~\ref{fig:fluxcomp}, the loss $-\Delta\Phi_T$ of the upper (pink) and lower (dashed purple) toroidal fluxes were found to be significantly larger than the other flux measurements owing to the relative strength of $B_z$ compared to the other magnetic field components. The ratio between the toroidal and poloidal fluxes displayed some minor temporal variation: the ratio gradually decreases from $\Phi_T/\Phi_P \approx 2.88$ at $t=0.0$ before reaching a quasi-stationary plateau by $t= 2.4$ (\textbf{e} in Figure~\ref{fig:fluxcomp}), with average ratio $\Phi_T/\Phi_P \approx 2.61$ over $t \in [2.4,5.0]$. Hence, the toroidal fluxes approximately tracked the poloidal fluxes during the SGTR phase. To aid comparison in Figure~\ref{fig:fluxcomp}, the toroidal flux losses are scaled down by this ratio, i.e., $-\Delta\Phi_T/2.61$. An explanation for this time dependence is that the initial flux ratio $\Phi_T/\Phi_P$ at $t=0.0$ corresponding with contour $B_z = B_{z0}$ is not constant for varying $B_{z0}$; the ratio ranges from a maximum $\Phi_T/\Phi_P \approx 2.878$ as $B_{z0} \to 0.75^+$ to a minimum of $\Phi_T/\Phi_P \approx 2.597$ at $B_{z0} \approx 0.8942$. Therefore, after the turbulent reconnection onset, while the laminar boundaries of each flux rope rapidly contract, the flux ratio $\Phi_T/\Phi_P$ appears to tend toward the sink $\Phi_T/\Phi_P \approx 2.597$ determined by the initial configuration.

The main conclusion of this comparison is that the ``reconnected flux'' measurements are largely in agreement, subject to applying a scaling factor to the toroidal fluxes. The most important practical difference comes from the consideration that the derivative $d\Phi/dt$ is sensitive to short term fluctuations in $\Phi$, hence it is desirable to identify a flux that is naturally as smooth as possible.  Due to the observed relative lack of fluctuations, we found that flux $\Phi_1$ used by \citet{HuangBhattacharjee2016} was the best candidate for this purpose, so we chose $V_{\mathrm{rec}} = d\Phi_1/dt$ in Section~\ref{sec:pathway}. The difference in smoothness appears to result from the reconnection process. Here, we conjecture that the laminar flux ropes are eroded through many small scale reconnection events that transfer magnetic flux into the turbulent reconnection layer (probed by $\Phi_2$, $-\Delta\Phi_P^{\mathrm{upper}}$ and $-\Delta\Phi_P^{\mathrm{lower}}$); however, later processing of magnetic flux within the reconnection layer smooths its output of globally-reconnected magnetic flux (probed by $\Phi_1$). In this scenario, the turbulent reconnection layer functions as a reservoir of reconnecting magnetic flux, similar to how a water reservoir converts a sporadic input of rainfall to a smoother output.

\section{Kink instability investigation}\label{sec:kink}

In this section, we provide some further analyses on the kink instability of the CFR that occurs at $t = 1.0$ (\textbf{c} in Figure~\ref{fig:Vrec}) described in Section~\ref{sec:transitionphase}. 

We first remark that exploring this kink instability analytically is nontrivial, since the CFR possesses a cat-eye cross-section, containing an elliptical core, within a magnetic shear. It is not a classical kink instability of a circular cylindrical plasma column within a vacuum. The stability of straight cylindrical flux ropes with circular cross-sections have been investigated for a variety of force-free configurations $\boldsymbol{\nabla \times B} = \alpha(\boldsymbol{x})\boldsymbol{B}$ for different force-free parameters $\alpha(\boldsymbol{x})$, e.g.,\  constant $\alpha$ \citep[e.g.,][]{Lundquist1951, LintonEA1996, FanEA1999}, piecewise-constant $\alpha(r)$ \citep[e.g.,][]{BrowningEA2008, HoodEA2009, BarefordEA2010, BarefordEA2013}, or continuous $\alpha(r)$ \citep[e.g.,][]{HuLi2002, HoodEA2009, RestanteEA2013, BarefordEA2015, HoodEA2016}. Numerous models of flux ropes with elliptical cross-sections have also been derived in the literature, both analytically \citep[e.g.,][]{MulliganRussell2001, HidalgoEA2002, VandasRomashets2003, ErdelyiMorton2009, VandasRomashets2017, Nieves-ChinchillaEA2018} and numerically \cite[e.g.,][]{CapKhalil1989, Tsuji1991}, for force-free or non-force-free fields. While kink instabilities of elliptical-cylindrical flux ropes have been investigated to some detail in several papers \citep{DewarEA1974, FreidbergHaas1974, GuQiu1980, ErdelyiMorton2009}, we do not attempt to apply their stability analysis results to the CFR here. Regarding cat-eye flux ropes, some analytical solutions are known \citep[e.g.,][]{ThroumoulopoulosEA2009, DewarEA2013}. 

The primary aims here are the following. Using topological tools, we investigate how closely the kink instability in our simulation resembles a classical kink instability and illuminate significant properties that would need to be considered within a potential analytical model. Namely, we inspect the local distribution of field line twist and writhe, and force-freeness of the CFR. Finally, we discuss the dominant kink mode. 

\paragraph{Twist and writhe} Twist-writhe conversion is characteristic of a classical kink instability, therefore it is insightful to evaluate the twist and writhe of field lines across the CFR cross-section. Since field lines traced from $z=-0.5$ to $z=0.5$ that make up the CFR form open curves in general, instead of closed loops, we consider the open definitions for these particular metrics for two disjoint curves, notated $\widetilde{\mathcal{T}}$ and $\widetilde{\mathcal{W}}$  \citep{BergerPrior2006, PriorYeates2014,  PriorYeates2018}. For our simulation, all field lines can be conveniently parameterised in the $z$ direction by system (\ref{eqn:odesys}) and we have no critical points since $dz_B(z)/ds = (B_z/B)(\boldsymbol{x}_B(z)) >0$, where $s$ is the local arc length of field lines. We only consider the CFR axis denoted $\boldsymbol{x}_{\mathrm{axis}}$, i.e.,\ the primary curve, and field lines $\boldsymbol{x}_B$, i.e., secondary curves, with foot points $(x_0,y_0)$ at $z_0 = -0.5$. The \textit{open twist} of a field line around the CFR axis, excluding $\boldsymbol{x}_{\mathrm{axis}}$ itself, is defined as
\begin{linenomath*}\begin{equation*}
\widetilde{\mathcal{T}}(x_0,y_0) = \frac{1}{2\pi} \int_{-0.5}^{0.5}  \boldsymbol{\widehat{T}}_{\mathrm{axis}} \boldsymbol{\cdot}  \boldsymbol{\widehat{n}}\boldsymbol{\times} \frac{d \boldsymbol{\widehat{n}}}{dz} \ dz.\label{eqn:Taxis}
 \end{equation*}\end{linenomath*}
Here, $\boldsymbol{\widehat{T}}_{\mathrm{axis}}(z) = (\boldsymbol{B}/B)(\boldsymbol{x}_{\mathrm{axis}}(z))$ is the unit tangent vector of $\boldsymbol{x}_{\mathrm{axis}}$, and $\boldsymbol{\widehat{n}}$ is the unit vector orthogonal to $\boldsymbol{\widehat{T}}_{\mathrm{axis}}$ along $\boldsymbol{x}_{\mathrm{axis}}$ towards $\boldsymbol{x}_B$, i.e., satisfying $\boldsymbol{\widehat{n}} \cdot \boldsymbol{\widehat{T}}_{\mathrm{axis}} = 0$ and the standard ribbon condition $\boldsymbol{x}_B = \boldsymbol{x}_{\mathrm{axis}} + \varepsilon \boldsymbol{\widehat{n}}$. This is simply the integral of the rotation rate of $\boldsymbol{x}_B$ about $\boldsymbol{x}_{\mathrm{axis}}$, where positive values signify right-handed twisting. The \textit{open writhe} of a field line is given by
 \begin{linenomath*}\begin{equation*}
\widetilde{\mathcal{W}}(x_0,y_0) = \frac{1}{2\pi} \int_{-0.5}^{0.5} \frac{1}{1+\bigl\vert\boldsymbol{e}_z\cdot\boldsymbol{\widehat{T}}_B\bigr\vert} \boldsymbol{e}_z \boldsymbol{\cdot} \boldsymbol{\widehat{T}}_B \boldsymbol{\times} \frac{d  \boldsymbol{\widehat{T}}_B }{dz} \ dz, \label{eqn:W} 
 \end{equation*}\end{linenomath*}
where $\boldsymbol{\widehat{T}}_B(z) = (\boldsymbol{B}/B)(\boldsymbol{x}_{B}(z))$ is the unit tangent vector of $\boldsymbol{x}_B$; this is a measure of the coiling and kinking of $\boldsymbol{x}_B$ only. We also consider the \textit{alternative twist} defined as
 \begin{equation}
\mathcal{T}(x_0,y_0) = \frac{1}{4\pi} \int_{\boldsymbol{x}_B}  \frac{j_{\parallel}}{B} \ ds,\label{eqn:Tw}
 \end{equation}
 where $j_{\parallel} = \boldsymbol{j} \boldsymbol{\cdot} \boldsymbol{e}_B$ is the parallel current density; this is the mean twist of two infinitesimally close field lines along curve $\boldsymbol{x}_B$ \citep{BergerPrior2006, LiuEA2016, PriorYeates2018, JiangEA2019}. \citet{LiuEA2016} showed that under special conditions, the equivalence $\widetilde{\mathcal{T}} = \mathcal{T}$ is obtained in the limit as $\varepsilon \rightarrow 0$; further, $\mathcal{T}$ is an effective approximation to $\widetilde{\mathcal{T}}$ within a flux rope if $\boldsymbol{x}_B$ is sufficiently close to $\boldsymbol{x}_{\mathrm{axis}}$ and the flux rope possesses approximate cylindrical symmetry (see their Appendix C.1.). However, in our case, $\widetilde{\mathcal{T}}$ and $\mathcal{T}$ differ significantly due to the highly elliptical core of the CFR cross-section.
 
We denote the alternative twist and open writhe evaluated at the CFR axis as $\mathcal{T}_{\mathrm{axis}}$ and $\widetilde{\mathcal{W}}_{\mathrm{axis}}$, respectively. For robustness, we approximated the open twist locally about the CFR axis by taking the mean $\widetilde{\mathcal{T}}$ of field lines seeded within a disc at $z_0 = -0.5$ of radius $r=5 \times 10^{-4}$ centred at $\boldsymbol{x}_{\mathrm{axis}}$, denoted $\bigl<\widetilde{\mathcal{T}}\bigr>$.

\begin{figure*} 
\centering
\hspace*{-1.9cm} \includegraphics[scale=0.84]{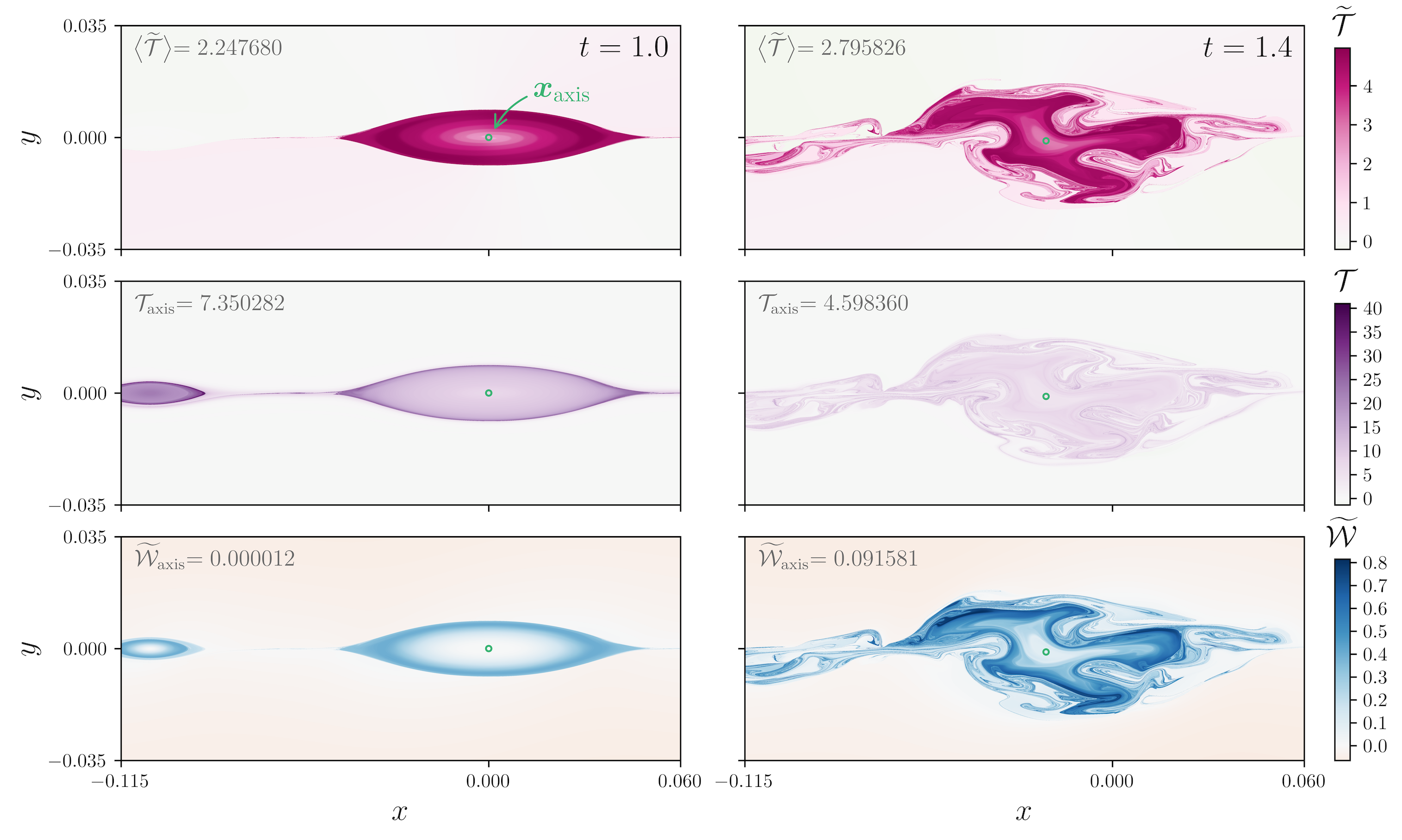} 
\caption{Closeups of the central flux rope (CFR) at the bottom boundary $z_0=-0.5$ during the 3D kink instability at $t = 1.0$ and $t = 1.4$. The CFR axis $\boldsymbol{x}_{\mathrm{axis}}$ (elliptic fixed point) is represented by the green circle (c.f., Figure~\ref{fig:colmap}). The top row shows the open twist of field lines (with respect to the CFR axis) $\widetilde{\mathcal{T}}$ and the mean value about the CFR axis $\big<\widetilde{\mathcal{T}}\bigr>$; the middle row shows the alternative twist of field lines $\mathcal{T}$ and the CFR axis $\mathcal{T}_{\mathrm{axis}}$; and the bottom row shows the open writhe of field lines $\widetilde{\mathcal{W}}$ and the CFR axis $\widetilde{\mathcal{W}}_{\mathrm{axis}}$. \label{fig:twistwrithe2D} }
\end{figure*}

\begin{figure} 
\centering
\hspace*{-1.4cm}\includegraphics[width=1.28\columnwidth]{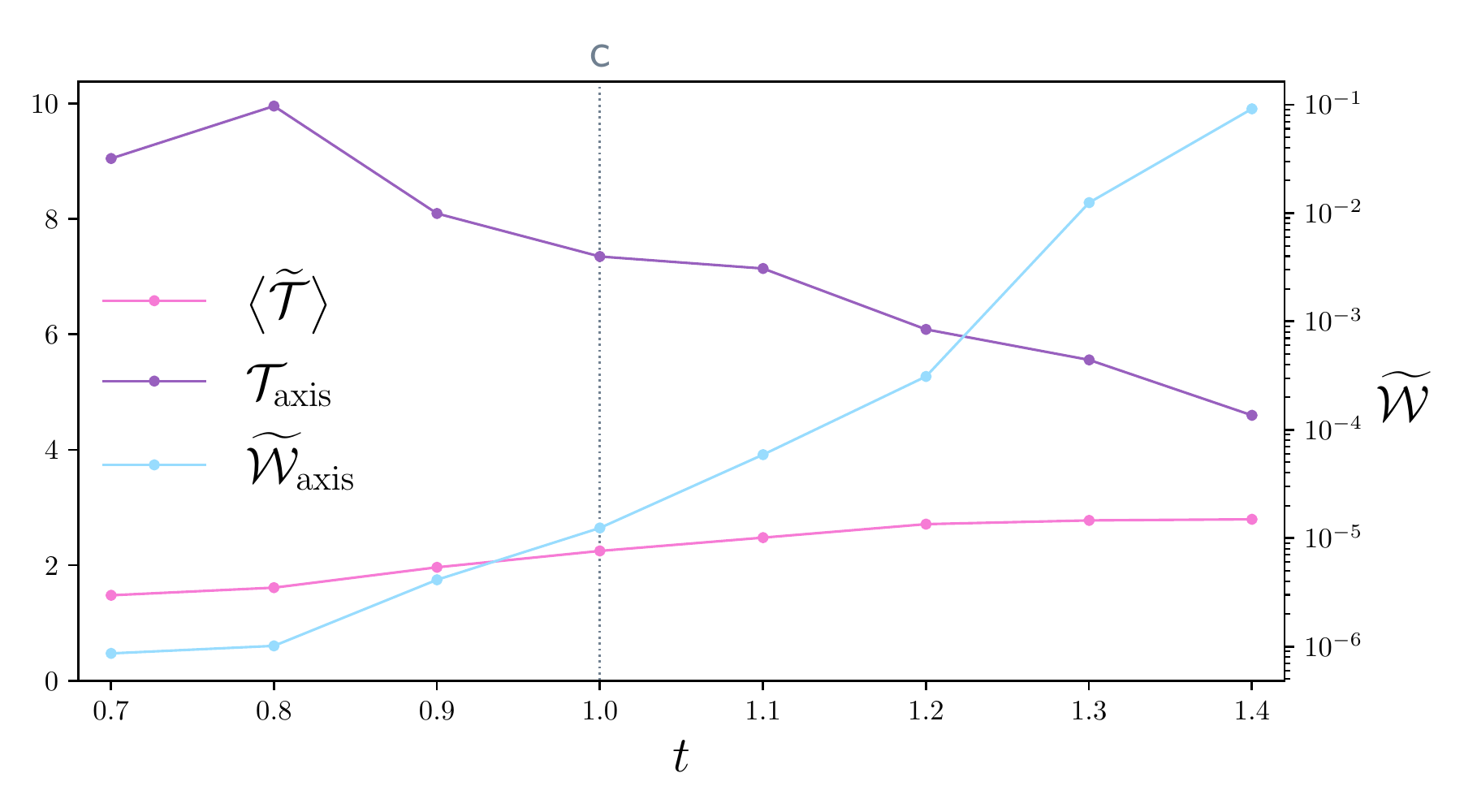}
\caption{Comparison plot of the mean open twist $\big<\widetilde{\mathcal{T}}\bigr>$ (pink), alternative twist $\mathcal{T}_{\mathrm{axis}}$ (purple), and open writhe $\widetilde{\mathcal{W}}_{\mathrm{axis}}$ (light blue) of/about the CFR axis over the interval $t = 0.7\mbox{--}1.4$ when the CFR axis can be detected. The vertical dotted line labelled \textbf{c} denotes the turbulent reconnection onset at $t= 1.0$.  \label{fig:twistwritheevol}  } 
\end{figure}

Figure~\ref{fig:twistwrithe2D} displays the 2D distribution of $\widetilde{\mathcal{T}}$, $\mathcal{T}$  and $\widetilde{\mathcal{W}}$ at $z_0=-0.5$ over the kink instability $t = 1.0\mbox{--}1.4$ (\textbf{c}--\textbf{d} in Figure~\ref{fig:Vrec}). The CFR axis $\boldsymbol{x}_{\mathrm{axis}}$, identified using the colour map (see Section~\ref{sec:transitionphase}), is represented by green circles. In the top left of each panel, the corresponding value of $\big<\widetilde{\mathcal{T}}\bigr>$, $\mathcal{T}_{\mathrm{axis}}$ or $\widetilde{\mathcal{W}}_{\mathrm{axis}}$ is provided. For $t \leq 1.0$, the open twist $\widetilde{\mathcal{T}}$ has a moderate drop towards the CFR axis; tests on a model elliptical flux rope, with uniform twist and same ellipticity as the CFR, indicate that this property is not due to small numerical errors towards $\boldsymbol{x}_{\mathrm{axis}}$. Otherwise, $\widetilde{\mathcal{T}}$ is level within the CFR away from the CFR axis and close to zero elsewhere. The alternative twist $\mathcal{T}$ is also indicator-like, but it is approximately flat within the CFR interior and has significantly larger values than $\widetilde{\mathcal{T}}$, especially near the CFR boundary. The open writhe $\widetilde{\mathcal{W}}$ is strong within the cat-eye away from the elliptical core; the relatively small numerical values do not imply an insignificant geometric effect \citep{PriorYeates2018}.

After the kink onset $t = 1.0$, we observe a rapid global decrease in $\mathcal{T}$ and global increase in $\widetilde{\mathcal{W}}$, while the CFR becomes more structurally complicated; for $\widetilde{\mathcal{T}}$, its maximum remains relatively constant, but its distribution becomes increasingly fragmented. Figure~\ref{fig:twistwritheevol} compares the evolution of the mean open twist about the CFR axis $\bigl<\widetilde{\mathcal{T}}\bigr>$ (pink) with the alternative twist $\mathcal{T}_{\mathrm{axis}}$ (purple) and open writhe $\widetilde{\mathcal{W}}_{\mathrm{axis}}$ (light blue) of the CFR axis over its detection interval $t = 0.7\mbox{--}1.4$. The right axis, corresponding with $\widetilde{\mathcal{W}}_{\mathrm{axis}}$, is shown with logarithmic scaling. We observe that $\bigl<\widetilde{\mathcal{T}}\bigr>$ increases by approximately $90\%$ and $\mathcal{T}_{\mathrm{axis}}$ decreases by approximately $50\%$ at roughly linear rates, while $\widetilde{\mathcal{W}}_{\mathrm{axis}}$ increases exponentially by over order $10^5$. Hence, the kink is not characterised by an exact twist-writhe conversion, which is not unexpected, but the results suggest that we have significant transfer between $\mathcal{T}$ and $\widetilde{\mathcal{W}}$ at a local and global level during the instability.

\paragraph{Force-freeness} Previous analyses on the kink instability of flux ropes frequently focus on the force-free parameter $\alpha(\boldsymbol{x})$, under the assumption of a force-free field $\boldsymbol{\nabla \times B} = \alpha(\boldsymbol{x})\boldsymbol{B}$. To determine how close the CFR is to a force-free equilibrium, we evaluate the parameters \citep{PontinEA2016}
\begin{linenomath*}\begin{equation*}
 \alpha^* = \frac{j_{\parallel}}{B}, \quad \varepsilon^* = \frac{j_{\perp}}{j}, 
 \end{equation*}\end{linenomath*}
where $j_{\perp} = \Vert \boldsymbol{j} \boldsymbol{\times} \boldsymbol{e}_B \Vert$. Parameter $\alpha^*$ converges to the force-free parameter $\alpha$ as a system tends towards force-free equilibrium; it can also be interpreted as the twist density along field lines  \citep{LiuEA2016}, since it is the integrand of the alternative twist $\mathcal{T}$ [definition (\ref{eqn:Tw})]. The normalised Lorentz force strength $\varepsilon^* \in [0,1]$ indicates the force-freeness of the field, ranging from an exact force-free equilibrium $\varepsilon^* = 0$ to a maximal non-force-free state $\varepsilon^* = 1$.

\begin{figure*} 
\centering
\gridline{\hspace*{-1.8cm}\fig{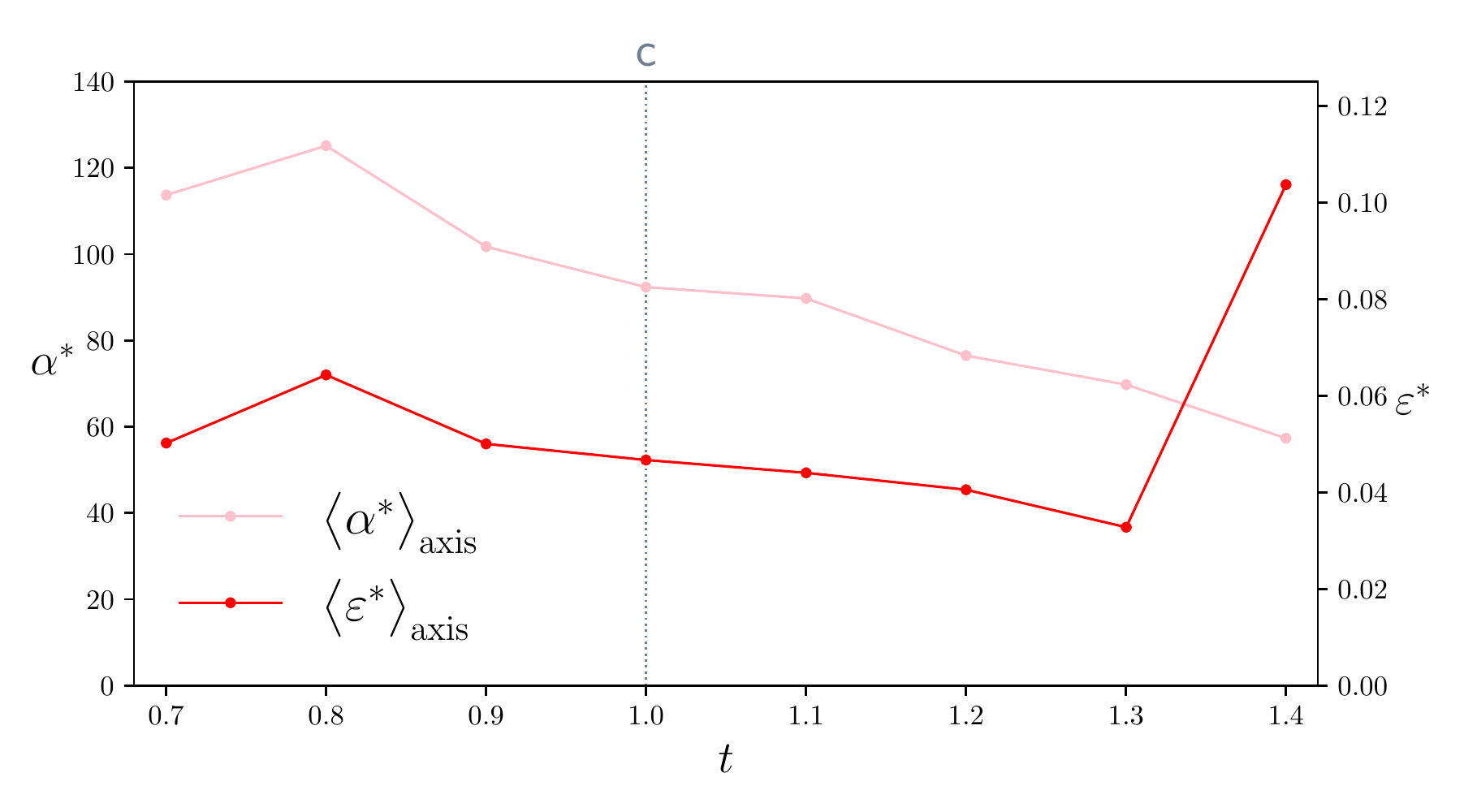}{0.6\textwidth}{(a)}
	     \fig{3Dplas_FFparFLmean1Dcomp_xslice_t=10000.eps}{0.6\textwidth}{(b)}}
\caption{Comparison plots of parameters $\alpha^*$ (pink) and $\varepsilon^*$ (red). Subfigure (a) shows the evolution of these parameters averaged over the CFR axis, denoted $\bigl< \alpha^* \bigr>_{\mathrm{axis}}$ and $\bigl< \varepsilon^* \bigr>_{\mathrm{axis}}$. The vertical dotted line labelled \textbf{c} denotes the turbulent reconnection onset at $t= 1.0$. Subfigure (b) is a sample plot comparing the field line averages of these parameters, denoted $\bigl< \alpha^* \bigr>_{B}$ and $\bigl< \varepsilon^* \bigr>_{B}$, at the turbulent reconnection onset $t= 1.0$. The 1D slice is taken at $z_0 =-0.5$ over $x = x_{\mathrm{axis}}$. The $y$ coordinate of the CFR axis $\boldsymbol{x}_{\mathrm{axis}}$ is indicated by the vertical dotted line. \label{fig:FFpar}} 
\end{figure*}

Figure~\ref{fig:FFpar}(a) shows the evolution of $\alpha^*$ and $\varepsilon^*$ averaged over the CFR axis, denoted $\bigl< \cdot \bigr>_{\mathrm{axis}}$, during the detection window $t = 0.7\mbox{--}1.4$. Figure~\ref{fig:FFpar}(b) displays the profiles of $\alpha^*$ and $\varepsilon^*$ at $t = 1.0$, averaged over field lines spanning $z \in [-0.5,0.5]$ with seed points on $z_0 = -0.5$;  these are denoted by $\bigl< \cdot \bigr>_{B}$ \citep{YeatesEA2021}. For simplicity, we have only provided the seed points along a slice at $x = x_{\mathrm{axis}}$ intersecting the CFR axis $\boldsymbol{x}_{\mathrm{axis}}$. We remark that $\bigl< \alpha^* \bigr>_{B}$ is equivalent to the alternative twist $\mathcal{T}$, up to approximate scaling $L/(4\pi)$ where $L$ is the total arc length of field lines. As the CFR grows before the kink instability at $t = 1.0$, there is a tendency for the $\bigl< \alpha^* \bigr>_{B}$ distribution to gradually flatten about the CFR axis and for the $\bigl< \varepsilon^* \bigr>_{B}$ distribution to decrease with a convex profile within the CFR interior. Hence, there is an evolution towards a force-free field with uniform $\alpha$, but the CFR does not achieve exact force-freeness at the kink instability onset $t = 1.0$ when $\bigl< \varepsilon^* \bigr>_{\mathrm{axis}} \approx 0.0467$. This is not unexpected, since flux ropes in astrophysical environments have been observed in general to be non-force-free, and therefore differ substantially from ideal models \citep{HuEA2014, HuEA2015}.

\paragraph{Dominant kink mode} Regarding the development of the largest mode $k_z/(2\pi) = 7$, it is clear that this has some dependence on the simulation grid resolution: our preliminary $500^3$ simulation for this paper exhibited a dominant $k_z/(2\pi) = 4$ mode instead, but the CFR cross-section was noticeably smaller with a higher ellipticity at the kink instability onset, suggesting that minor variations in the field geometry are causing this discrepancy. A similar kink instability along initial plasmoids, in addition to a 3D slab-type kink along the current layer, was observed in \citet{LapentaBettarini2011} and \citet{OishiEA2015} with a different dominant wavenumber of $k_z/(2\pi) = 12$; \citet{OishiEA2015} found that the dominant mode might have a dependence on the underlying wavenumber spectrum for the initial velocity perturbations used in their simulation. 

A rough explanation for the development of the dominant $k_z/(2\pi) = 7$ mode can be attempted by applying results from simple force-free circular-cylindrical models. In this context, the helical field line pitch $q = B_{\theta}/(rB_z)$, related to the twist by $q = 2\pi \widetilde{\mathcal{T}}$, has been found to be a crucial parameter. \citet{LintonEA1996} studied the kink instability of an isolated circular cylindrical flux rope with a uniform pitch $q$ and a vanishing external magnetic field, and found that the most unstable kink mode has helical pitch $k_z \approx q$. This geometry is clearly a poor approximation to our CFR configuration, but applying this result to our simulation, for twist values towards the CFR axis at the kink onset $t = 1.0$, leads to the approximate dominant mode $k_z/(2\pi) \approx \bigl<\widetilde{\mathcal{T}}\bigr> \approx 2.25$ or $k_z/(2\pi) \approx \mathcal{T}_{\mathrm{axis}} \approx 7.35$; the latter result is satisfactory, but it is unclear if this coincidental or not.


\bibliography{references}{}

\begin{thebibliography}{}
\expandafter\ifx\csname natexlab\endcsname\relax\def\natexlab#1{#1}\fi

\bibitem[{Agudelo~Rueda {et~al.}(2021)Agudelo~Rueda, Verscharen, Wicks, Owen,
  Nicolaou, Walsh, Zouganelis, Germaschewski, \&
  Vargas~Domínguez}]{AgudeloRuedaEA2021}
Agudelo~Rueda, J.~A., Verscharen, D., Wicks, R.~T., {et~al.} 2021, Journal of
  Plasma Physics, 87, 905870228

\bibitem[{{Alfv{\'e}n}(1942)}]{Alfven1942}
{Alfv{\'e}n}, H. 1942, NAT, 150, 405

\bibitem[{{Arber} {et~al.}(2001){Arber}, {Longbottom}, {Gerrard}, \&
  {Milne}}]{ArberEA2001}
{Arber}, T.~D., {Longbottom}, A.~W., {Gerrard}, C.~L., \& {Milne}, A.~M. 2001,
  Journal of Computational Physics, 171, 151

\bibitem[{{Baalrud} {et~al.}(2012){Baalrud}, {Bhattacharjee}, \&
  {Huang}}]{BaalrudEA2012}
{Baalrud}, S.~D., {Bhattacharjee}, A., \& {Huang}, Y.~M. 2012, Physics of
  Plasmas, 19, 022101

\bibitem[{{Bareford} {et~al.}(2010){Bareford}, {Browning}, \& {van der
  Linden}}]{BarefordEA2010}
{Bareford}, M.~R., {Browning}, P.~K., \& {van der Linden}, R.~A.~M. 2010, \aap,
  521, A70

\bibitem[{{Bareford} \& {Hood}(2015)}]{BarefordEA2015}
{Bareford}, M.~R., \& {Hood}, A.~W. 2015, Philosophical Transactions of the
  Royal Society of London Series A, 373, 20140266

\bibitem[{{Bareford} {et~al.}(2013){Bareford}, {Hood}, \&
  {Browning}}]{BarefordEA2013}
{Bareford}, M.~R., {Hood}, A.~W., \& {Browning}, P.~K. 2013, \aap, 550, A40

\bibitem[{{Beresnyak}(2017)}]{Beresnyak2017}
{Beresnyak}, A. 2017, \apj, 834, 47

\bibitem[{{Berger} \& {Prior}(2006)}]{BergerPrior2006}
{Berger}, M.~A., \& {Prior}, C. 2006, Journal of Physics A Mathematical
  General, 39, 8321

\bibitem[{{Boozer}(2012)}]{Boozer2012}
{Boozer}, A.~H. 2012, Physics of Plasmas, 19, 112901

\bibitem[{Borgogno {et~al.}(2015)Borgogno, Califano, Faganello, \&
  Pegoraro}]{BorgognoEA2015}
Borgogno, D., Califano, F., Faganello, M., \& Pegoraro, F. 2015, Physics of
  Plasmas, 22, 032301

\bibitem[{{Borgogno} {et~al.}(2008){Borgogno}, {Grasso}, {Pegoraro}, \&
  {Schep}}]{BorgognoEA2008}
{Borgogno}, D., {Grasso}, D., {Pegoraro}, F., \& {Schep}, T.~J. 2008, Physics
  of Plasmas, 15, 102308

\bibitem[{{Borgogno} {et~al.}(2011{\natexlab{a}}){Borgogno}, {Grasso},
  {Pegoraro}, \& {Schep}}]{BorgognoEA2011a}
---. 2011{\natexlab{a}}, Physics of Plasmas, 18, 102307

\bibitem[{{Borgogno} {et~al.}(2011{\natexlab{b}}){Borgogno}, {Grasso},
  {Pegoraro}, \& {Schep}}]{BorgognoEA2011b}
---. 2011{\natexlab{b}}, Physics of Plasmas, 18, 102308

\bibitem[{{Borgogno} {et~al.}(2017){Borgogno}, {Perona}, \&
  {Grasso}}]{BorgognoEA2017}
{Borgogno}, D., {Perona}, A., \& {Grasso}, D. 2017, Physics of Plasmas, 24,
  122303

\bibitem[{Bowers \& Li(2007)}]{BowersLi2007}
Bowers, K., \& Li, H. 2007, Phys. Rev. Lett., 98, 035002

\bibitem[{{Browning} {et~al.}(2008){Browning}, {Gerrard}, {Hood}, {Kevis}, \&
  {Van der Linden}}]{BrowningEA2008}
{Browning}, P.~K., {Gerrard}, C., {Hood}, A.~W., {Kevis}, R., \& {Van der
  Linden}, R. A.~M. 2008, A\&A, 485, 837

\bibitem[{Browning {et~al.}(2014)Browning, Stanier, Ashworth, McClements, \&
  Lukin}]{BrowningEA2014}
Browning, P.~K., Stanier, A., Ashworth, G., McClements, K.~G., \& Lukin, V.~S.
  2014, Plasma Physics and Controlled Fusion, 56, 064009

\bibitem[{{Burrell}(1997)}]{Burrell1997}
{Burrell}, K.~H. 1997, Physics of Plasmas, 4, 1499

\bibitem[{Cap \& Khalil(1989)}]{CapKhalil1989}
Cap, F., \& Khalil, S. 1989, Nuclear Fusion, 29, 1166

\bibitem[{{Chian} {et~al.}(2014){Chian}, {Rempel}, {Aulanier}, {Schmieder},
  {Shadden}, {Welsch}, \& {Yeates}}]{ChianEA2014}
{Chian}, A. C.~L., {Rempel}, E.~L., {Aulanier}, G., {et~al.} 2014, \apj, 786,
  51

\bibitem[{{Chitta} \& {Lazarian}(2020)}]{ChittaLazarian2020}
{Chitta}, L.~P., \& {Lazarian}, A. 2020, APJL, 890, L2

\bibitem[{{Comisso} {et~al.}(2018){Comisso}, {Huang}, {Lingam}, {Hirvijoki}, \&
  {Bhattacharjee}}]{ComissoEA2018}
{Comisso}, L., {Huang}, Y.~M., {Lingam}, M., {Hirvijoki}, E., \&
  {Bhattacharjee}, A. 2018, APJ, 854, 103

\bibitem[{{Comisso} {et~al.}(2017){Comisso}, {Lingam}, {Huang}, \&
  {Bhattacharjee}}]{ComissoEA2017}
{Comisso}, L., {Lingam}, M., {Huang}, Y.~M., \& {Bhattacharjee}, A. 2017, APJ,
  850, 142

\bibitem[{{Connor} {et~al.}(2004){Connor}, {Fukuda}, {Garbet}, {Gormezano},
  {Mukhovatov}, {Wakatani}, {ITB Database Group}, {ITPA Topical Group on
  Transport}, \& {Barrier Physics}}]{ConnorEA2004}
{Connor}, J.~W., {Fukuda}, T., {Garbet}, X., {et~al.} 2004, Nuclear Fusion, 44,
  R1

\bibitem[{Dahlburg {et~al.}(2003)Dahlburg, Klimchuk, \&
  Antiochos}]{DahlburgEA2003}
Dahlburg, R., Klimchuk, J., \& Antiochos, S. 2003, Advances in Space Research,
  32, 1029, connections and Reconnections in Solar and Stellar Coronae

\bibitem[{Dahlburg {et~al.}(1992)Dahlburg, Antiochos, \& Zang}]{DahlburgEA1992}
Dahlburg, R.~B., Antiochos, S.~K., \& Zang, T.~A. 1992, Physics of Fluids B:
  Plasma Physics, 4, 3902

\bibitem[{{Dahlburg} {et~al.}(2005){Dahlburg}, {Klimchuk}, \&
  {Antiochos}}]{DahlburgEA2005}
{Dahlburg}, R.~B., {Klimchuk}, J.~A., \& {Antiochos}, S.~K. 2005, \apj, 622,
  1191

\bibitem[{Dahlin {et~al.}(2015)Dahlin, Drake, \& Swisdak}]{DahlinEA2015}
Dahlin, J.~T., Drake, J.~F., \& Swisdak, M. 2015, Physics of Plasmas, 22,
  100704

\bibitem[{Dahlin {et~al.}(2017)Dahlin, Drake, \& Swisdak}]{DahlinEA2017}
---. 2017, Physics of Plasmas, 24, 092110

\bibitem[{{Daughton}(2003)}]{Daughton2003}
{Daughton}, W. 2003, Physics of Plasmas, 10, 3103

\bibitem[{{Daughton} {et~al.}(2014){Daughton}, {Nakamura}, {Karimabadi},
  {Roytershteyn}, \& Loring}]{DaughtonEA2014}
{Daughton}, W., {Nakamura}, T., {Karimabadi}, H., {Roytershteyn}, V., \&
  Loring, B. 2014, Physics of Plasmas, 21, 052307

\bibitem[{{Daughton} {et~al.}(2011){Daughton}, {Roytershteyn}, {Karimabadi},
  {Yin}, {Albright}, {Bergen}, \& {Bowers}}]{DaughtonEA2011}
{Daughton}, W., {Roytershteyn}, V., {Karimabadi}, H., {et~al.} 2011, Nature
  Physics, 7, 539

\bibitem[{Daughton {et~al.}(2006)Daughton, Scudder, \&
  Karimabadi}]{DaughtonEA2006}
Daughton, W., Scudder, J., \& Karimabadi, H. 2006, Physics of Plasmas, 13,
  072101

\bibitem[{Dewar {et~al.}(2013)Dewar, Bhattacharjee, Kulsrud, \&
  Wright}]{DewarEA2013}
Dewar, R.~L., Bhattacharjee, A., Kulsrud, R.~M., \& Wright, A.~M. 2013, Physics
  of Plasmas, 20, 082103

\bibitem[{Dewar {et~al.}(1974)Dewar, Grimm, Johnson, Frieman, Greene, \&
  Rutherford}]{DewarEA1974}
Dewar, R.~L., Grimm, R.~C., Johnson, J.~L., {et~al.} 1974, The Physics of
  Fluids, 17, 930

\bibitem[{{Di Giannatale} {et~al.}(2021){Di Giannatale}, {Bonfiglio},
  {Cappello}, {Chac{\'o}n}, \& {Veranda}}]{DiGiannataleEA2021}
{Di Giannatale}, G., {Bonfiglio}, D., {Cappello}, S., {Chac{\'o}n}, L., \&
  {Veranda}, M. 2021, Nuclear Fusion, 61, 076013

\bibitem[{{Di Giannatale} {et~al.}(2018{\natexlab{a}}){Di Giannatale},
  {Falessi}, {Grasso}, {Pegoraro}, {Schep}, {Veranda}, {Bonfiglio}, \&
  {Cappello}}]{DiGiannataleEA2018}
{Di Giannatale}, G., {Falessi}, M., {Grasso}, D., {et~al.} 2018{\natexlab{a}},
  Journal of Physics: Conference Series, 1125, 012008

\bibitem[{{Di Giannatale} {et~al.}(2018{\natexlab{b}}){Di Giannatale},
  {Falessi}, {Grasso}, {Pegoraro}, \& {Schep}}]{DiGiannataleEA2017a}
{Di Giannatale}, G., {Falessi}, M.~V., {Grasso}, D., {Pegoraro}, F., \&
  {Schep}, T.~J. 2018{\natexlab{b}}, Physics of Plasmas, 25, 052306

\bibitem[{{Di Giannatale} {et~al.}(2018{\natexlab{c}}){Di Giannatale},
  {Falessi}, {Grasso}, {Pegoraro}, \& {Schep}}]{DiGiannataleEA2017b}
---. 2018{\natexlab{c}}, Physics of Plasmas, 25, 052307

\bibitem[{{Dong} {et~al.}(2018){Dong}, {Wang}, {Huang}, {Comisso}, \&
  {Bhattacharjee}}]{DongEA2018}
{Dong}, C., {Wang}, L., {Huang}, Y.-M., {Comisso}, L., \& {Bhattacharjee}, A.
  2018, PRL, 121, 165101

\bibitem[{Edmondson {et~al.}(2010)Edmondson, Antiochos, DeVore, \&
  Zurbuchen}]{EdmondsonEA2010}
Edmondson, J.~K., Antiochos, S.~K., DeVore, C.~R., \& Zurbuchen, T.~H. 2010,
  The Astrophysical Journal, 718, 72

\bibitem[{Edmondson \& Lynch(2017)}]{EdmondsonLynch2017}
Edmondson, J.~K., \& Lynch, B.~J. 2017, The Astrophysical Journal, 849, 28

\bibitem[{{Erd{\'e}lyi} \& {Morton}(2009)}]{ErdelyiMorton2009}
{Erd{\'e}lyi}, R., \& {Morton}, R.~J. 2009, \aap, 494, 295

\bibitem[{Eyink {et~al.}(2013)Eyink, Vishniac, Lalescu, Aluie, Kanov, Bürger,
  Burns, Meneveau, \& Szalay}]{EyinkEA2013}
Eyink, G., Vishniac, E., Lalescu, C., {et~al.} 2013, Nature, 497, 466

\bibitem[{Eyink {et~al.}(2011)Eyink, Lazarian, \& Vishniac}]{EyinkEA2011}
Eyink, G.~L., Lazarian, A., \& Vishniac, E.~T. 2011, The Astrophysical Journal,
  743, 51

\bibitem[{{Faganello} {et~al.}(2014){Faganello}, {Califano}, {Pegoraro}, \&
  {Retin{\`o}}}]{FaganelloEA2014}
{Faganello}, M., {Califano}, F., {Pegoraro}, F., \& {Retin{\`o}}, A. 2014, EPL
  (Europhysics Letters), 107, 19001

\bibitem[{{Falessi} {et~al.}(2015){Falessi}, {Pegoraro}, \&
  {Schep}}]{FalessiEA2015}
{Falessi}, M.~V., {Pegoraro}, F., \& {Schep}, T.~J. 2015, Journal of Plasma
  Physics, 81, 495810505

\bibitem[{{Fan} {et~al.}(1999){Fan}, {Zweibel}, {Linton}, \&
  {Fisher}}]{FanEA1999}
{Fan}, Y., {Zweibel}, E.~G., {Linton}, M.~G., \& {Fisher}, G.~H. 1999, \apj,
  521, 460

\bibitem[{{Fletcher} {et~al.}(2011){Fletcher}, {Dennis}, {Hudson}, {Krucker},
  {Phillips}, {Veronig}, {Battaglia}, {Bone}, {Caspi}, {Chen}, {Gallagher},
  {Grigis}, {Ji}, {Liu}, {Milligan}, \& {Temmer}}]{FletcherEA2011}
{Fletcher}, L., {Dennis}, B.~R., {Hudson}, H.~S., {et~al.} 2011, \ssr, 159, 19

\bibitem[{Freidberg \& Haas(1974)}]{FreidbergHaas1974}
Freidberg, J.~P., \& Haas, F.~A. 1974, The Physics of Fluids, 17, 440

\bibitem[{Gekelman {et~al.}(2020)Gekelman, DeHaas, Prior, \&
  Yeates}]{GekelmanEA2020}
Gekelman, W., DeHaas, T., Prior, C., \& Yeates, A. 2020, SN Applied Sciences,
  2, doi:10.1007/s42452-020-03896-4

\bibitem[{{Gold} \& {Hoyle}(1960)}]{GoldHoyle1960}
{Gold}, T., \& {Hoyle}, F. 1960, \mnras, 120, 89

\bibitem[{{Goldreich} \& {Sridhar}(1995)}]{GoldreichSridhar1995}
{Goldreich}, P., \& {Sridhar}, S. 1995, APJ, 438, 763

\bibitem[{{Grad} \& {Rubin}(1958)}]{GradRubin1958}
{Grad}, H., \& {Rubin}, H. 1958, Journal of Nuclear Energy, 7, 284

\bibitem[{{Gu} \& {Qiu}(1980)}]{GuQiu1980}
{Gu}, Y.~N., \& {Qiu}, N.~X. 1980, Acta Physica Sinica, 29, 1367

\bibitem[{{Guo} {et~al.}(2021){Guo}, {Li}, {Daughton}, {Li}, {Kilian}, {Liu},
  {Zhang}, \& {Zhang}}]{GuoEA2021}
{Guo}, F., {Li}, X., {Daughton}, W., {et~al.} 2021, \apj, 919, 111

\bibitem[{{Guo} {et~al.}(2015){Guo}, {Liu}, {Daughton}, \& {Li}}]{GuoEA2015}
{Guo}, F., {Liu}, Y.-H., {Daughton}, W., \& {Li}, H. 2015, \apj, 806, 167

\bibitem[{Haller(2000)}]{Haller2000}
Haller, G. 2000, Chaos: An Interdisciplinary Journal of Nonlinear Science, 10,
  99

\bibitem[{Haller(2005)}]{Haller2005}
---. 2005, Journal of Fluid Mechanics, 525, 1–26

\bibitem[{Haller(2011)}]{Haller2011}
---. 2011, Physica D: Nonlinear Phenomena, 240, 574

\bibitem[{Haller(2015)}]{Haller2015}
---. 2015, Annual Review of Fluid Mechanics, 47, 137

\bibitem[{Henri {et~al.}(2012)Henri, Califano, Faganello, \&
  Pegoraro}]{HenriEA2012}
Henri, P., Califano, F., Faganello, M., \& Pegoraro, F. 2012, Physics of
  Plasmas, 19, 072908

\bibitem[{{Hidalgo} {et~al.}(2002){Hidalgo}, {Nieves-Chinchilla}, \&
  {Cid}}]{HidalgoEA2002}
{Hidalgo}, M.~A., {Nieves-Chinchilla}, T., \& {Cid}, C. 2002, \grl, 29, 1637

\bibitem[{{Hood} {et~al.}(2009){Hood}, {Browning}, \& {van der
  Linden}}]{HoodEA2009}
{Hood}, A.~W., {Browning}, P.~K., \& {van der Linden}, R.~A.~M. 2009, \aap,
  506, 913

\bibitem[{{Hood} {et~al.}(2016){Hood}, {Cargill}, {Browning}, \&
  {Tam}}]{HoodEA2016}
{Hood}, A.~W., {Cargill}, P.~J., {Browning}, P.~K., \& {Tam}, K.~V. 2016, \apj,
  817, 5

\bibitem[{{Hu} {et~al.}(2014){Hu}, {Qiu}, {Dasgupta}, {Khare}, \&
  {Webb}}]{HuEA2014}
{Hu}, Q., {Qiu}, J., {Dasgupta}, B., {Khare}, A., \& {Webb}, G.~M. 2014, \apj,
  793, 53

\bibitem[{Hu {et~al.}(2015)Hu, Qiu, \& Krucker}]{HuEA2015}
Hu, Q., Qiu, J., \& Krucker, S. 2015, Journal of Geophysical Research, 120,
  5266

\bibitem[{Hu \& Li(2002)}]{HuLi2002}
Hu, Y.~Q., \& Li, L. 2002, Journal of Plasma Physics, 67, 139–147

\bibitem[{{Huang} \& {Bhattacharjee}(2010)}]{HuangBhattacharjee2010}
{Huang}, Y.-M., \& {Bhattacharjee}, A. 2010, Physics of Plasmas, 17, 062104

\bibitem[{{Huang} \& {Bhattacharjee}(2012)}]{HuangBhattacharjee2012}
---. 2012, PRL, 109, 265002

\bibitem[{{Huang} \& {Bhattacharjee}(2016)}]{HuangBhattacharjee2016}
---. 2016, \apj, 818, 20

\bibitem[{{Huang} {et~al.}(2014){Huang}, {Bhattacharjee}, \&
  {Boozer}}]{HuangEA2014}
{Huang}, Y.-M., {Bhattacharjee}, A., \& {Boozer}, A.~H. 2014, The Astrophysical
  Journal, 793, 106

\bibitem[{{Huang} {et~al.}(2011){Huang}, {Bhattacharjee}, \&
  {Sullivan}}]{HuangEA2011}
{Huang}, Y.-M., {Bhattacharjee}, A., \& {Sullivan}, B.~P. 2011, Physics of
  Plasmas, 18, 072109

\bibitem[{{Huang} {et~al.}(2017){Huang}, {Comisso}, \&
  {Bhattacharjee}}]{HuangEA2017}
{Huang}, Y.-M., {Comisso}, L., \& {Bhattacharjee}, A. 2017, APJ, 849, 75

\bibitem[{{Huang} {et~al.}(2019){Huang}, {Comisso}, \&
  {Bhattacharjee}}]{HuangEA2019}
---. 2019, Physics of Plasmas, 26, 092112

\bibitem[{{Hudson} \& {Breslau}(2008)}]{HudsonBreslau2008}
{Hudson}, S.~R., \& {Breslau}, J. 2008, \prl, 100, 095001

\bibitem[{Hudson \& Suzuki(2014)}]{HudsonSuzuki2014}
Hudson, S.~R., \& Suzuki, Y. 2014, Physics of Plasmas, 21, 102505

\bibitem[{{Jafari} {et~al.}(2018){Jafari}, {Vishniac}, {Kowal}, \&
  {Lazarian}}]{JafariEA2018}
{Jafari}, A., {Vishniac}, E.~T., {Kowal}, G., \& {Lazarian}, A. 2018, APJ, 860,
  52

\bibitem[{{Ji} {et~al.}(2022){Ji}, {Daughton}, {Jara-Almonte}, {Le}, {Stanier},
  \& {Yoo}}]{JiEA2022}
{Ji}, H., {Daughton}, W., {Jara-Almonte}, J., {et~al.} 2022, Nature Reviews
  Physics, 4, 263

\bibitem[{Jiang {et~al.}(2019)Jiang, Duan, Feng, Zou, Zuo, \&
  Wang}]{JiangEA2019}
Jiang, C., Duan, A., Feng, X., {et~al.} 2019, Frontiers in Astronomy and Space
  Sciences, 6, 63

\bibitem[{Kantz \& Schreiber(2003)}]{KantzSchreiber2003}
Kantz, H., \& Schreiber, T. 2003, Nonlinear Time Series Analysis, 2nd edn.
  (Cambridge University Press), doi:10.1017/CBO9780511755798

\bibitem[{{Karlick{\'y}} {et~al.}(2012){Karlick{\'y}}, {B{\'a}rta}, \&
  {Nickeler}}]{KarlickyEA2012}
{Karlick{\'y}}, M., {B{\'a}rta}, M., \& {Nickeler}, D. 2012, AAP, 541, A86

\bibitem[{{Karpen} {et~al.}(2012){Karpen}, {Antiochos}, \&
  {DeVore}}]{KarpenEA2012}
{Karpen}, J.~T., {Antiochos}, S.~K., \& {DeVore}, C.~R. 2012, \apj, 760, 81

\bibitem[{Kida \& Orszag(1992)}]{KidaOrszag1992}
Kida, S., \& Orszag, S.~A. 1992, Journal of Scientific Computing, 7, 1

\bibitem[{{Klimchuk}(2015)}]{Klimchuk2015}
{Klimchuk}, J.~A. 2015, Philosophical Transactions of the Royal Society of
  London Series A, 373, 20140256

\bibitem[{{Kowal} {et~al.}(2017){Kowal}, {Falceta-Gon{\c{c}}alves}, {Lazarian},
  \& {Vishniac}}]{KowalEA2017}
{Kowal}, G., {Falceta-Gon{\c{c}}alves}, D.~A., {Lazarian}, A., \& {Vishniac},
  E.~T. 2017, \apj, 838, 91

\bibitem[{{Kowal} {et~al.}(2020){Kowal}, {Falceta-Gon{\c{c}}alves}, {Lazarian},
  \& {Vishniac}}]{KowalEA2020}
---. 2020, \apj, 892, 50

\bibitem[{{Kowal} {et~al.}(2009){Kowal}, {Lazarian}, {Vishniac}, \&
  {Otmianowska-Mazur}}]{KowalEA2009}
{Kowal}, G., {Lazarian}, A., {Vishniac}, E.~T., \& {Otmianowska-Mazur}, K.
  2009, APJ, 700, 63

\bibitem[{{Kowal} {et~al.}(2012){Kowal}, {Lazarian}, {Vishniac}, \&
  {Otmianowska-Mazur}}]{KowalEA2012}
---. 2012, Nonlinear Processes in Geophysics, 19, 297

\bibitem[{Lapenta \& Bettarini(2011)}]{LapentaBettarini2011}
Lapenta, G., \& Bettarini, L. 2011, {EPL} (Europhysics Letters), 93, 65001

\bibitem[{{Lapenta} \& {Lazarian}(2012)}]{LapentaLazarian2012}
{Lapenta}, G., \& {Lazarian}, A. 2012, Nonlinear Processes in Geophysics, 19,
  251

\bibitem[{Lazarian {et~al.}(2015)Lazarian, Eyink, Vishniac, \&
  Kowal}]{LazarianEA2015}
Lazarian, A., Eyink, G., Vishniac, E., \& Kowal, G. 2015, Philosophical
  Transactions of the Royal Society A: Mathematical, Physical and Engineering
  Sciences, 373, 20140144

\bibitem[{Lazarian {et~al.}(2020)Lazarian, Eyink, Jafari, Kowal, Li, Xu, \&
  Vishniac}]{LazarianEA2020}
Lazarian, A., Eyink, G.~L., Jafari, A., {et~al.} 2020, Physics of Plasmas, 27,
  012305

\bibitem[{Lazarian \& Vishniac(2009)}]{LazarianVishniac2009}
Lazarian, A., \& Vishniac, E. 2009, Revista Mexicana de Astronomía y
  Astrofísica : Universidad Nacional Autónoma de México. Instituto de
  Astronomía, 36

\bibitem[{{Lazarian} \& {Vishniac}(1999)}]{LazarianVishniac1999}
{Lazarian}, A., \& {Vishniac}, E.~T. 1999, \apj, 517, 700

\bibitem[{{Le} {et~al.}(2018){Le}, {Daughton}, {Ohia}, {Chen}, {Liu}, {Wang},
  {Nystrom}, \& {Bird}}]{LeEA2018}
{Le}, A., {Daughton}, W., {Ohia}, O., {et~al.} 2018, Physics of Plasmas, 25,
  062103

\bibitem[{Leake {et~al.}(2020)Leake, Daldorff, \& Klimchuk}]{LeakeEA2020}
Leake, J.~E., Daldorff, L. K.~S., \& Klimchuk, J.~A. 2020, The Astrophysical
  Journal, 891, 62

\bibitem[{{Li} {et~al.}(2019){Li}, {Guo}, {Li}, {Stanier}, \&
  {Kilian}}]{LiEA2019}
{Li}, X., {Guo}, F., {Li}, H., {Stanier}, A., \& {Kilian}, P. 2019, \apj, 884,
  118

\bibitem[{Lichtenberg {et~al.}(1992)Lichtenberg, Lieberman, John, \&
  Marsden}]{LichtenbergLieberman1992}
Lichtenberg, A., Lieberman, M., John, F., \& Marsden, J. 1992, Regular and
  Chaotic Dynamics, Applied Mathematical Sciences (Springer)

\bibitem[{{Lingam} \& {Comisso}(2018)}]{LingamComisso2018}
{Lingam}, M., \& {Comisso}, L. 2018, Physics of Plasmas, 25, 012114

\bibitem[{{Linton} {et~al.}(1996){Linton}, {Longcope}, \&
  {Fisher}}]{LintonEA1996}
{Linton}, M.~G., {Longcope}, D.~W., \& {Fisher}, G.~H. 1996, \apj, 469, 954

\bibitem[{{Liu} {et~al.}(2016){Liu}, {Kliem}, {Titov}, {Chen}, {Wang}, {Wang},
  {Liu}, {Xu}, \& {Wiegelmann}}]{LiuEA2016}
{Liu}, R., {Kliem}, B., {Titov}, V.~S., {et~al.} 2016, \apj, 818, 148

\bibitem[{Liu {et~al.}(2011)Liu, Li, Yin, Albright, Bowers, \&
  Liang}]{LiuEA2011}
Liu, W., Li, H., Yin, L., {et~al.} 2011, Physics of Plasmas, 18, 052105

\bibitem[{Liu {et~al.}(2018)Liu, Wilson, Green, \& Hughes}]{LiuEA2018}
Liu, Y., Wilson, C., Green, M., \& Hughes, C. 2018, Journal of Geophysical
  Research: Oceans, 123, doi:10.1002/2017JC013390

\bibitem[{Liu {et~al.}(2013)Liu, Daughton, Karimabadi, Li, \&
  Roytershteyn}]{LiuEA2013}
Liu, Y.-H., Daughton, W., Karimabadi, H., Li, H., \& Roytershteyn, V. 2013,
  Phys. Rev. Lett., 110, 265004

\bibitem[{Loureiro {et~al.}(2009)Loureiro, Uzdensky, Schekochihin, Cowley, \&
  Yousef}]{LoureiroEA2009}
Loureiro, N.~F., Uzdensky, D.~A., Schekochihin, A.~A., Cowley, S.~C., \&
  Yousef, T.~A. 2009, Monthly Notices of the Royal Astronomical Society:
  Letters, 399, L146

\bibitem[{Lundquist(1951)}]{Lundquist1951}
Lundquist, S. 1951, Phys. Rev., 83, 307

\bibitem[{MacKay {et~al.}(1984)MacKay, Meiss, \& Percival}]{MacKayEA1984}
MacKay, R.~S., Meiss, J.~D., \& Percival, I.~C. 1984, Phys. Rev. Lett., 52, 697

\bibitem[{{Madrid} \& {Mancho}(2009)}]{MadridMancho2009}
{Madrid}, J.~A.~J., \& {Mancho}, A.~M. 2009, Chaos, 19, 013111

\bibitem[{{Matthaeus} \& {Lamkin}(1985)}]{MatthaeusLamkin1985}
{Matthaeus}, W.~H., \& {Lamkin}, S.~L. 1985, Physics of Fluids, 28, 303

\bibitem[{{Matthaeus} \& {Lamkin}(1986)}]{MatthaeusLamkin1986}
---. 1986, Physics of Fluids, 29, 2513

\bibitem[{{Mentink} {et~al.}(2005){Mentink}, {Bergmans}, {Kamp}, \&
  {Schep}}]{MentinkEA2005}
{Mentink}, J.~H., {Bergmans}, J., {Kamp}, L.~P.~J., \& {Schep}, T.~J. 2005,
  Physics of Plasmas, 12, 052311

\bibitem[{{Mulligan} \& {Russell}(2001)}]{MulliganRussell2001}
{Mulligan}, T., \& {Russell}, C.~T. 2001, \jgr, 106, 10581

\bibitem[{{Nakamura} {et~al.}(2013){Nakamura}, {Daughton}, {Karimabadi}, \&
  {Eriksson}}]{NakamuraEA2013}
{Nakamura}, T.~K.~M., {Daughton}, W., {Karimabadi}, H., \& {Eriksson}, S. 2013,
  Journal of Geophysical Research (Space Physics), 118, 5742

\bibitem[{{Nakamura} {et~al.}(2017){Nakamura}, {Hasegawa}, {Daughton},
  {Eriksson}, {Li}, \& {Nakamura}}]{NakamuraEA2017}
{Nakamura}, T.~K.~M., {Hasegawa}, H., {Daughton}, W., {et~al.} 2017, Nature
  Communications, 8, 1582

\bibitem[{{Nieves-Chinchilla} {et~al.}(2018){Nieves-Chinchilla}, {Linton},
  {Hidalgo}, \& {Vourlidas}}]{Nieves-ChinchillaEA2018}
{Nieves-Chinchilla}, T., {Linton}, M.~G., {Hidalgo}, M.~A., \& {Vourlidas}, A.
  2018, \apj, 861, 139

\bibitem[{Nolan {et~al.}(2020)Nolan, Serra, \& Ross}]{NolanEA2020}
Nolan, P.~J., Serra, M., \& Ross, S.~D. 2020, Nonlinear Dynamics, 100, 3825

\bibitem[{{Oishi} {et~al.}(2015){Oishi}, {Mac Low}, {Collins}, \&
  {Tamura}}]{OishiEA2015}
{Oishi}, J.~S., {Mac Low}, M.-M., {Collins}, D.~C., \& {Tamura}, M. 2015,
  \apjl, 806, L12

\bibitem[{Onu {et~al.}(2015)Onu, Huhn, \& Haller}]{OnuEA2015}
Onu, K., Huhn, F., \& Haller, G. 2015, Journal of Computational Science, 7, 26

\bibitem[{{Parker}(1957)}]{Parker1957}
{Parker}, E.~N. 1957, JGR, 62, 509

\bibitem[{Parnell {et~al.}(2010)Parnell, Haynes, \& Galsgaard}]{ParnellEA2010}
Parnell, C.~E., Haynes, A.~L., \& Galsgaard, K. 2010, Journal of Geophysical
  Research, 115

\bibitem[{{Pegoraro} {et~al.}(2019){Pegoraro}, {Bonfiglio}, {Cappello},
  {Giannatale}, {Falessi}, {Grasso}, \& {Veranda}}]{PegoraroEA2019}
{Pegoraro}, F., {Bonfiglio}, D., {Cappello}, S., {et~al.} 2019, Plasma Physics
  and Controlled Fusion, 61, 044003

\bibitem[{{Polymilis} {et~al.}(2003){Polymilis}, {Servizi}, {Skokos},
  {Turchetti}, \& {Vrahatis}}]{PolymilisEA2003}
{Polymilis}, C., {Servizi}, G., {Skokos}, C., {Turchetti}, G., \& {Vrahatis},
  M.~N. 2003, Chaos, 13, 94

\bibitem[{Pontin(2011)}]{Pontin2011}
Pontin, D. 2011, Advances in Space Research, 47, 1508

\bibitem[{{Pontin} {et~al.}(2016){Pontin}, {Candelaresi}, {Russell}, \&
  {Hornig}}]{PontinEA2016}
{Pontin}, D., {Candelaresi}, S., {Russell}, A., \& {Hornig}, G. 2016, Plasma
  Physics and Controlled Fusion, 58, doi:10.1088/0741-3335/58/5/054008

\bibitem[{Pontin {et~al.}(2013)Pontin, Wilmot-Smith, \& Hornig}]{PontinEA2013}
Pontin, D., Wilmot-Smith, A., \& Hornig, G. 2013, Procedia IUTAM, 9, 110–120

\bibitem[{{Pontin} \& {Hornig}(2015)}]{PontinHornig2015}
{Pontin}, D.~I., \& {Hornig}, G. 2015, \apj, 805, 47

\bibitem[{{Pontin} {et~al.}(2017){Pontin}, {Janvier}, {Tiwari}, {Galsgaard},
  {Winebarger}, \& {Cirtain}}]{PontinEA2017}
{Pontin}, D.~I., {Janvier}, M., {Tiwari}, S.~K., {et~al.} 2017, \apj, 837, 108

\bibitem[{{Pontin} {et~al.}(2011){Pontin}, {Wilmot-Smith}, {Hornig}, \&
  {Galsgaard}}]{PontinEA2011}
{Pontin}, D.~I., {Wilmot-Smith}, A.~L., {Hornig}, G., \& {Galsgaard}, K. 2011,
  A\&A, 525, A57

\bibitem[{{Potter} {et~al.}(2019){Potter}, {Browning}, \&
  {Gordovskyy}}]{PotterEA2019}
{Potter}, M.~A., {Browning}, P.~K., \& {Gordovskyy}, M. 2019, AAP, 623, A15

\bibitem[{Priest \& Forbes(2000)}]{PriestForbes2000}
Priest, E., \& Forbes, T. 2000, Magnetic Reconnection: MHD Theory and
  Applications (Cambridge University Press), doi:10.1017/CBO9780511525087

\bibitem[{Prior \& Yeates(2014)}]{PriorYeates2014}
Prior, C., \& Yeates, A.~R. 2014, The Astrophysical Journal, 787, 100

\bibitem[{Prior \& Yeates(2018)}]{PriorYeates2018}
---. 2018, Phys. Rev. E, 98, 013204

\bibitem[{{Pritchett}(2013)}]{Pritchett2013}
{Pritchett}, P.~L. 2013, Physics of Plasmas, 20, 080703

\bibitem[{{Rempel} {et~al.}(2013){Rempel}, {Chian}, {Brandenburg}, {Mu{\~n}oz},
  \& {Shadden}}]{RempelEA2013}
{Rempel}, E.~L., {Chian}, A. C.~L., {Brandenburg}, A., {Mu{\~n}oz}, P.~R., \&
  {Shadden}, S.~C. 2013, Journal of Fluid Mechanics, 729, 309

\bibitem[{{Restante} {et~al.}(2013){Restante}, {Markidis}, {Lapenta}, \&
  {Intrator}}]{RestanteEA2013}
{Restante}, A.~L., {Markidis}, S., {Lapenta}, G., \& {Intrator}, T. 2013,
  Physics of Plasmas, 20, 082501

\bibitem[{{Rubino} {et~al.}(2015){Rubino}, {Borgogno}, {Veranda}, {Bonfiglio},
  {Cappello}, \& {Grasso}}]{RubinoEA2015}
{Rubino}, G., {Borgogno}, D., {Veranda}, M., {et~al.} 2015, Plasma Physics and
  Controlled Fusion, 57, 085004

\bibitem[{{Scott} {et~al.}(2017){Scott}, {Pontin}, \& {Hornig}}]{ScottEA2017}
{Scott}, R., {Pontin}, D., \& {Hornig}, G. 2017, The Astrophysical Journal,
  848, 117

\bibitem[{Serra \& Haller(2016)}]{SerraHaller2016}
Serra, M., \& Haller, G. 2016, Chaos: An Interdisciplinary Journal of Nonlinear
  Science, 26, 053110

\bibitem[{Servidio {et~al.}(2009)Servidio, Matthaeus, Shay, Cassak, \&
  Dmitruk}]{ServidioEA2009}
Servidio, S., Matthaeus, W.~H., Shay, M.~A., Cassak, P.~A., \& Dmitruk, P.
  2009, Phys. Rev. Lett., 102, 115003

\bibitem[{Servidio {et~al.}(2010)Servidio, Matthaeus, Shay, Dmitruk, Cassak, \&
  Wan}]{ServidioEA2010}
Servidio, S., Matthaeus, W.~H., Shay, M.~A., {et~al.} 2010, Physics of Plasmas,
  17, 032315

\bibitem[{Servidio {et~al.}(2011)Servidio, Dmitruk, Greco, Wan, Donato, Cassak,
  Shay, Carbone, \& Matthaeus}]{ServidioEA2011a}
Servidio, S., Dmitruk, P., Greco, A., {et~al.} 2011, Nonlinear Processes in
  Geophysics, 18, 675

\bibitem[{Shadden {et~al.}(2005)Shadden, Lekien, \& Marsden}]{ShaddenEA2005}
Shadden, S.~C., Lekien, F., \& Marsden, J.~E. 2005, Physica D: Nonlinear
  Phenomena, 212, 271

\bibitem[{{Shafranov}(1966)}]{Shafranov1966}
{Shafranov}, V.~D. 1966, Reviews of Plasma Physics, 2, 103

\bibitem[{{Singh} {et~al.}(2019){Singh}, {Pucci}, {Tenerani}, {Shibata},
  {Hillier}, \& {Velli}}]{SinghEA2019}
{Singh}, K.~A.~P., {Pucci}, F., {Tenerani}, A., {et~al.} 2019, \apj, 881, 52

\bibitem[{Sironi {et~al.}(2016)Sironi, Giannios, \& Petropoulou}]{SironiEA2016}
Sironi, L., Giannios, D., \& Petropoulou, M. 2016, Monthly Notices of the Royal
  Astronomical Society, 462, 48

\bibitem[{{Sisti} {et~al.}(2019){Sisti}, {Faganello}, {Califano}, \&
  {Lavraud}}]{SistiEA2019}
{Sisti}, M., {Faganello}, M., {Califano}, F., \& {Lavraud}, B. 2019, \grl, 46,
  11,597

\bibitem[{Sisti {et~al.}(2021)Sisti, Finelli, Pedrazzi, Faganello, Califano, \&
  Ponti}]{SistiEA2021b}
Sisti, M., Finelli, F., Pedrazzi, G., {et~al.} 2021, The Astrophysical Journal,
  908, 107

\bibitem[{{Sisti, M.} {et~al.}(2021){Sisti, M.}, {Fadanelli, S.}, {Cerri, S.
  S.}, {Faganello, M.}, {Califano, F.}, \& {Agullo, O.}}]{SistiEA2021a}
{Sisti, M.}, {Fadanelli, S.}, {Cerri, S. S.}, {et~al.} 2021, A\&A, 655, A107

\bibitem[{Stanier {et~al.}(2019)Stanier, Daughton, Le, Li, \&
  Bird}]{StanierEA2019}
Stanier, A., Daughton, W., Le, A., Li, X., \& Bird, R. 2019, Physics of
  Plasmas, 26, 072121

\bibitem[{{Strauss}(1988)}]{Strauss1988}
{Strauss}, H.~R. 1988, APJ, 326, 412

\bibitem[{{Striani} {et~al.}(2016){Striani}, {Mignone}, {Vaidya}, {Bodo}, \&
  {Ferrari}}]{StrianiEA2016}
{Striani}, E., {Mignone}, A., {Vaidya}, B., {Bodo}, G., \& {Ferrari}, A. 2016,
  MNRAS, 462, 2970

\bibitem[{{Sturrock}(1966)}]{Sturrock1966}
{Sturrock}, P.~A. 1966, \nat, 211, 695

\bibitem[{{Sweet}(1958)}]{Sweet1958}
{Sweet}, P.~A. 1958, The Observatory, 78, 30

\bibitem[{Tala \& Garbet(2006)}]{TalaEA2006}
Tala, T., \& Garbet, X. 2006, Comptes Rendus Physique, 7, 622, turbulent
  transport in fusion magnetised plasmas

\bibitem[{Temam(1988)}]{Temam1988}
Temam, R. 1988, Infinite-dimensional dynamical systems in mechanics and
  physics, Applied mathematical sciences ; v. 68 (New York London
  Springer-Verlag)

\bibitem[{{Throumoulopoulos} {et~al.}(2009){Throumoulopoulos}, {Tasso}, \&
  {Poulipoulis}}]{ThroumoulopoulosEA2009}
{Throumoulopoulos}, G.~N., {Tasso}, H., \& {Poulipoulis}, G. 2009,
  Magnetohydrodynamic ``cat eyes'' and stabilizing effects of plasma flow,
  arXiv, doi:10.48550/ARXIV.0901.2522

\bibitem[{{Titov} {et~al.}(2002){Titov}, {Hornig}, \&
  {Démoulin}}]{TitovEA2002}
{Titov}, V.~S., {Hornig}, G., \& {Démoulin}, P. 2002, Journal of Geophysical
  Research: Space Physics, 107, SSH 3

\bibitem[{{Tsuji}(1991)}]{Tsuji1991}
{Tsuji}, Y. 1991, Physics of Fluids B, 3, 3379

\bibitem[{Uritsky {et~al.}(2010)Uritsky, Pouquet, Rosenberg, Mininni, \&
  Donovan}]{UritskyEA2010}
Uritsky, V.~M., Pouquet, A., Rosenberg, D., Mininni, P.~D., \& Donovan, E.~F.
  2010, Phys. Rev. E, 82, 056326

\bibitem[{Vandas \& Romashets(2003)}]{VandasRomashets2003}
Vandas, M., \& Romashets, E. 2003, A\&A, 398, 801

\bibitem[{Vandas \& Romashets(2017)}]{VandasRomashets2017}
---. 2017, Solar Physics, 292, doi:10.1007/s11207-017-1149-5

\bibitem[{{Veranda} {et~al.}(2020{\natexlab{a}}){Veranda}, {Bonfiglio},
  {Cappello}, {Chac\`on}, {Escande}, \& {Di Giannatale}}]{VerandaEA2020a}
{Veranda}, M., {Bonfiglio}, D., {Cappello}, S., {et~al.} 2020{\natexlab{a}},
  EPJ Web Conf., 230, 00013

\bibitem[{{Veranda} {et~al.}(2020{\natexlab{b}}){Veranda}, {Bonfiglio},
  {Cappello}, {di Giannatale}, \& {Escande}}]{VerandaEA2020b}
{Veranda}, M., {Bonfiglio}, D., {Cappello}, S., {di Giannatale}, G., \&
  {Escande}, D.~F. 2020{\natexlab{b}}, Nuclear Fusion, 60, 016007

\bibitem[{Vrahatis(1995)}]{Vrahatis1995}
Vrahatis, M.~N. 1995, Journal of Computational Physics, 119, 105

\bibitem[{{Wan} {et~al.}(2013){Wan}, {Matthaeus}, {Servidio}, \&
  {Oughton}}]{WanEA2013}
{Wan}, M., {Matthaeus}, W.~H., {Servidio}, S., \& {Oughton}, S. 2013, Physics
  of Plasmas, 20, 042307

\bibitem[{{Wilmot-Smith} {et~al.}(2010){Wilmot-Smith}, {Pontin}, \&
  {Hornig}}]{Wilmot-SmithEA2010}
{Wilmot-Smith}, A.~L., {Pontin}, D.~I., \& {Hornig}, G. 2010, A\&A, 516, A5

\bibitem[{Wyper {et~al.}(2017)Wyper, Antiochos, \& DeVore}]{WyperEA2017}
Wyper, P.~F., Antiochos, S.~K., \& DeVore, C.~R. 2017, Nature, 544, 452

\bibitem[{Wyper \& Hesse(2015)}]{WyperHesse2015}
Wyper, P.~F., \& Hesse, M. 2015, Physics of Plasmas, 22, 042117

\bibitem[{Wyper \& Pontin(2014)}]{WyperPontin2014}
Wyper, P.~F., \& Pontin, D.~I. 2014, Physics of Plasmas, 21, 082114

\bibitem[{{Yang} {et~al.}(2020){Yang}, {Li}, {Guo}, {Li}, {Li}, {He}, {Zhang},
  \& {Feng}}]{YangEA2020}
{Yang}, L., {Li}, H., {Guo}, F., {et~al.} 2020, The Astrophysical Journal, 901,
  L22

\bibitem[{Yeates \& Hornig(2011{\natexlab{a}})}]{YeatesHornig2011a}
Yeates, A.~R., \& Hornig, G. 2011{\natexlab{a}}, Journal of Physics A:
  Mathematical and Theoretical, 44, 265501

\bibitem[{Yeates \& Hornig(2011{\natexlab{b}})}]{YeatesHornig2011b}
---. 2011{\natexlab{b}}, Physics of Plasmas, 18, 102118

\bibitem[{Yeates \& Hornig(2013)}]{YeatesHornig2013}
---. 2013, Physics of Plasmas, 20, 012102

\bibitem[{{Yeates} {et~al.}(2012){Yeates}, {Hornig}, \&
  {Welsch}}]{YeatesEA2012}
{Yeates}, A.~R., {Hornig}, G., \& {Welsch}, B.~T. 2012, A\&A, 539, A1

\bibitem[{{Yeates} {et~al.}(2010){Yeates}, {Hornig}, \&
  {Wilmot-Smith}}]{YeatesEA2010}
{Yeates}, A.~R., {Hornig}, G., \& {Wilmot-Smith}, A.~L. 2010, PRL, 105, 085002

\bibitem[{{Yeates} {et~al.}(2015){Yeates}, {Russell}, \&
  {Hornig}}]{YeatesEA2015}
{Yeates}, A.~R., {Russell}, A.~J.~B., \& {Hornig}, G. 2015, Proceedings of the
  Royal Society of London Series A, 471, 50012

\bibitem[{Yeates {et~al.}(2021)Yeates, Russell, \& Hornig}]{YeatesEA2021}
Yeates, A.~R., Russell, A. J.~B., \& Hornig, G. 2021, Physics of Plasmas, 28,
  082904

\bibitem[{{Zenitani} \& {Hoshino}(2005)}]{ZenitaniHoshino2005}
{Zenitani}, S., \& {Hoshino}, M. 2005, \prl, 95, 095001

\bibitem[{{Zenitani} \& {Hoshino}(2007)}]{ZenitaniHoshino2007}
---. 2007, \apj, 670, 702

\bibitem[{{Zenitani} \& {Hoshino}(2008)}]{ZenitaniHoshino2008}
---. 2008, \apj, 677, 530

\bibitem[{Zhang {et~al.}(2021)Zhang, Sironi, \& Giannios}]{ZhangEA2021}
Zhang, H., Sironi, L., \& Giannios, D. 2021, The Astrophysical Journal, 922,
  261

\bibitem[{{Zhang} {et~al.}(2021){Zhang}, {Guo}, {Daughton}, {Li}, \&
  {Li}}]{Zhang2EA2021}
{Zhang}, Q., {Guo}, F., {Daughton}, W., {Li}, H., \& {Li}, X. 2021, \prl, 127,
  185101

\bibitem[{Zhdankin {et~al.}(2013)Zhdankin, Uzdensky, Perez, \&
  Boldyrev}]{ZhdankinEA2013}
Zhdankin, V., Uzdensky, D.~A., Perez, J.~C., \& Boldyrev, S. 2013, The
  Astrophysical Journal, 771, 124

\bibitem[{Zweibel \& Yamada(2016)}]{ZweibelYamada2016}
Zweibel, E.~G., \& Yamada, M. 2016, Proceedings of the Royal Society A:
  Mathematical, Physical and Engineering Sciences, 472, 20160479

\end{thebibliography}
\bibliographystyle{aasjournal}



\end{document}